\documentclass[aps,prb,twocolumn,showpacs,a4paper]{revtex4}

\usepackage[T1]{fontenc}
\usepackage[utf8]{inputenc}

\usepackage{graphicx}
\usepackage{amssymb}%
\usepackage{epsfig}
\usepackage{amsmath}
\usepackage{subeqnarray}
\usepackage{ulem}
\usepackage{color}
\usepackage{subfigure}
\graphicspath{{FIG_ARTICLE_PSG_classique/}}

\usepackage{pgf}
\usepackage{tikz}
\usepackage[pdftex,breaklinks=true,colorlinks=true]{hyperref}
\usepackage[active]{srcltx}

\newcommand{\journal}[4]
{\ifthenelse{\equal{#1}{pr}}{
Phys Rev. {\bf #2}, \href{http://link.aps.org/abstract/PR/v#2/e#3}{#3} (#4)}
{\ifthenelse{\equal{#1}{prl}}{
\prl {\bf #2}, \href{http://link.aps.org/abstract/PRL/v#2/e#3}{#3} (#4)}
{\ifthenelse{\equal{#1}{prb}}{
\prb {\bf #2}, \href{http://link.aps.org/abstract/PRB/v#2/e#3}{#3} (#4)}
{\ifthenelse{\equal{#1}{pra}}{
\pra {\bf #2}, \href{http://link.aps.org/abstract/PRA/v#2/e#3}{#3} (#4)}
{\ifthenelse{\equal{#1}{arxiv}}{preprint
\href{http://arxiv.org/abs/#2.#3}{arXiv:#2.#3}}
{\ifthenelse{\equal{#1}{rmp}}{
\rmp {\bf #2}, \href{http://link.aps.org/abstract/RMP/v#2/e#3}{#3} (#4)}
{\ifthenelse{\equal{#1}{cond-mat}}{preprint
\href{http://arxiv.org/abs/cond-mat/#2}{cond-mat/#2}}
{\ifthenelse{\equal{#1}{pre}}{
\pre {\bf #2}, \href{http://link.aps.org/abstract/PRE/v#2/e#3}{#3} (#4)}
{#1 {\bf #2}, #3 (#4)}}}}}}}}}

\begin{document}

\title{Lattice symmetries and regular states in classical  frustrated antiferromagnets}
\author{L.~Messio,$^1$ C.~Lhuillier,$^2$ and G.~Misguich,$^3$}
\affiliation{
$1.$Institute of Theoretical Physics, Ecole Polytechnique F\'ed\'erale de Lausanne, CH-1015 Lausanne, Switzerland.
\\
$2.$Laboratoire de Physique Th\'eorique de la Mati\`ere Condens\'ee, UMR 7600 CNRS, Universit\'e Pierre et Marie Curie, Paris VI, 75252 Paris Cedex 05, France.
\\
$3.$Institut de Physique Th\'eorique, CNRS, URA 2306, CEA, IPhT, 91191 Gif-sur-Yvette, France.}
\date{\today}

\begin{abstract}
We consider some classical and frustrated lattice spin models with global $O(3)$ spin symmetry.
There is no general analytical method to find a ground-state if the spin dependence of the Hamiltonian is more than quadratic (i.e. beyond the Heisenberg model) and/or if the lattice has more than one site per unit cell.
To deal with these situations, we introduce a family of variational spin configurations, dubbed ``regular states'', which respect  all  the lattice symmetries {\it modulo global $O(3)$ spin transformations} (rotations and/or spin flips).
The construction of  these states is explicited through a group theoretical approach, and all the regular states  on the square, triangular, honeycomb and kagome lattices are listed.
Their equal time structure factors and powder-averages are shown for comparison with experiments. 
All the well known N\'eel states with 2 or 3 sublattices  appear amongst regular states on various lattices, but the regular states also encompass exotic non-planar states with cubic, tetrahedral or cuboctahedral geometry of the $T=0$ order parameter.
Whatever the details of the Hamiltonian (with the same symmetry group), a large fraction of these regular states are energetically stationary with respect to small deviations of the spins.
In fact these regular states appear as exact ground-states in a very large range of parameter space of the simplest models that we have been looking at.
As examples, we display the variational phase diagrams of the $J_1$-$J_2$-$J_3$ Heisenberg model on all the previous lattices as well as that of the $J_1$-$J_2$-$K$ ring-exchange model on square and triangular lattices.
\end{abstract}


\maketitle

\section{Introduction}

Finding the ground-state (GS) of an antiferromagnetic {\it quantum} spin model is a notoriously difficult problem.
Moreover, even {\it classical} spin models at zero temperature can be non-trivial to solve, unless one carries some extensive numerical investigation.
In particular there is no general method to determine the lowest energy configurations for a simple Heisenberg $O(3)$ model of the type
\begin{equation}
 E=\sum_{i,j} J(|{\bf x}_i-{\bf x}_j|) \mathbf S_i \cdot \mathbf S_j
\label{eq:Heisenberg}
\end{equation}
if the lattice sites $\{{\bf x}_i\}$ do {\it not} form a Bravais lattice.
It is only if there is a single site per unit cell (Bravais lattice) that one can easily construct some GS\cite{villain77} (see Sec.~\ref{ssec:Espiral}).

Another situation where the classical energy minimization is not simple is that of multiple-spin interactions, where the energy is not quadratic in the spin components.
Finding the GS in presence of interactions of the type $(\mathbf S_i\cdot \mathbf S_j) (\mathbf S_k\cdot \mathbf S_l)$ can be difficult and,  in general, has to be done numerically even on simple lattices with a single site per unit cell.
Such terms arise in the classical limit of ring-exchange interactions.
For instance, the -- apparently simple -- classical model with Heisenberg interactions competing with four-spin ring-exchange on the triangular lattice is not completely solved.\cite{km97}

In this study, we introduce and construct  a family of spin configurations, dubbed ``regular states''.
These configurations are those which respect all  the symmetries of a given lattice  {\it modulo global spin transformations} (rotations and/or spin flips).
This property is obeyed by most usual N\'eel states.
For instance, a two- (resp. three-) sublattice Néel state on the square (resp. triangular) lattice  respects the lattice symmetries  provided each symmetry operation is ``compensated'' by the  appropriate global spin rotation of angle $0$ or $\pi$ (resp. $0$, $\pm2\pi/3$).

By definition, the set of regular states only depends on the symmetries of the model -- the lattice symmetries  and the spin  symmetries -- and therefore does not depend
on the strength of the different interactions ($J({|\bf x|})$ in the example of Eq.~(\ref{eq:Heisenberg})).
These states comprise the well-known structures, like the two and three sublattice N\'eel states mentioned above, but also some new states, like non-planar structures on the kagome lattice that will be discussed in Sec.~\ref{ssec:kag}.

The reason why these states are interesting for the study of frustrated antiferromagnets is that they are good ``variational candidates`` to be the ground-state (GS) of many specific models.
In fact, rather surprisingly, we found that these states (together with spiral states) exhaust all the GS in a large range of parameters of the frustrated spin models we have investigated.
For instance, in the case of an Heisenberg model on the kagome lattice (studied in Sec.~\ref{ssec:j1j2j3}) with competing interactions between first, second and third neighbors, some non-planar spin structures (based on cuboctahedron) turn out to be stable phases.
In other words, the set of regular states and spiral states form a  good starting point to determine the phase diagram of a classical $O(3)$ model, without having to resort to lengthy numerical minimizations.\footnote{Once all the regular states have been constructed for given lattice and spin symmetries (using a simple group theoretic construction, as explained in Sec.~\ref{sec:reg_construction}), one can directly compare the energies they have for a given microscopic Hamiltonian.}
In several cases, we even observed that one of the regular states reaches an exact energy lower bound, therefore proving that it is one (maybe not unique) GS of the model.

These states may also be used when analyzing experimental data on magnetic compounds where the lattice structure is known, but where the values (and range) of the magnetic interactions are not known. In such a case, the (equal time) magnetic correlations -- measured by neutron scattering -- can directly be compared to those of the regular states. If these correlations match those of one regular state, this may be used, in turn, to get some information about the couplings.
With this application in mind, we provide the magnetic structure factors of all the regular states we construct and powder-averages of some of them (see App.~\ref{App:powder}).

The organization of the paper is as follows: in Sec.~\ref{sec:notdef} we present the definition of a regular state, a state that weakly breaks the lattice symmetry and all the notations needed for the group theoretical approach.
In Sec.~\ref{ssec:asg}, we explain the algebraic structure of the group of joint space- and spin-transformations that leave a regular spin configuration invariant (Algebraic symmetry group) and then explain how to construct regular states (Sec.~\ref{ssec:comp_states}).
This approach is algebraically very similar to Wen's construction of symmetric spin liquids,\cite{Wen_PSG} but there are also strong differences in the invariance requirements (see App.~\ref{App:PSG_quantique}): whereas the symmetric spin liquids do not break lattice symmetries (they are ``liquids''), our regular states indeed break lattice symmetries but in a ``weak'' way.
These sub-sections are self-contained, but can be skipped by readers interested essentially in the results.
In Sec.~\ref{ssec:ex_tri}, we give an example of such a contruction on the triangular lattice and list the regular states on this lattice.
In sections Sec.~\ref{ssec:kag} and \ref{ssec:hexa} we list the regular states on the kagome and  honeycomb lattices (which have the same algebraic symmetry group as the triangular lattice), and with a minimum of algebra we present the regular states on the square lattice (Sec.~\ref{ssec:square}).
We then show that spiral states can be seen in this picture as regular states with a lattice symmetry group reduced to the translation group (Sec.~\ref{ssec:spiral}).
In Sec.~\ref{sec:geom} we discuss geometrical properties of regular states and the relationship between regular states and representations of the lattice symmetry group.
This section can be skipped by readers more interested in physics than in geometry.
In Sec.~\ref{sec:energetics} we study the energetics of these regular states and therefore their interest for the variational description of the $T=0$ phase diagrams of frustrated spin models.
We first show in Sec.~\ref{ssec:stat} that all regular states which do not belong to a continuous family are energetically stationary with respects to small spin deviations and thus good GS candidates for a large family of Hamiltonians.
After having given a lower bound on the energy of Heisenberg models (Sec.~\ref{ssec:Espiral}), we then show that over a large range of coupling constants the  regular states are indeed exact GSs of the $J_1$-$J_2$-$J_3$ model on the  honeycomb and kagome lattices (Sec.~\ref{ssec:j1j2j3}).
We then display in Sec.~\ref{ssec:MSE} a  variational phase diagram of the  $J_1$-$J_2$-$K$ model on square and triangular lattices.
In Sec.~\ref{ssec:transition} we discuss finite temperature phase transitions: the non planar states are chiral and should give rise to a $T\neq 0$ phase transtion.
Sec.~\ref{sec:CCL} is our conclusion.
Powder-averages of the structure factors of the regular states  on triangular and kagome lattices are displayed in App.~\ref{App:powder}.
Analogies and differences between the present analysis and Wen's analysis of quantum spin models are explained in App.~\ref{App:PSG_quantique}.

\section{Notations and definitions}
\label{sec:notdef}

We will mostly concentrate on Heisenberg-like models where on each lattice site $i$, the spin $\mathbf S_{i}$ is a three component unit vector.
But the concept of regular state can be easily extended to the  general situation where $\mathbf S_i$ belongs to an other manifold ${\mathcal A}$ (as for example for nematic or quadrupolar order parameters).

We note by $S_S$ the group of the ``global spin symmetries'' of the Hamiltonian.
In the  general framework, an element of $S_S$ is
a mapping of ${\mathcal A}$ onto itself which does not change the energy of the spin configurations.
For an Heisenberg model without applied magnetic field, $S_S$ is simply (isomorphic to) the orthogonal
group $O(3)$. In a similar way, we note by $S_L$ the lattice symmetry group of the Hamiltonian.
An element of $S_L$ acts on spin configurations by mapping
the lattice $L$ onto itself and is the identity in the spin space ${\mathcal A}$.

In this paper, we will restrict ourselves to the (rather common) situation where  the full symmetry group $S_H$ of the model is the direct product $S_S\times S_L$.\footnote{A case where $S_H\neq S_S\times S_L$ is the antiferromagnetic square lattice with a site-dependent magnetic field taking two opposite values on each sublattice.
The spin inversion $\mathbf S_i\to -\mathbf S_i$ is not in $S_S$, the translation by one lattice spacing is not in $S_L$, but the composition of both is in $S_H$.
The theory developed in this paper can however be used in this case by replacing $S_L$ by $S_H/S_S$. }

Let $\mathcal G$ be the set
of all the applications from the lattice symmetry group $S_L$ to the spin symmetry group $S_S$.
An element $G$ of $\mathcal G$ associates a spin symmetry $G_X$ to each lattice symmetry $X$:
\begin{equation}
 \begin{array}{cc}
   G : & S_L  \to     S_S \\
       & X    \mapsto G_X
 \end{array}
\end{equation}

We now concentrate on a fixed spin configuration $c$.
We note $H_c$ its stabilizer, that is the subgroup of $S_H$  which elements do not modify $c$.
Its spin symmetry group $H_c^S$ is the group of unbroken spin symmetries: $H_c^S=S_S\cap H_c$.

Definitions:
\begin{itemize} \item {\it A mapping $G\in {\mathcal G}$ is said to be  compatible with a spin configuration $c$ if the composition of an element of $S_L$ with its image by $G$ leaves $c$ unchanged}:
\begin{equation}
 \forall X \in S_L, \;\;\;G_X X\in H_c.
\end{equation}

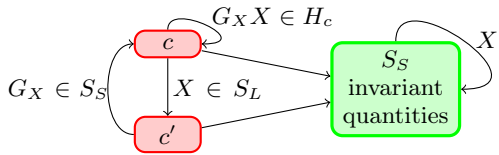
\begin{figure}
\begin{center}
\begin{tikzpicture}[scale=0.3,auto,
block/.style   ={rectangle, draw=red, thick,
               text width=10em, fill=red!20, text centered, rounded corners},
block2/.style   ={rectangle, draw=green, very thick,
               text width=12em, fill=green!20, text centered, rounded corners},
block3/.style   ={rectangle, draw=white, very thick,
               text width=12em, fill=white, text centered, rounded corners}]
\node (Ansatz1) at (0,5) [block,draw,text width=2em] {$c$};
\node (Ansatz2) at (0,1) [block,draw,text width=2em] {$c'$};
\node (Mil) at (2.2,3) [block3,draw,text width=6em]{$X \in S_L$};
\node (Mil) at (-4.8,3) [block3,draw,text width=5em]{$G_X \in S_S$};
\node (PhysState) at (10,3) [block2,draw,text width=4.5em] {$S_S$ invariant quantities};
\node (Mil3) at (14,5.2) [block3,draw,text width=1em]{$X$};
\node (Mil4) at (4.6,6.2) [block3,draw,text width=5em]{$G_X X\in H_c$};
\draw[->] (PhysState) .. controls +(up:5) and +(right:7) .. (PhysState) node [above]{};
\draw[->] (Ansatz1) .. controls +(up:2) and +(right:4) .. (Ansatz1) node [above]{};
\draw[->] (Ansatz1) -- (Ansatz2) node [above left] {};
\draw[->] (Ansatz2) .. controls +(left:3) and +(left:3) .. (Ansatz1) node [below right] {} ;
\draw[->] (Ansatz1) -- (PhysState) ;
\draw[->] (Ansatz2) -- (PhysState) ;
\end{tikzpicture}
\end{center}
\caption{(Color online) A lattice symmetry $X\in S_L$ acts on a spin configuration $c$ to give a new configuration $c'=X c$. If $c$ is regular, there is a spin symmetry $G_X\in S_S$ such that one gets back the initial state: $G_X c'=c$.
}
\label{fig1}
\end{figure}

\item
{\it A configuration $c$ is said to be regular if any lattice symmetry $X\in S_L$
can be ``compensated'' by an appropriate spin symmetry  $G_X \in S_S$, which
means $G_XX |c\rangle=|c\rangle$ (that is $G_XX\in H_c$).}
In other words, {\it $c$ is regular if there exists a mapping $G\in {\mathcal G}$
such that $G$ and $c$ are compatible}.
\end{itemize}
In a regular state, the observables which are invariant under $S_S$ are therefore invariant under all  lattice symmetries.
These definitions are summarized in Fig.~\ref{fig1}.

The simplest regular states are those which are already invariant under lattice symmetries (i.e. $S_L\subset H_c$), without the need to perform any spin symmetry.
This is the case of a ferromagnetic (F) configuration, with all spins oriented in the same way.
But less trivial possibilities exist, as the classical GS of the antiferromagnetic (AF) first neighbor Heisenberg interaction on the square lattice.
This GS possesses two sublattices with opposite spin orientations.
Each lattice symmetry $X$ either conserves the spin orientations, or reverses them, so we can choose as $G_X$ either the identity or the spin inversion $\mathbf S_i\to -\mathbf S_i$.

If the subgroup $H_c^S=S_S\cap H_c$ of unbroken spin symmetries contains more than the identity, there are several elements of $\mathcal G$ compatible with $c$.
For each $X$, they are as many $G_X$ as elements in $H_c^S$.
In the previous example of the GS of the AF square lattice, $H_c^S$ is the set of spin transformations that preserve the two opposite spins orientations: this group is isomorph to $O(2)$.
Beginning with a compatible $G$, each $G_X$ can be composed with an element of $H_c^S$ to give an other compatible element of $\mathcal G$.

To summarize, regular states are not restricted to states strictly respecting the lattice symmetries, but to states that in some way \textit{weakly} respect them.
We will now explain how to construct {\it all} the regular spin configurations on a given lattice.

\section{Construction of regular states}
\label{sec:reg_construction}

To construct the regular states, we proceed in two steps.
{\it In the first step, we fix a given unbroken spin symmetry group $H_c^S$}, and consider the algebraic constraints that the lattice symmetry group $S_L$ imposes
on a mapping $G \in {\mathcal G}$, assuming that some (so far unknown) spin configuration $c$ is compatible with $G$.
These constraints lead to a selection of a subset ${\mathcal G}^A$ of ${\mathcal G}$, composed of the mappings $G$ which are compatible with the lattice symmetries.
For an element $G$ of ${\mathcal G}^A$, the group
\begin{equation}
H^G=\{G_XX,X\in S_L\}\times H_c^S,
\end{equation}
is dubbed  the {\it algebraic symmetry group} associated to $G$.

{\it In the second step}, one determines the configurations (if any) which are compatible with a given algebraic symmetry group.

\subsection{Algebraic symmetry groups}
\label{ssec:asg}

We fix the spin symmetry group $H_c^S$ (to be exhaustive, we will  consecutively consider each possible $H_c^S$).
Let $X$, $Y$ and $Z$, three elements of $S_L$ such that $XY=Z$.
We will see that this algebraic relation imposes
some constraints on the mappings $G$ which are compatible with a spin configuration.
Indeed, we assume that there exists a configuration $c$ compatible with $G$.
Then, $G_ZZ$ and $G_XXG_YY$ are in $H_c$.
This implies that $G_XXG_YYZ^{-1}G_Z^{-1}$ is also in $H_c$.
Elements of $S_L$ and $S_S$ commute, so we have $G_XXG_YYZ^{-1}G_Z^{-1}=G_XG_YG_Z^{-1}$, which is a pure spin transformation.
We deduce that
\begin{equation}
 \forall X,Y \in S_L, \;\;\;G_XG_YG_{(XY)}^{-1}\in H_c^S.
 \label{eq:algebraic_cond}
\end{equation}
If the constraint above is not satisfied, $G$ must be excluded from the set ${\mathcal G}^A$ of the algebraically compatible mappings.
${\mathcal G}^A$ contains only elements of $\mathcal G$ verifying Eq.~(\ref{eq:algebraic_cond}).

\begin{figure}
\begin{center}
 \includegraphics[width=.18\textwidth,trim=2cm 6.5cm 2cm 6cm,clip]{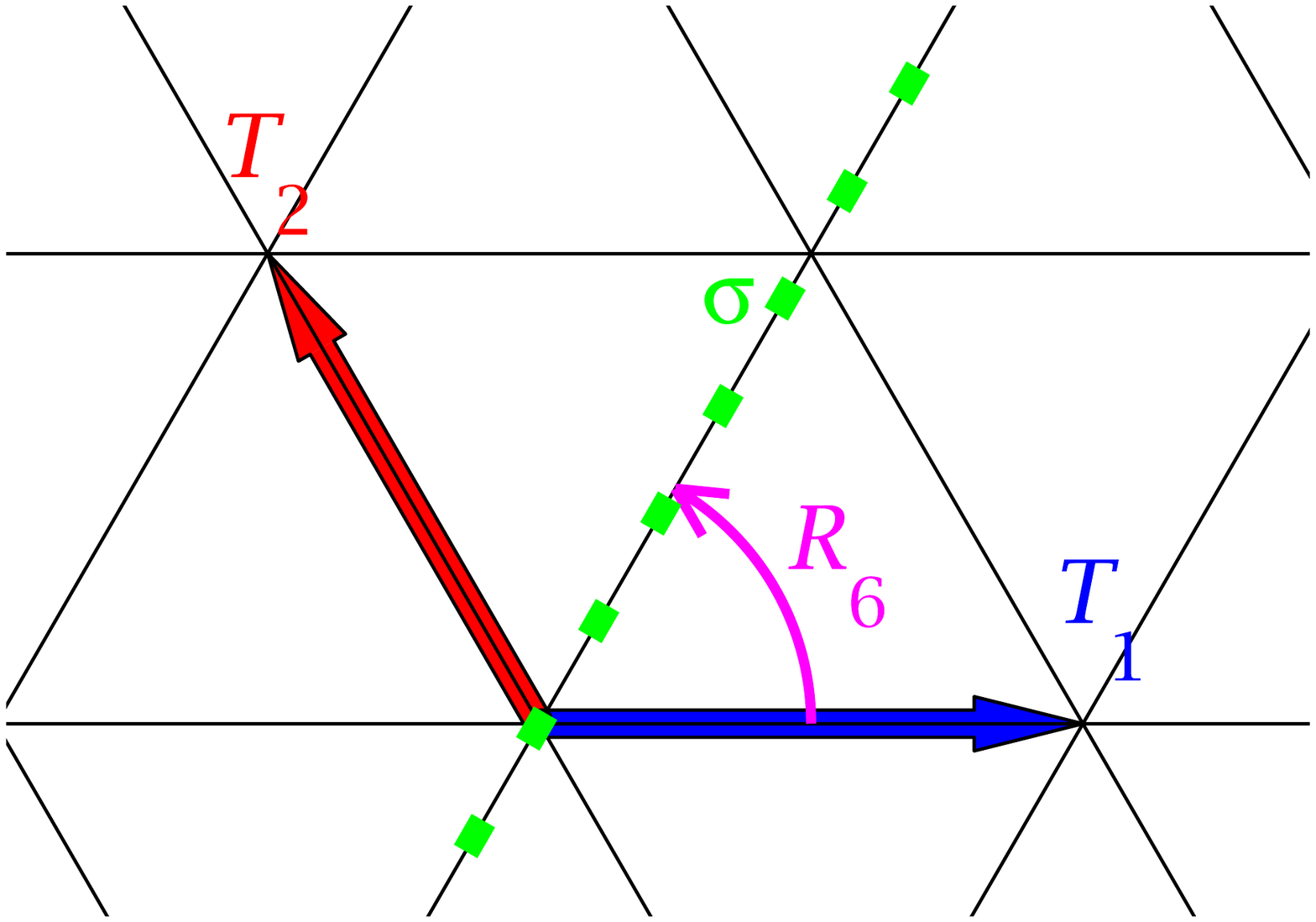}
 \includegraphics[width=.18\textwidth,trim=1cm 6cm 1cm 6cm,clip]{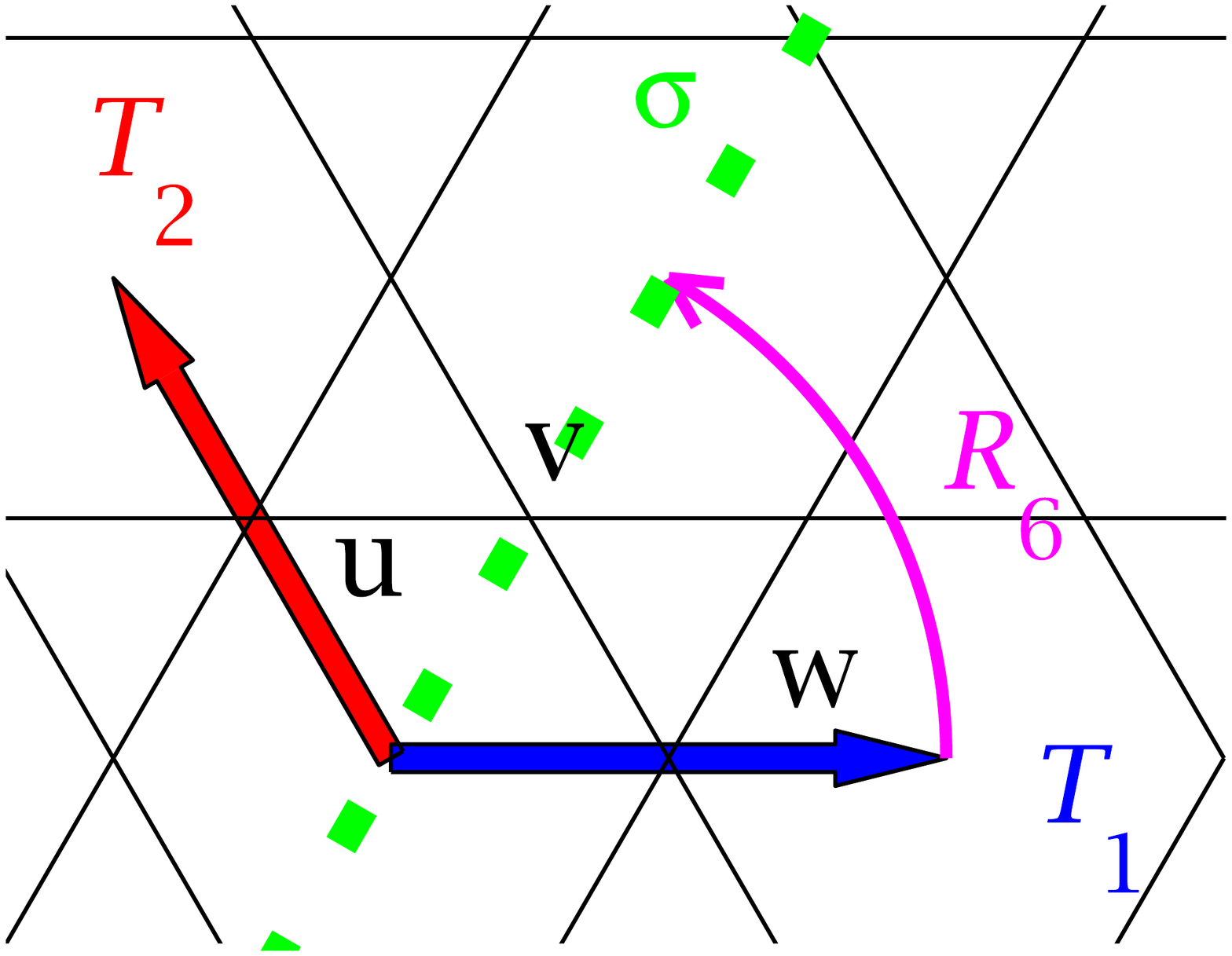}\\
 \includegraphics[width=.18\textwidth,trim=2cm 7cm 2cm 7cm,clip]{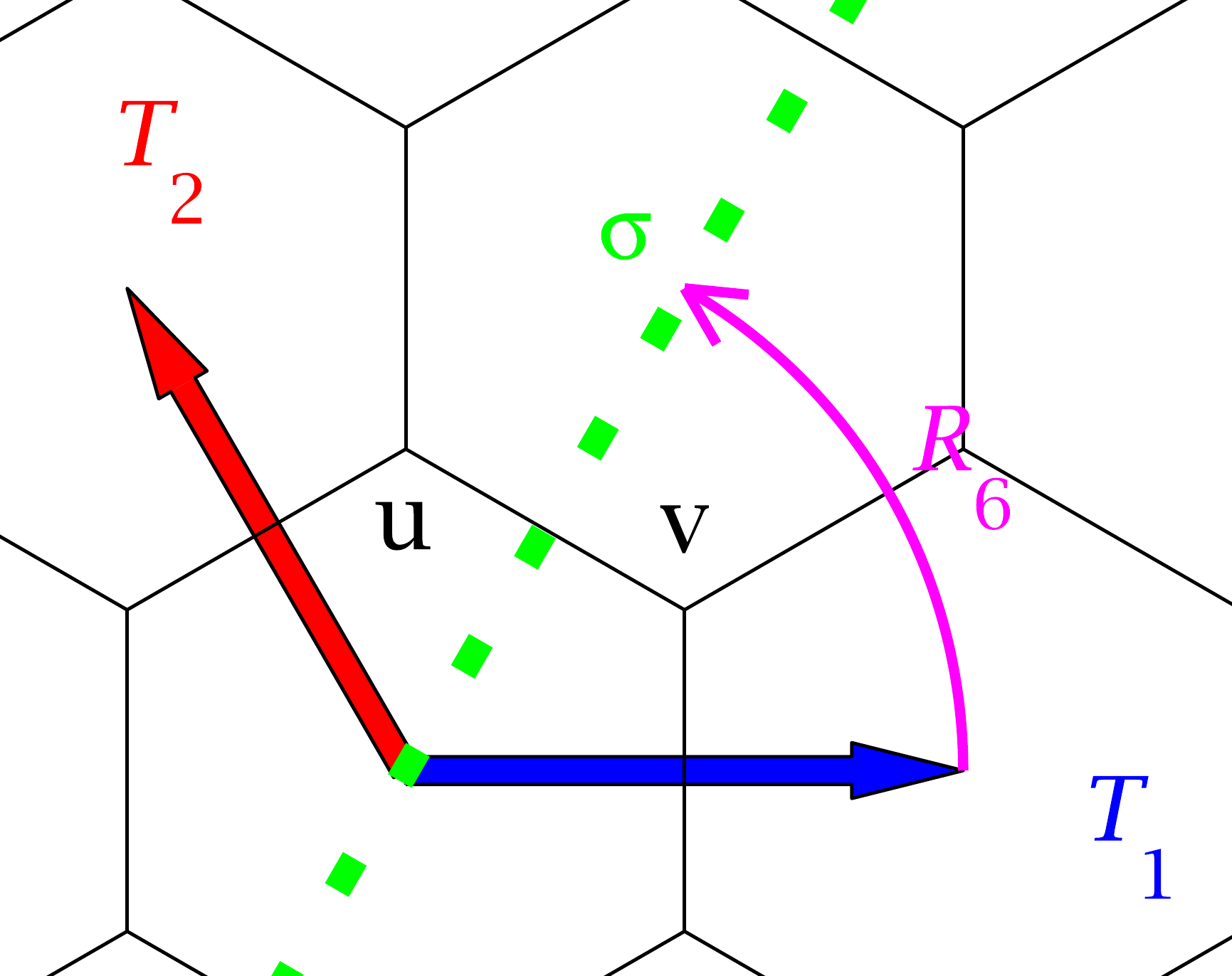}
 \includegraphics[width=.18\textwidth,trim=0cm 5cm 0cm 5cm,clip]{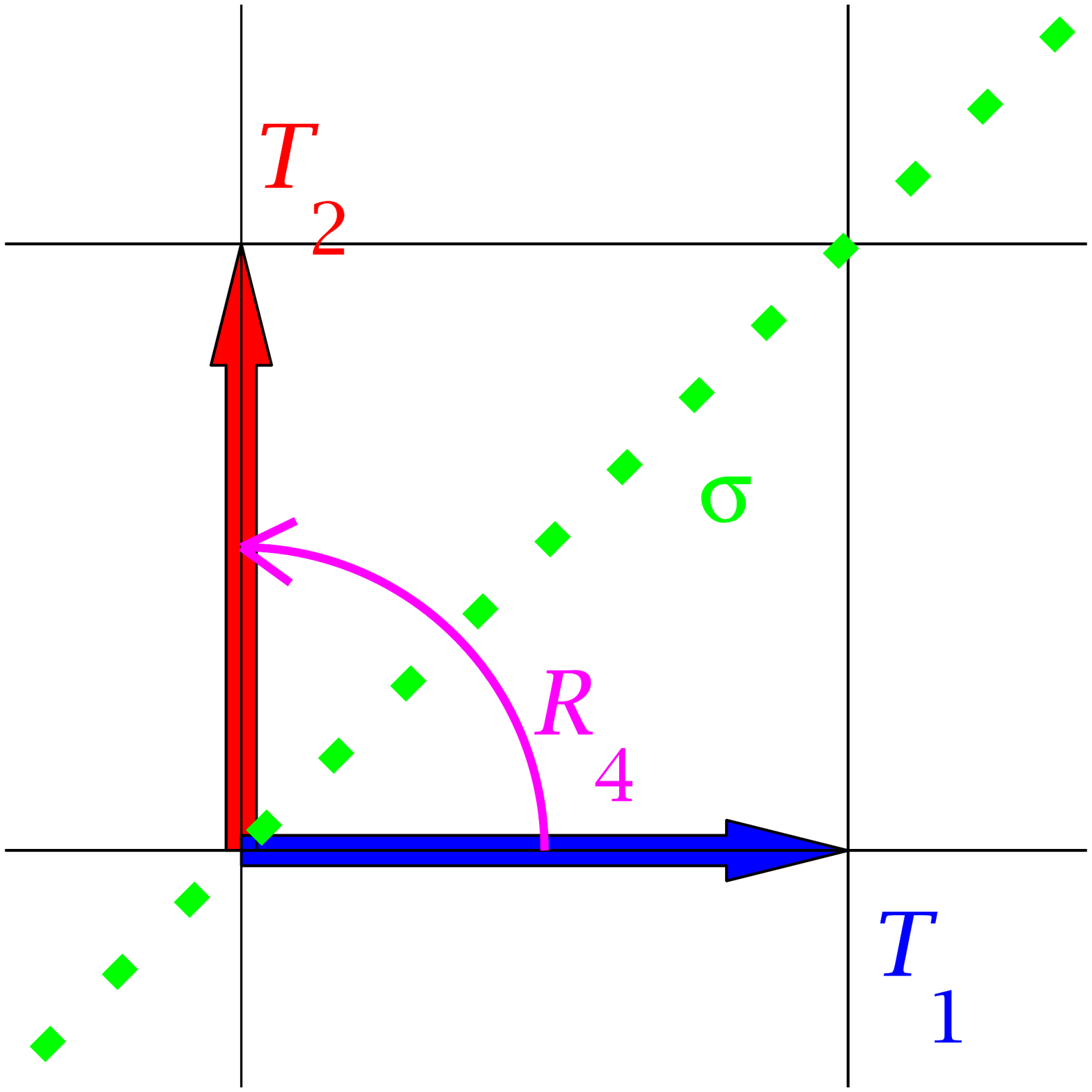}
 \caption{(Colour online) Generators of the lattice symmetries  group $S_L$ for the triangular, kagome, honeycomb and square lattices. For the first three lattices : the two translations $T_1$ and $T_2$ (along the two basis vectors $\mathbf T_1$ and $\mathbf T_2$), the reflexion $\sigma$ an the rotation $R_6$ of angle $\pi/3$.
 For the square lattice, generators of $S_L$ are $T_1$, $T_2$, $\sigma$ an the rotation $R_4$ of angle $\pi/2$. }
 \label{fig:sym_latt}
\end{center}
\end{figure}

Now we illustrate these general considerations using the following example:
$L$ is an infinite triangular lattice and
the spin space $\mathcal A$ is the two-dimensional sphere ${\mathcal S}_2$ (Heisenberg spins).
$S_L$ is generated by two translations $T_1$ and $T_2$ along vectors $\mathbf T_1$ and $\mathbf T_2$, a reflexion $\sigma$ and a rotation $R_6$ of angle $\pi/3$, described in Fig.~\ref{fig:sym_latt} and defined in the $(\mathbf T_1,\mathbf T_2)$ basis as: 
\begin{subeqnarray}
 T_1:(r_1,r_2)&\mapsto&(r_1+1,r_2)\\
 T_2:(r_1,r_2)&\mapsto&(r_1,r_2+1)\\
 \sigma:(r_1,r_2)&\mapsto&(r_2,r_1)\\
 R_6:(r_1,r_2)&\mapsto&(r_1-r_2,r_1).
\end{subeqnarray}

The spin symmetry group $S_S$ is chosen to be $O(3)$ (as for an Heisenberg model).
In such a system, the unbroken symmetry group $H_c^S$ is either isomorph to $\{I\}$, $\mathbb Z_2$ or $O(2)$, depending on the orientations of the spins (non-coplanar, coplanar or colinear respectively).
The non-planar case, $H_c^S=\{I\}$, is the most interesting case and we choose it for this example.
The two other cases can be treated by reducing $\mathcal A$ to the circle ${\mathcal S}_1$ or ${\mathcal S}_0=\{1,-1\}$ ($XY$ or Ising spins) and $S_S$ to $O(2) $ or $O(1)$ in order to have $H_c^S=\{I\}$, which considerably simplifies the calculations.

We assume that a mapping $G$ belongs to ${\mathcal G}^A$ (algebraically compatible).
As $H_c^S=\{I\}$,   Eq.~(\ref{eq:algebraic_cond}) allows to construct
the full mapping $G$ simply from the images of the generators of the lattice symmetry group $S_L$.
As several combinations of generators can produce the same element of $S_L$, the images by $G$ of the $S_L$ generators must satisfy some algebraic relations.
These relations where needed in a similar algebraic study in Ref.~\onlinecite{wangVishwanath2006} and consist in all the relations necessary to put each product of generators  in the form
$\sigma^sR_6^rT_1^{t_1}T_2^{t_2}$, where $s=0,1$, $r=0,1, . . . ,5$ and $t_1$, $t_2\in\mathbb Z$.
These relations are:
\begin{subeqnarray}
 \label{eq:algebraic_system0}
 T_1T_2& =&T_2T_1 \\
 T_1R_6T_2 &=&R_6\\
 R_6T_1T_2 &=&T_2R_6  \\
 T_1\sigma&=&\sigma T_2 \\
 R_6^6  &=&I \\
 \sigma^2&=&I  \\
 R_6\sigma R_6&=&\sigma.
\end{subeqnarray}

From these equations and from Eq.~(\ref{eq:algebraic_cond}) we get:
\begin{subeqnarray}
 \label{eq:algebraic_system}
 G_{T_1}G_{T_2}& =&G_{T_2}G_{T_1} \slabel{eq:algebraic_system_a}\\
 G_{T_1}G_{R_6}G_{T_2} &=&G_{R_6} \\
 G_{R_6}G_{T_1}G_{T_2} &=& G_{T_2}G_{R_6} \slabel{eq:algebraic_system_c}\\
 G_{T_1}G_{\sigma}&=&G_{\sigma} G_{T_2} \slabel{eq:algebraic_system_d}\\
 G_{R_6}^6  &=&I \\
 G_{\sigma^2}&=&I  \\
 G_{R_6}G_{\sigma} G_{R_6}&=&G_{\sigma}.
\end{subeqnarray}

To solve this system, we first remark from Eq.~(\ref{eq:algebraic_system_a}) and (\ref{eq:algebraic_system_d}) that $G_{T_1}$ and $G_{T_2}$
commute and can be obtained from each other by a similarity transformation.
Thus, we are in one of these four cases
\begin{subequations}
 \begin{gather}
  G_{T_1}=G_{T_2}=I, \label{eq:case_a}\\
  \theta_{T_1}=\theta_{T_2}=\pi\textrm{ and }\mathbf n_{T_1}\perp\mathbf n_{T_2}, \label{eq:case_b}\\
  G_{T_1}=G_{T_2}\neq I, \label{eq:case_c}\\
  G_{T_1}=G_{T_2}^{-1}\neq I\textrm{ and }\theta_{T_1}\neq\pi.\label{eq:case_d}
 \end{gather}
\end{subequations}
There, we have labeled the elements of  $O(3)$ by a
rotation of axis $\mathbf n$ and an angle $\theta\in\lbrack 0,\pi\rbrack$ (times a determinant $\varepsilon=\pm1$, not appearing here).
Up to a global similarity relation, we obtain 28 solutions of the system of Eqs.~(\ref{eq:algebraic_system}) in the case of Eq.~(\ref{eq:case_a}), 4 for Eq.~(\ref{eq:case_b}), 8 for Eq.~(\ref{eq:case_c}) and 0 for Eq.~(\ref{eq:case_d}) (it contradicts Eq.~(\ref{eq:algebraic_system_c})).
The 40 solutions are listed below :
\begin{subequations}
\label{eq:solution}
 \begin{gather}
G_{T_1}=G_{T_2}=I,
G_\sigma=\varepsilon_\sigma I,
G_{R_6}=\varepsilon_R I,
\label{eq:solution_a}
\\
G_{T_1}=G_{T_2}=I,
G_\sigma=\varepsilon_\sigma I,
G_{R_6}=\varepsilon_R R_{\mathbf z\pi},
\label{eq:solution_b}
\\
G_{T_1}=G_{T_2}=I,
G_\sigma=\varepsilon_\sigma R_{\mathbf z\pi},
G_{R_6}=\varepsilon_R I,
\label{eq:solution_c}
\\
G_{T_1}=G_{T_2}=I,
G_\sigma=\varepsilon_\sigma R_{\mathbf z\pi},
G_{R_6}=\varepsilon_R R_{\mathbf z\pi},
\label{eq:solution_d}
\\
G_{T_1}=G_{T_2}=I,
G_\sigma=\varepsilon_\sigma R_{\mathbf z\pi},
G_{R_6}=\varepsilon_R R_{\mathbf x\pi},
\label{eq:solution_e}
\\
G_{T_1}=G_{T_2}=I,
G_\sigma=\varepsilon_\sigma R_{\mathbf z\pi},
G_{R_6}=\varepsilon_R R_{\mathbf x\theta},
\label{eq:solution_f}
\\
\begin{array}{ll}
G_{T_1}= R_{\mathbf x\pi},
G_{T_2}= R_{\mathbf y\pi},
G_\sigma=-\varepsilon_\sigma\begin{pmatrix}
0 &  1 & 0\\
 1 & 0 & 0\\
0 & 0 & 1
\end{pmatrix},\\
\qquad
\qquad
\qquad
\qquad
G_{R_6}=\begin{pmatrix}
0 & \varepsilon_R & 0\\
0 & 0 & 1\\
1 & 0 & 0
\end{pmatrix},
\end{array}
\label{eq:solution_g}
\\
G_{T_1}=G_{T_2}=R_{\mathbf z\frac{2\pi}{3}},
G_\sigma=\varepsilon_\sigma I,
G_{R_6}=\varepsilon_R R_{\mathbf x\pi},
\\
G_{T_1}=G_{T_2}= R_{\mathbf z\frac{2\pi}{3}},
G_\sigma=\varepsilon_\sigma R_{\mathbf z\pi},
G_{R_6}=\varepsilon_R R_{\mathbf x\pi}.
\label{eq:solution_h}
 \end{gather}
\end{subequations}
where $\mathbf x\perp \mathbf z$, $\varepsilon_\sigma$, $\varepsilon_R\in\{-1,1\}$ and $\theta\in\left\{ \frac{\pi}{3}, \frac{2\pi}{3}\right\}$.
Each line corresponds to 4 solutions, except Eq.~(\ref{eq:solution_f}) with 8 solutions.
We stress that the algebraic symmetry groups depend on $S_S$, $H_c^S$ and on the algebraic properties of $S_L$, but not directly on the lattice $L$. In particular, different lattices can have the same algebraic symmetry groups.
The results Eqs.~\ref{eq:solution} are exactly the same on a honeycomb or a kagome lattice with symmetries of Fig.~\ref{fig:sym_latt} because the algebraic equations Eqs.~\ref{eq:algebraic_system0} stay the same.

\subsection{Compatible states}
\label{ssec:comp_states}

The second step consists in taking each element of ${\mathcal G}^A$ and finding all the compatible states.
This last step is fully lattice dependent.

To construct a regular state compatible with some mapping $G\in{\mathcal G}^A$, one first chooses the direction of the spin on a site $i$.
Then, by applying all the transformations of $S_L$, we deduce the spin directions on the other sites.
A constraint appears when two different transformations $X$ and $Y$ lead to the same site $X(i)=Y(i)$.
The image spins have to be the same: $G_X(\mathbf S_i)=G_Y(\mathbf S_i)$.
It can either give a constraint on the direction of $\mathbf S_i$, either indicate that no $G-$compatible state exists.

To find these constraints, we divide the lattice sites in orbits under the action of $S_L$ (if all the sites are equivalent, there is a single orbit).
In each orbit, we choose a site $i$.
Each non trivial transformation $X$ that does not displace $i$ gives a constraint: $G_X(\mathbf S_i)=\mathbf S_i$.
For each $G\in{\mathcal G}^A$, the associated regular states are obtained by choosing a site in each orbit, a spin direction respecting the site constraints and then propagating the spin directions through the lattice using the symmetries in $S_L$.

\subsection{Example of regular state construction: the triangular lattice}
\label{ssec:ex_tri}

Let us apply this method to the example of the triangular lattice.
There is a single orbit, and only two distinct and non trivial transformations leave invariant the site of coordinates $(0,0)$ in the $(\mathbf T_1,\mathbf T_2)$ basis (see Fig.~\ref{fig:sym_latt}): $\sigma$ and ${R_6}^n$ for $1\leq n\leq5$, giving the two constraints $G_\sigma(\mathbf S(0,0))=G_{R_6}(\mathbf S(0,0))=\mathbf S(0,0)$.

\begin{figure}
\begin{center}
 \subfigure[\;Ferromagnetic (F) state. $E=6J_1+6J_2+6J_3+42K$.\label{fig:reg_tri_a}]{
   \includegraphics[trim = 15mm 81mm 22mm 81mm, clip,width=.2\textwidth]{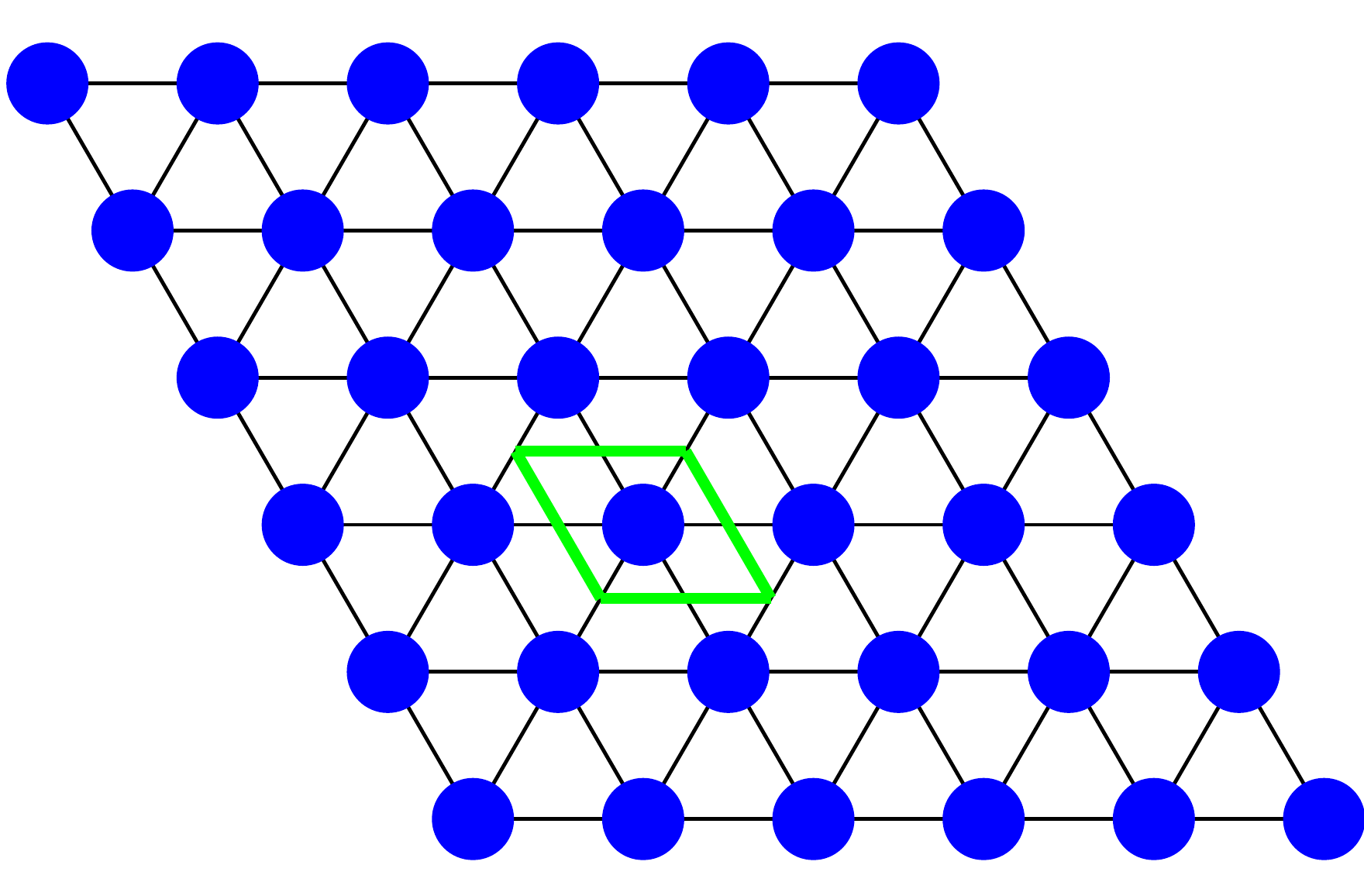}
   \includegraphics[trim = 12mm -70mm 19mm 60mm, clip,width=.08\textwidth]{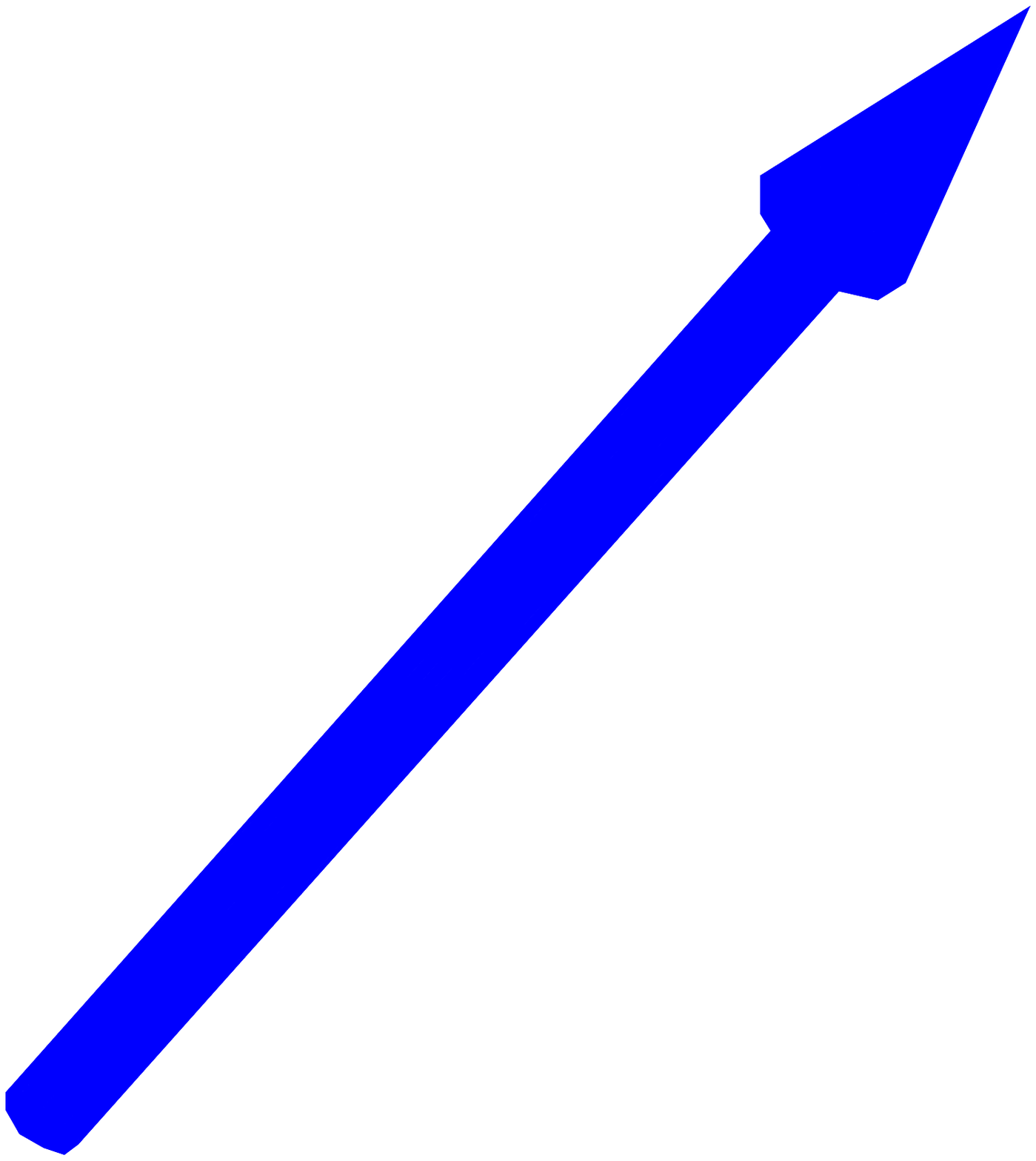}
   \includegraphics[trim = 86mm 63mm 76mm 51mm, clip,width=.14\textwidth]{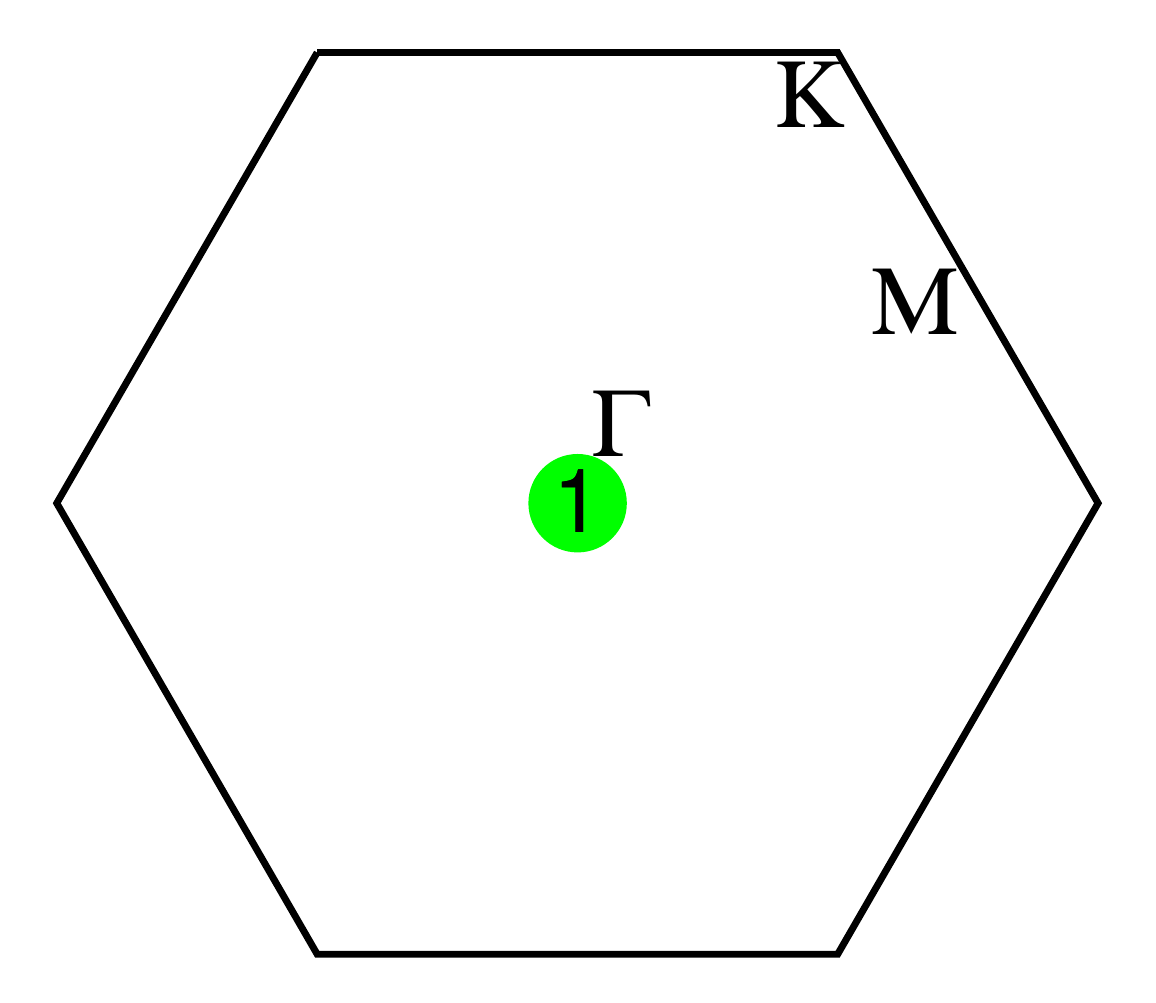}}
 \subfigure[\;Tetrahedral state. $E=-2J_1-2J_2+6J_3-34K/3$.\label{fig:reg_tri_b}]{
   \includegraphics[trim = 15mm 81mm 22mm 81mm, clip,width=.2\textwidth]{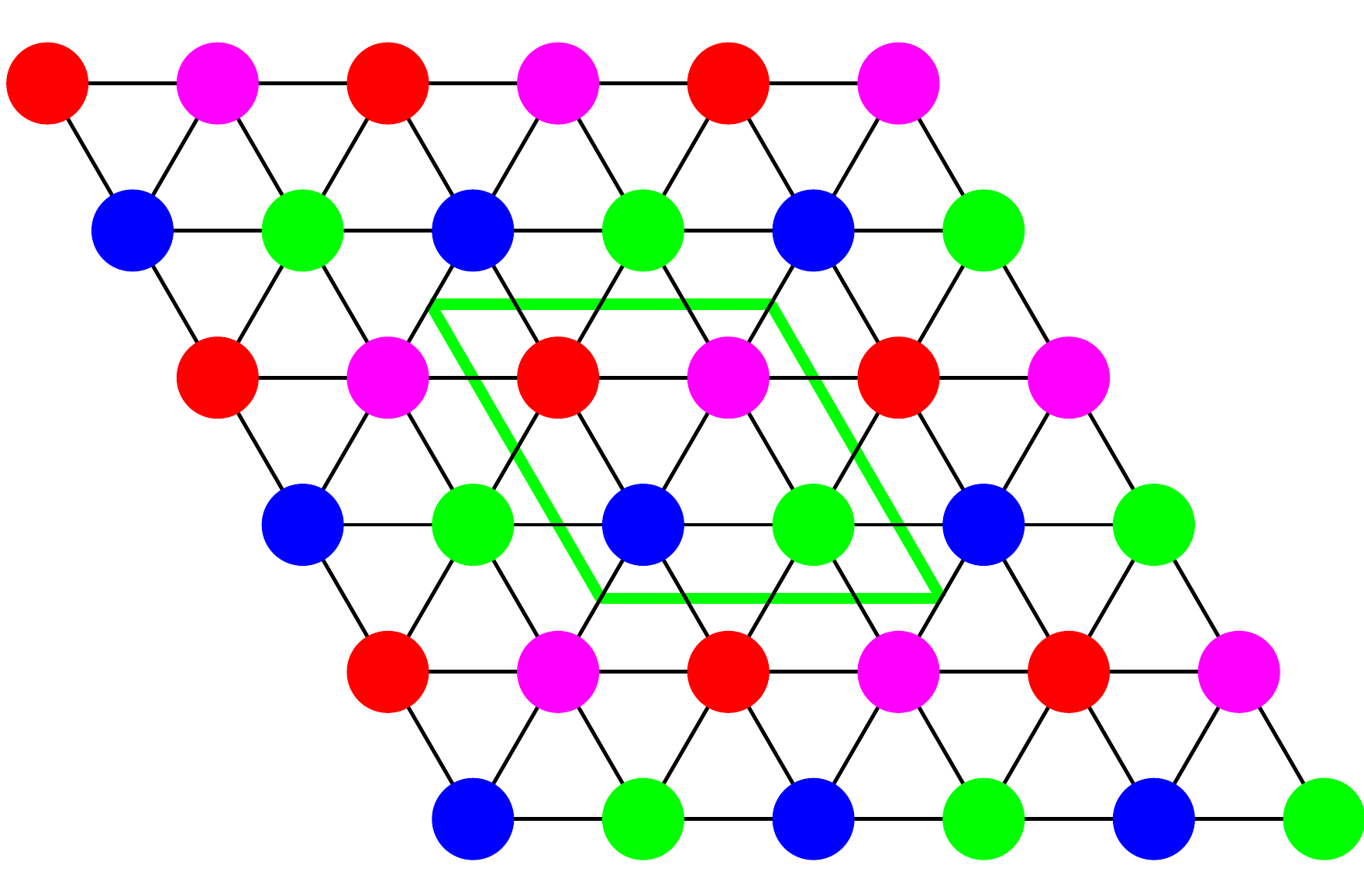}
   \includegraphics[trim = 12mm -70mm 19mm 60mm, clip,width=.08\textwidth]{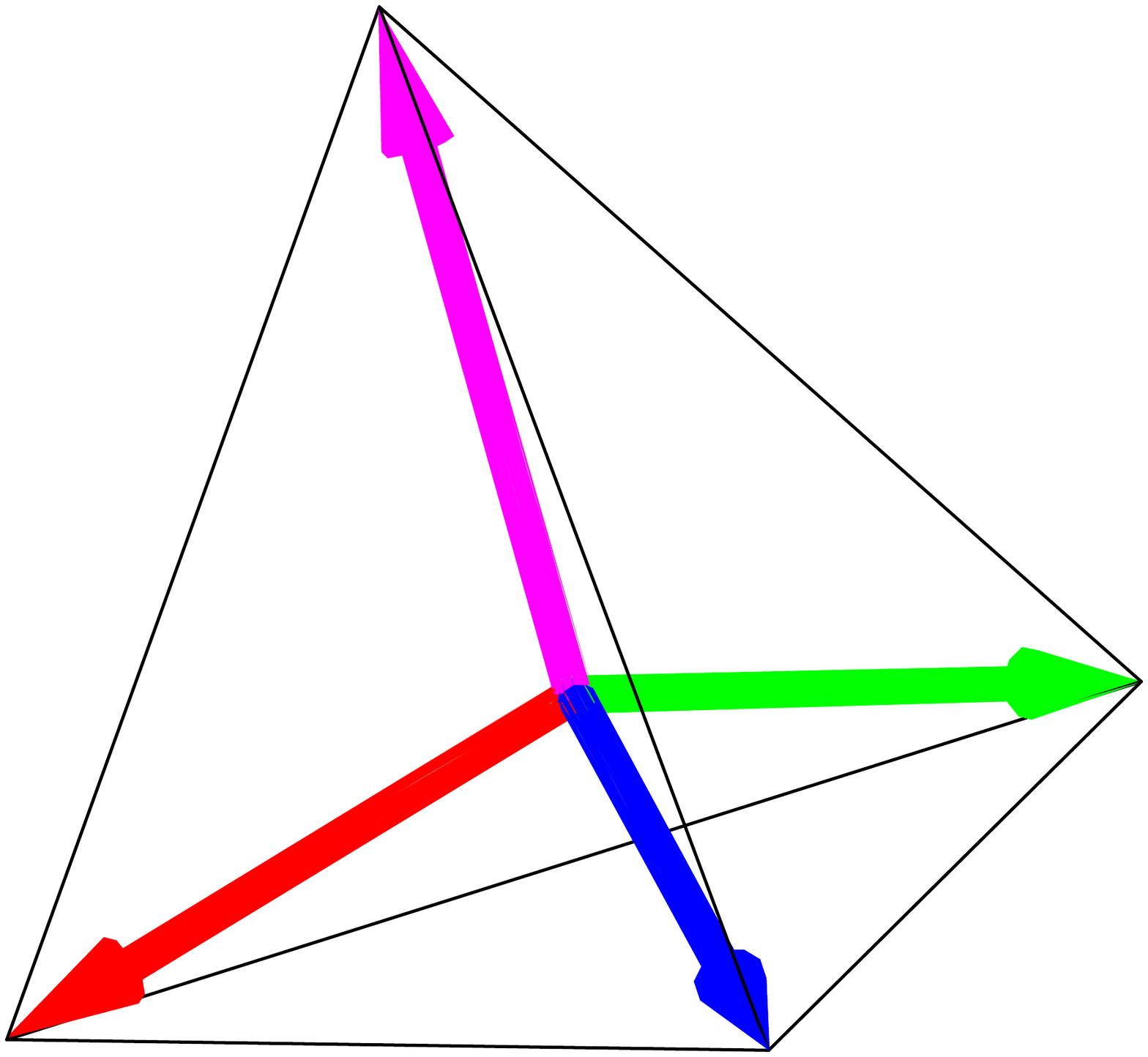}
   \includegraphics[trim = 86mm 63mm 76mm 51mm, clip,width=.14\textwidth]{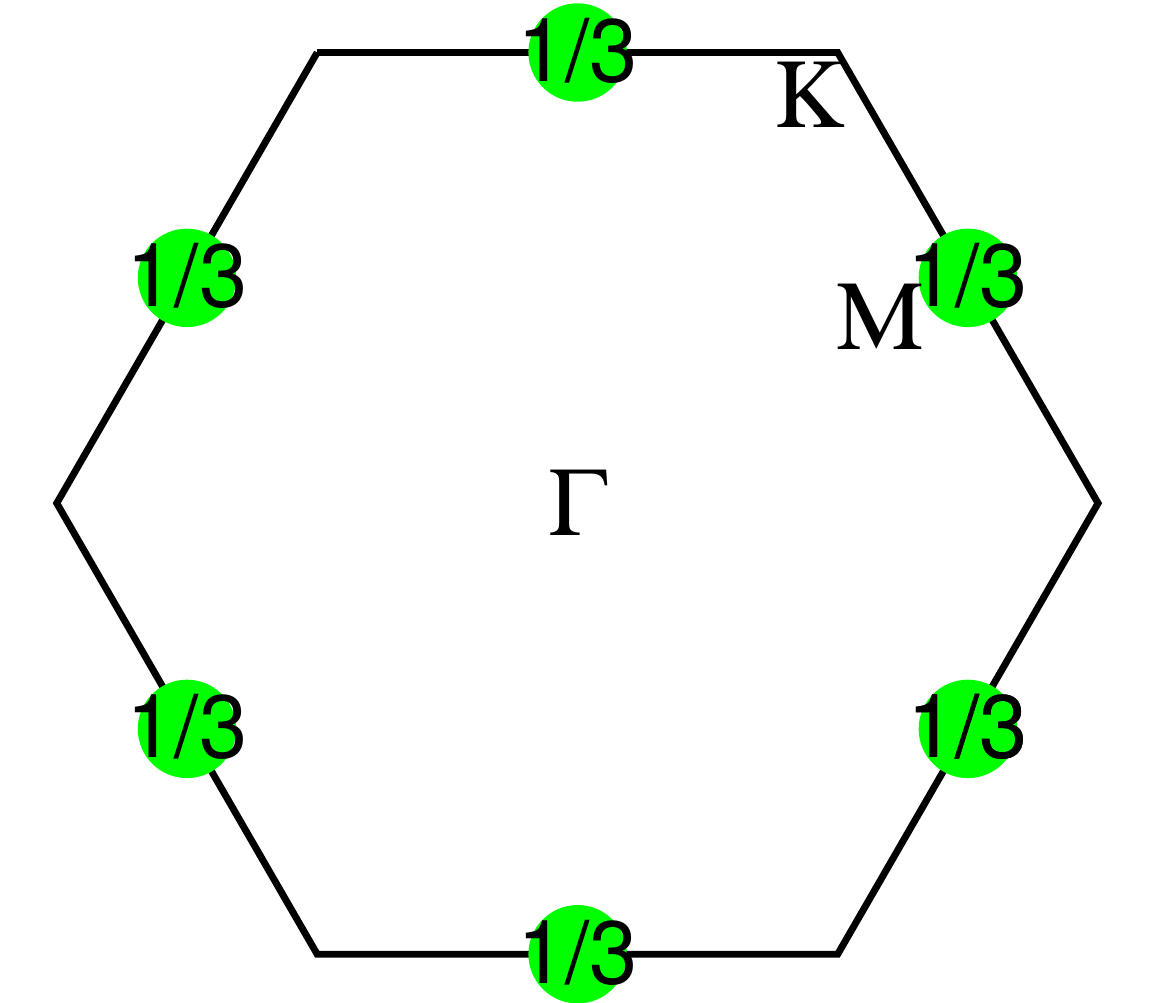}}
 \subfigure[\;Coplanar state. $E=-3J_1+6J_2-3J_3-3K$.\label{fig:reg_tri_c}]{
   \includegraphics[trim = 15mm 81mm 22mm 81mm, clip,width=.2\textwidth]{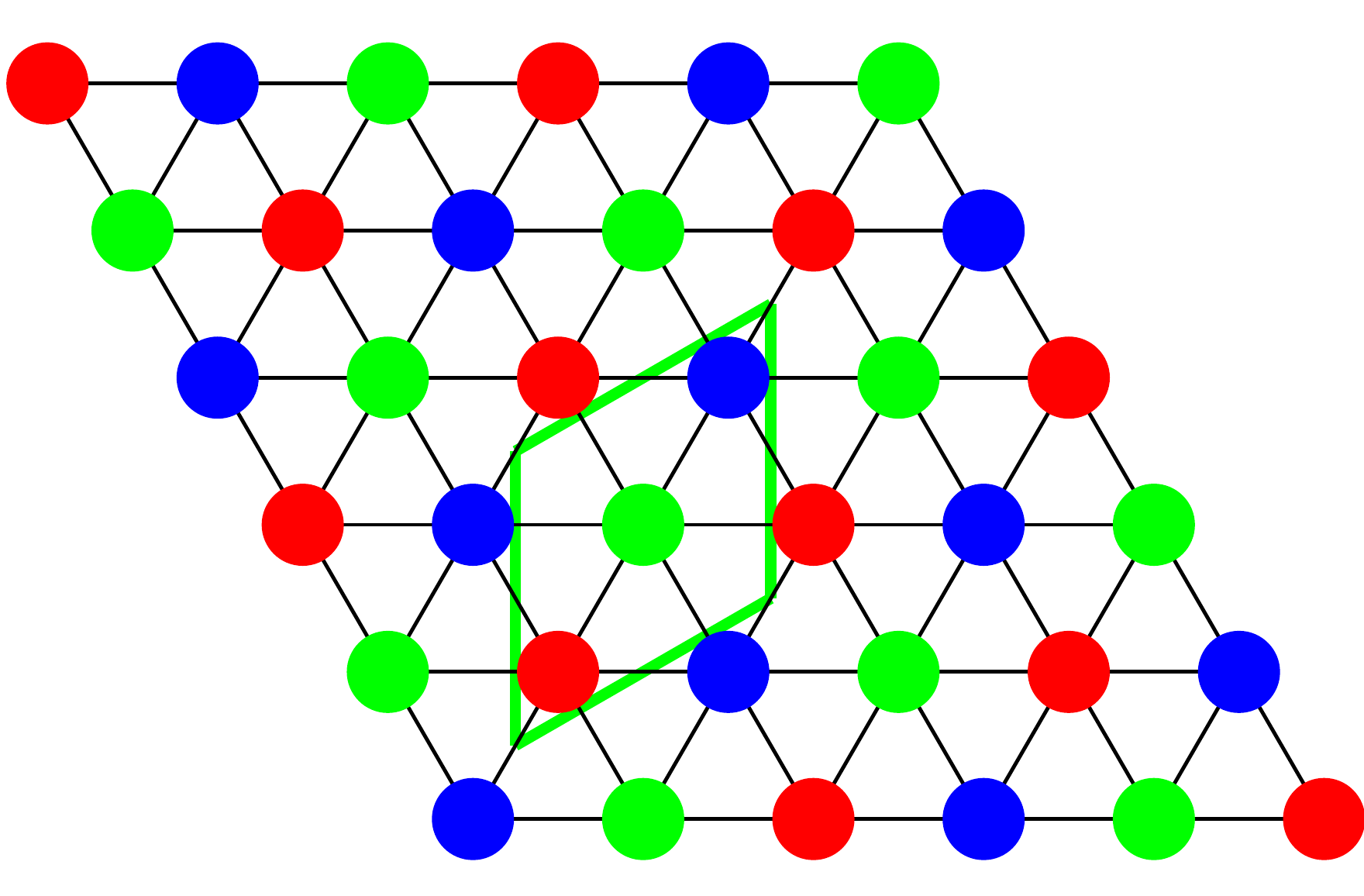}
   \includegraphics[trim = 12mm -120mm 19mm 60mm, clip,width=.059\textwidth]{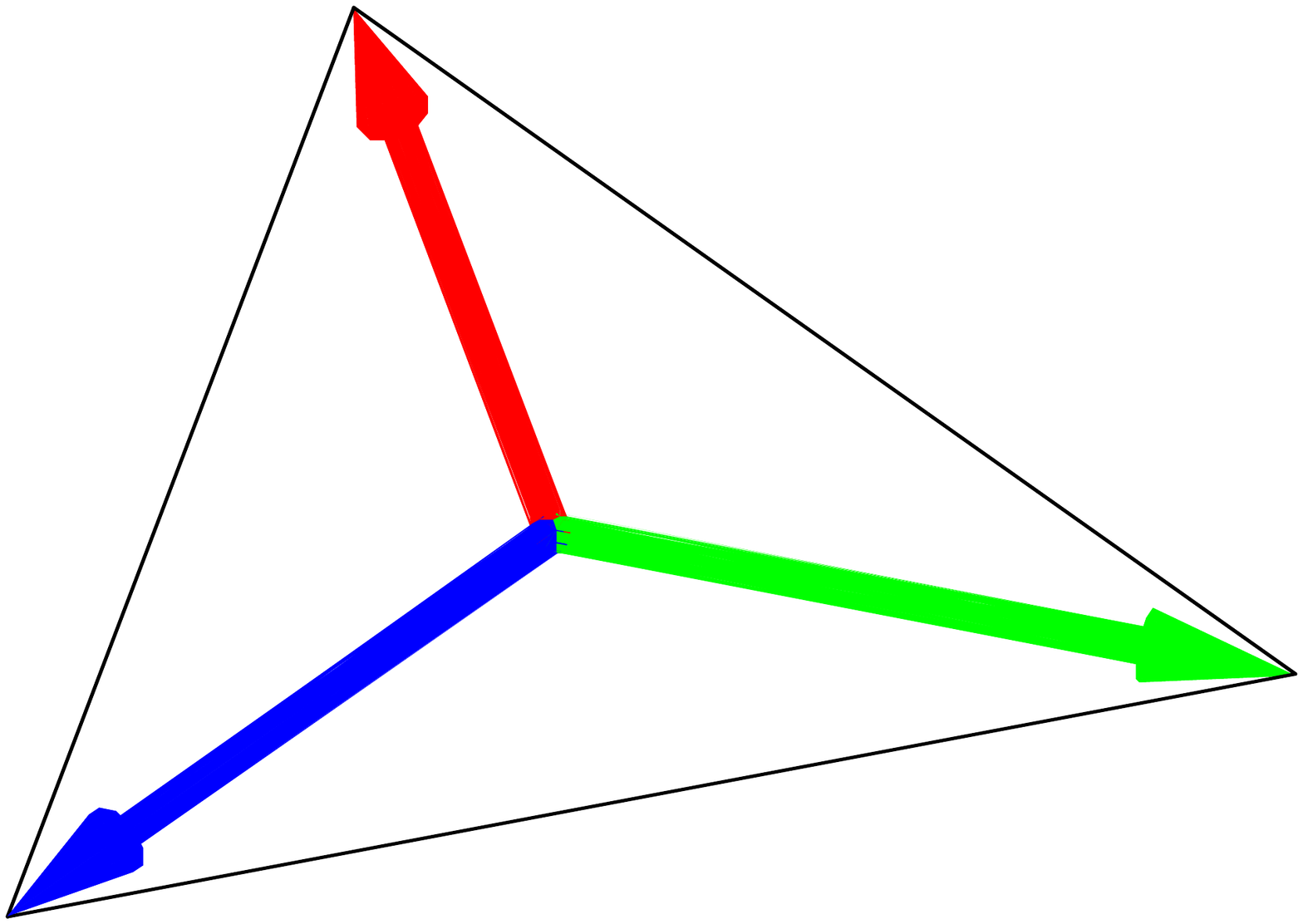}
   \includegraphics[trim = 86mm 63mm 76mm 51mm, clip,width=.14\textwidth]{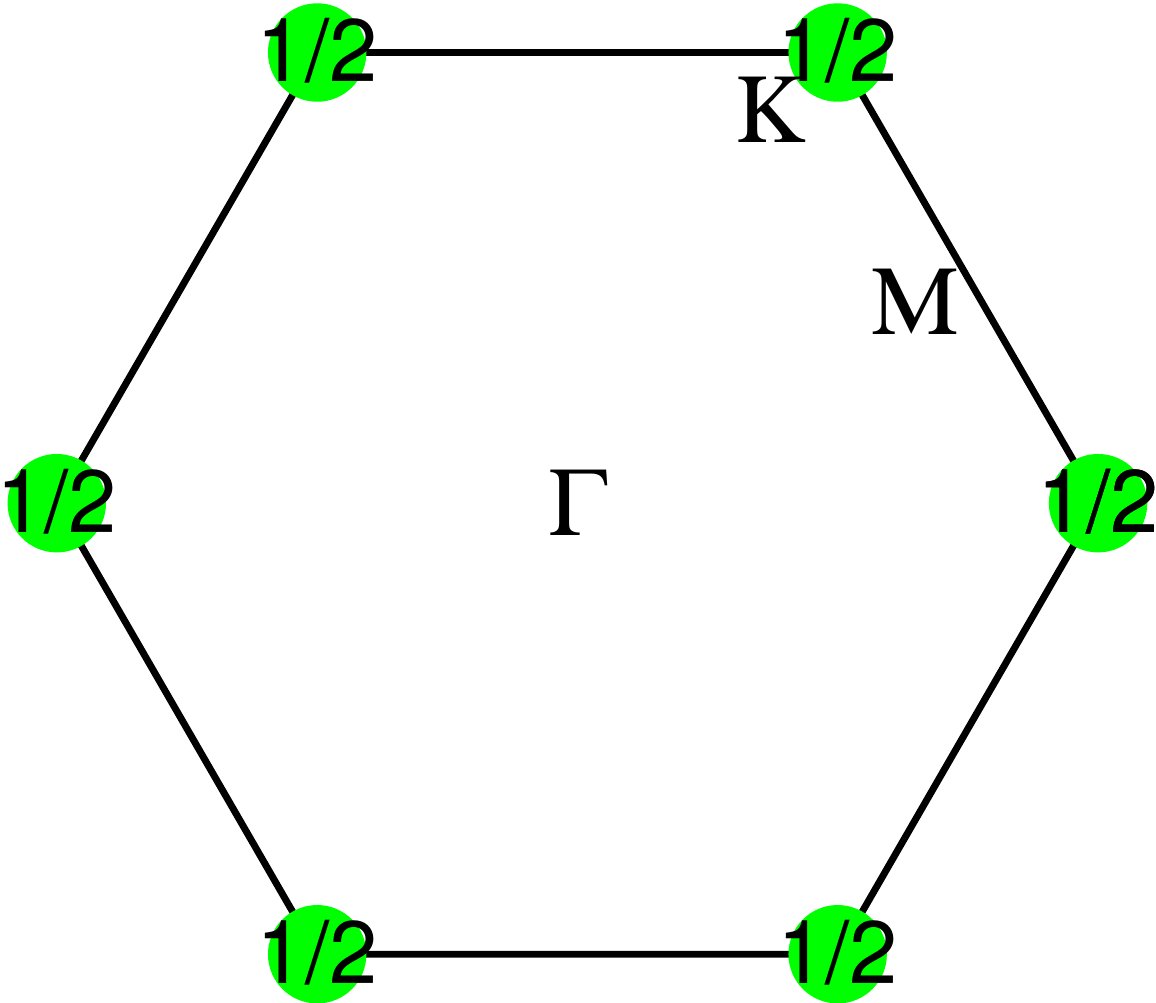}}
 \subfigure[\;F umbrella states\label{fig:reg_tri_d}]{
   \includegraphics[trim = 15mm 81mm 22mm 81mm, clip,width=.2\textwidth]{tri_3ssr.pdf}
   \includegraphics[trim = 12mm -120mm 19mm 60mm, clip,width=.059\textwidth]{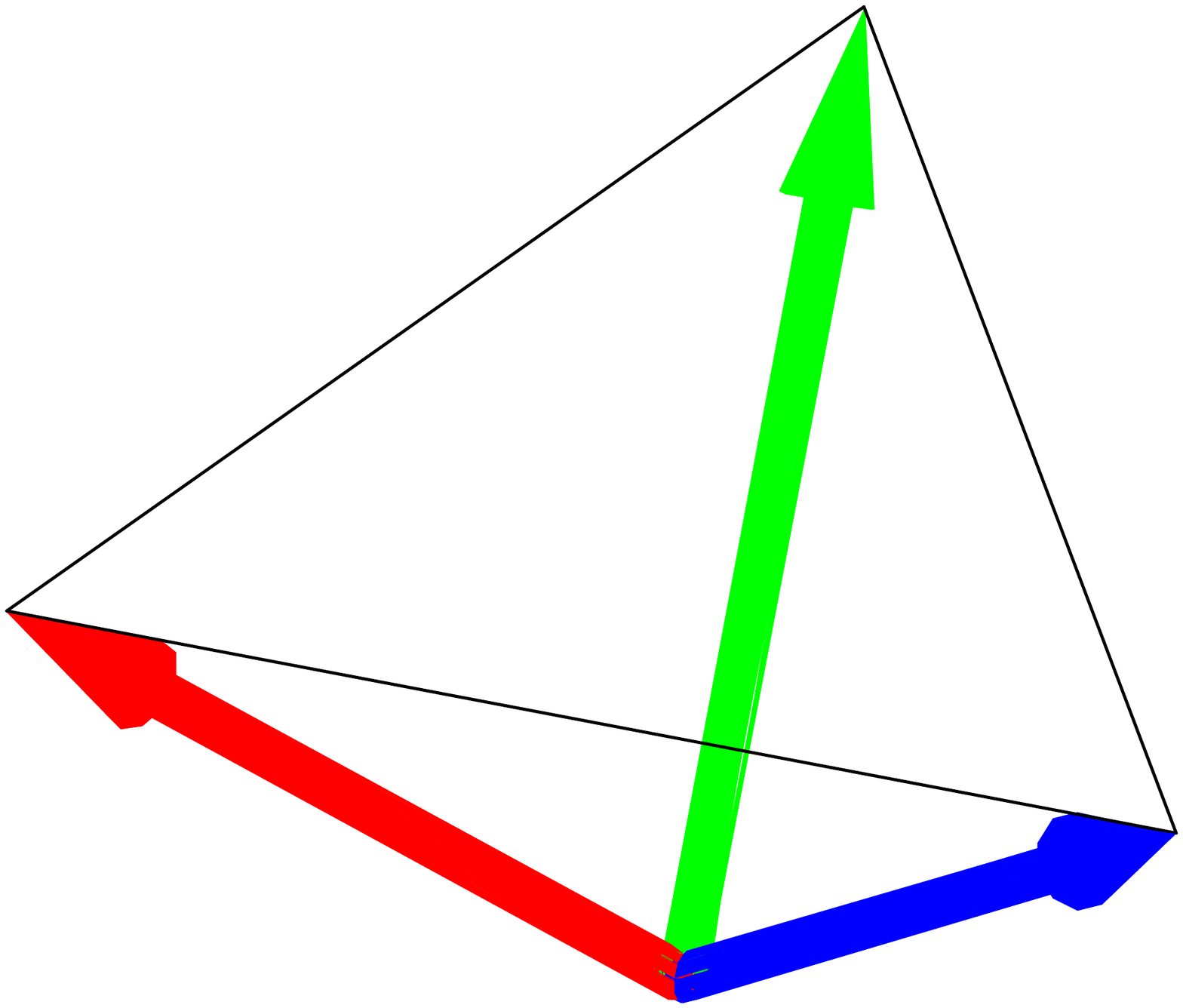}}\\
 \caption{(Color online) Regular states on the triangular lattice.
The sublattice arrangements (labelled by colors) and the spin directions on each sublattice are displayed in the left and center columns.
A spin unit cell is surrounded with green lines.
The positions and weights of the Bragg peaks in the hexagonal Brillouin zone of the lattice are in the right column.
The energy  per site of each structure is given as a function of the parameters of the models described in Sec~\ref{sec:energetics}.}
 \label{fig:reg_tri}
 \end{center}
 \end{figure}

The mapping of Eq.~(\ref{eq:solution_a}) has compatible states only for $\varepsilon_R=\varepsilon_\sigma=1$.
They are ferromagnetic (F) states, as shown in Fig.~\ref{fig:reg_tri_a}.
Since $G_{T_{1-2}}=I$ for the Eqs.~(\ref{eq:solution_b})-(\ref{eq:solution_f}), no new regular states can be compatible with any of them.

The mapping of Eq.~(\ref{eq:solution_g}) has compatible states only for $\varepsilon_R=1$ and $\varepsilon_\sigma=-1$. Then $\mathbf S(0,0)=\pm(1,1,1)/\sqrt{3}$ and the state is the tetrahedron state depicted in Fig.~\ref{fig:reg_tri_b}, where the spins of four sublattices point toward the corners of a tetrahedron.
The sign of $\mathbf S(0,0)$ determines the chirality of the configuration.

The next regular state is the coplanar state of Fig.~\ref{fig:reg_tri_c}, which is compatible with Eq.~(\ref{eq:solution_h}) for $\varepsilon_R=\varepsilon_\sigma=1$ and $\mathbf S(0,0)=\pm(1,0,0)$.
The three sublattices are coplanar with relative angles of $120^\circ$.
This state is not chiral because the configurations obtained with the two possible $\mathbf S(0,0)$ are related by a global spin rotation in $SO(3)$.

A continuum of \textit{umbrella} states are compatible with Eq.~(\ref{eq:solution_h}) with $\varepsilon_\sigma=1$ and $\varepsilon_R=-1$.
They are depicted in Fig.~\ref{fig:reg_tri_d}, where the sublattices are the same than for the coplanar states but the relative angles between the spin orientations are all identical and $\leq 120^\circ$.
This family interpolates between the F and the coplanar states.

We started by choosing $H_c^S=\{I\}$, but states with $H_c^S=\{I\}$, $\mathbf Z_2$ (for the coplanar state) or $O(2)$ (for the F state) have been obtained anyway.
One can check that choosing another $H_c^S$ would not give any new regular state.
All the regular states are thus those gathered in Fig.~\ref{fig:reg_tri}.

The Bragg peaks of these states are displayed in the hexagonal Brillouin zone in the right column of Fig.~\ref{fig:reg_tri} and their powder-averaged structure factors in App.~\ref{App:powder} together with the formulas for these quantities.

\section{Regular states for Heisenberg spins on several simple lattices}
\label{sec:reg_states}

In the following we enumerate the regular states on the kagome and honeycomb lattices, two lattices which have a symmetry group $S_L$ isomorphic to that of the triangular lattice.
To be complete, we also present the regular states on the square lattice and discuss the spiral states that may be seen as regular states when $S_L$ reduces to the translation group.

\subsection{Kagome lattice}
\label{ssec:kag}

\begin{figure}
\begin{center}
 \subfigure[\;Ferromagnetic (F) state.  $E=4J_1+4J_2+2J_3'+4J_3$.\label{fig:reg_kag_a}]{
   \includegraphics[trim = 15mm 81mm 22mm 81mm, clip,width=.185\textwidth]{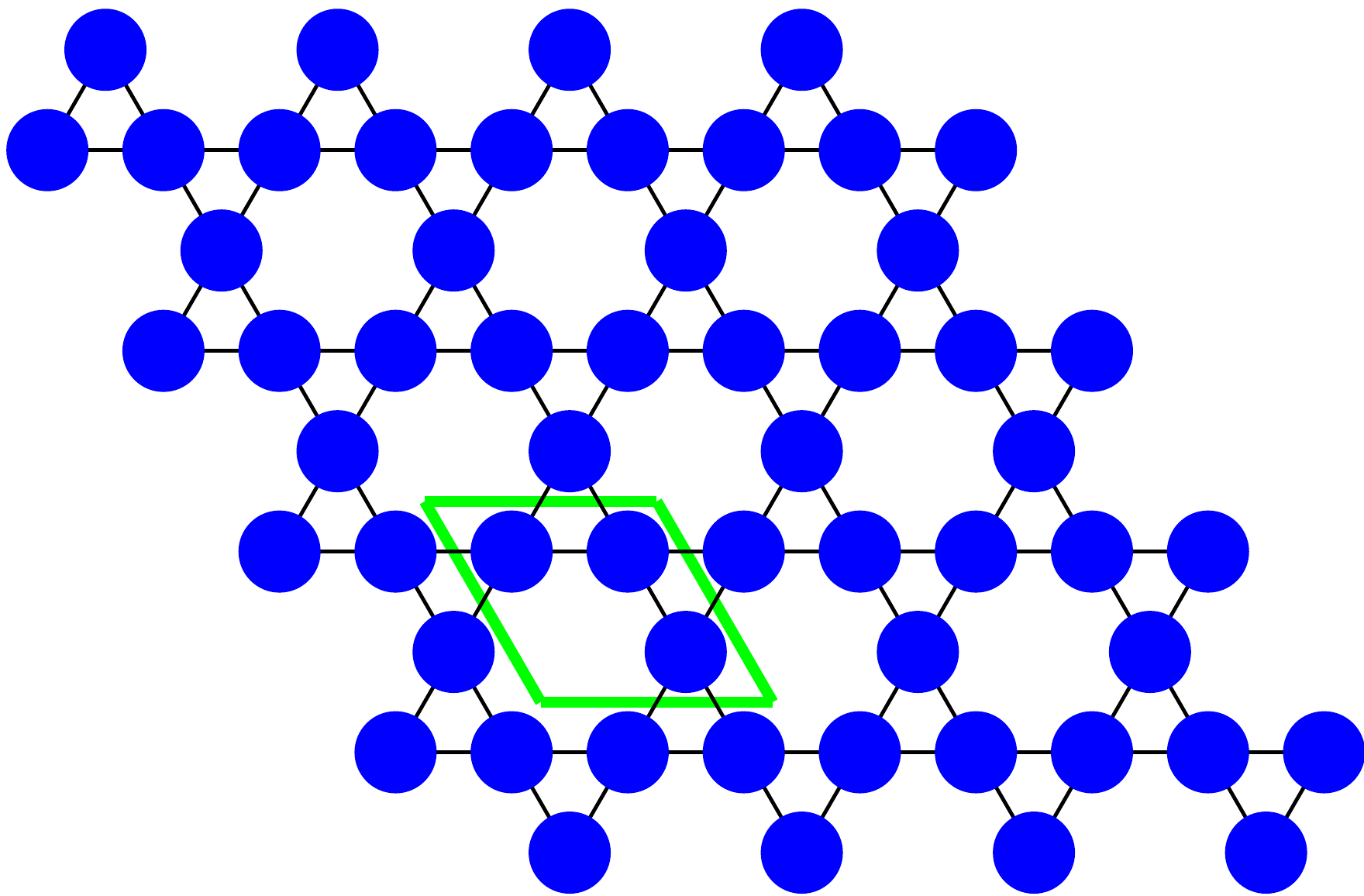}
   \includegraphics[trim = 25mm -50mm 42mm 60mm, clip,width=.06\textwidth]{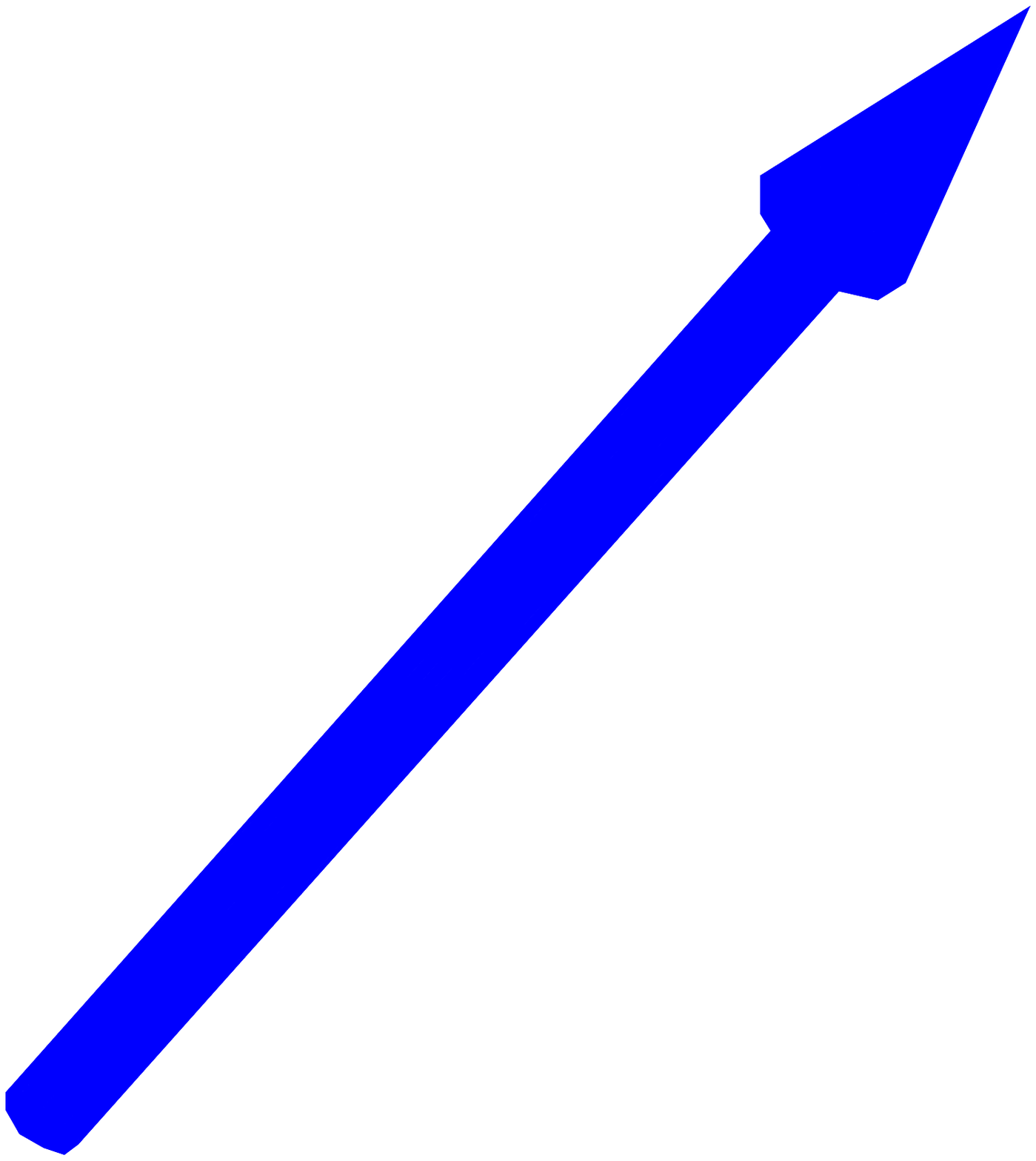}
   \includegraphics[trim = 86mm 63mm 76mm 51mm, clip,width=.14\textwidth]{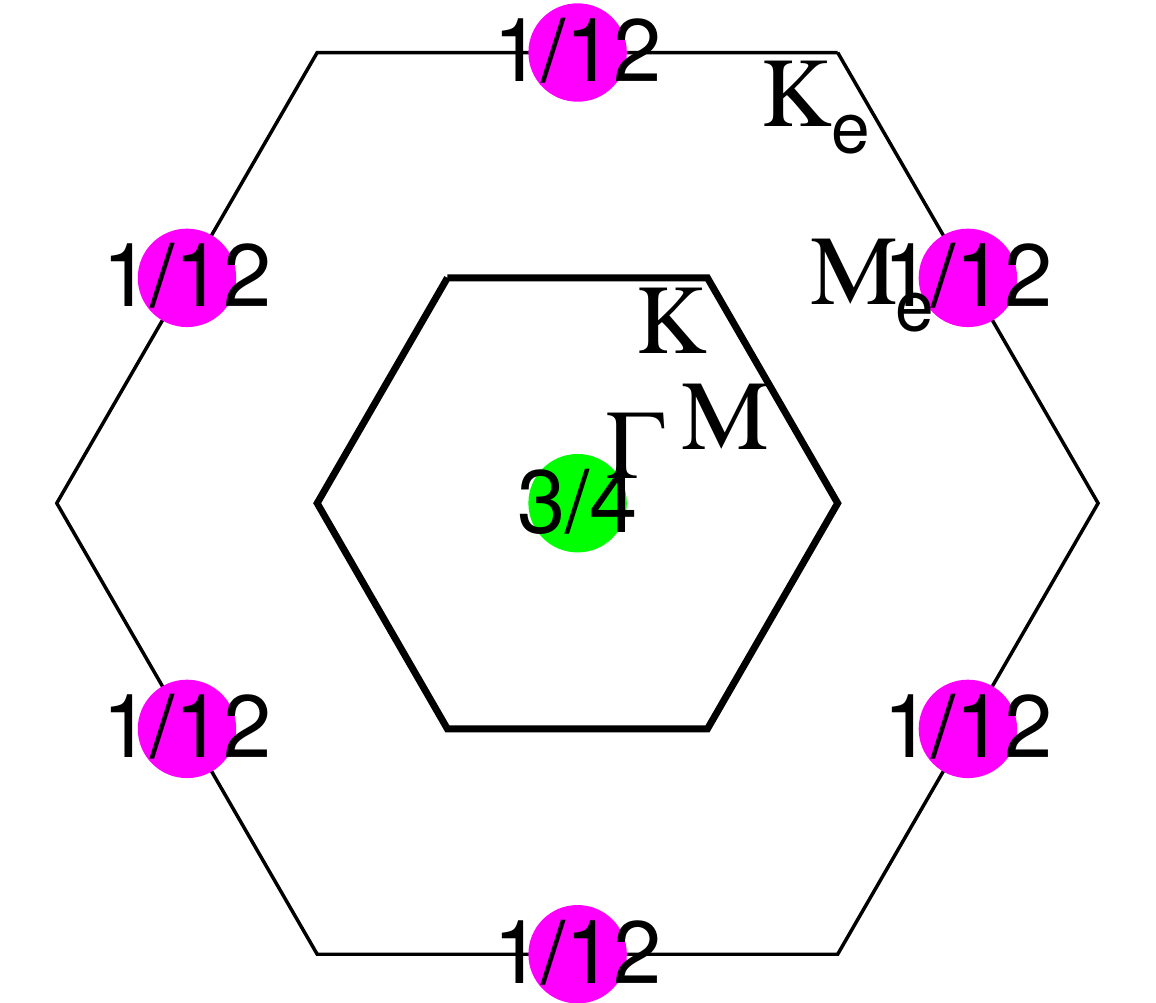}}
 \subfigure[\;$\mathbf q=\mathbf 0$ state. $E=-2J_1-2J_2+2J_3'+4J_3$.\label{fig:reg_kag_b}]{
   \includegraphics[trim = 15mm 81mm 22mm 81mm, clip,width=.185\textwidth]{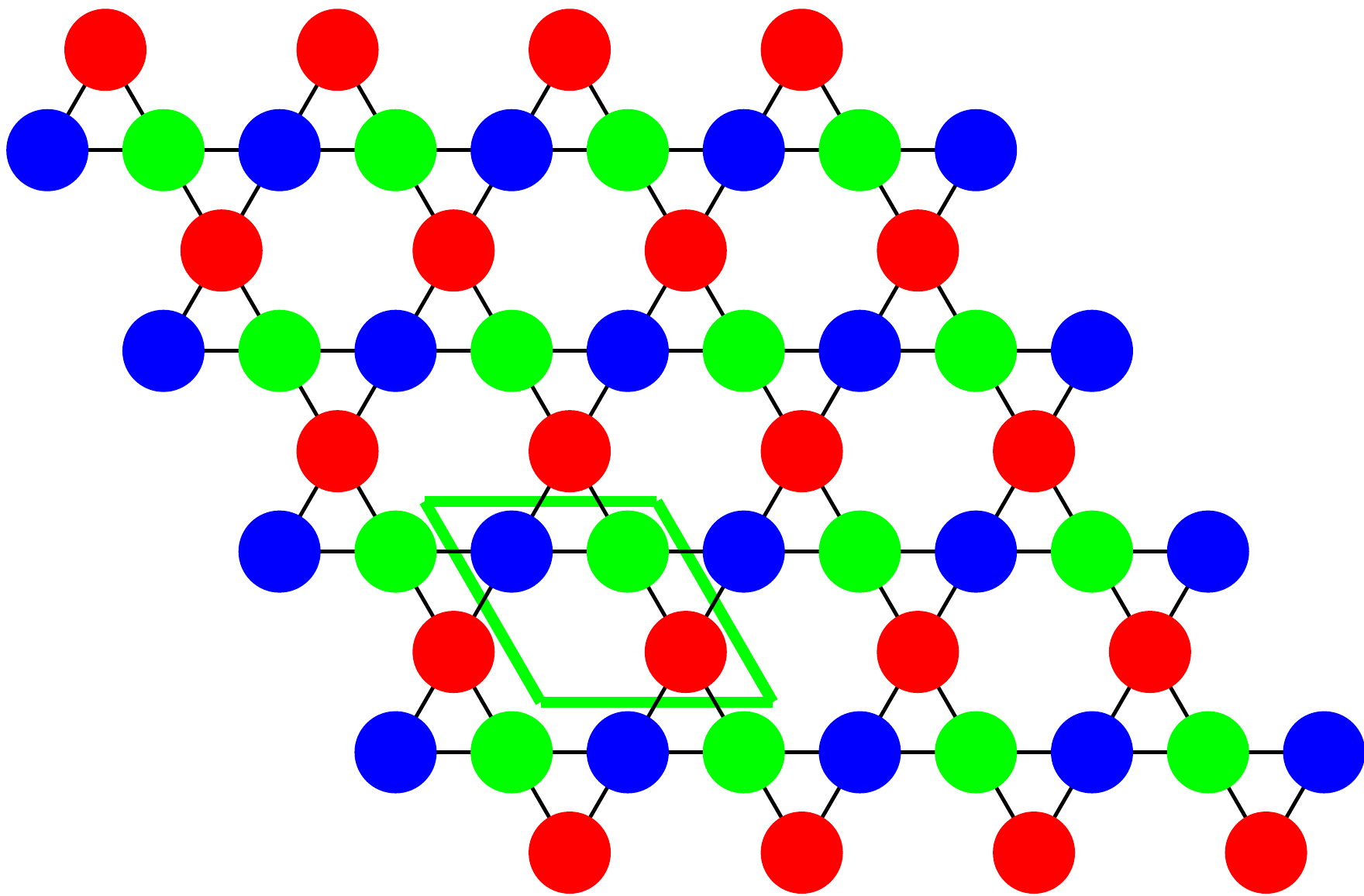}
   \includegraphics[trim = 12mm -70mm 19mm 60mm, clip,width=.08\textwidth]{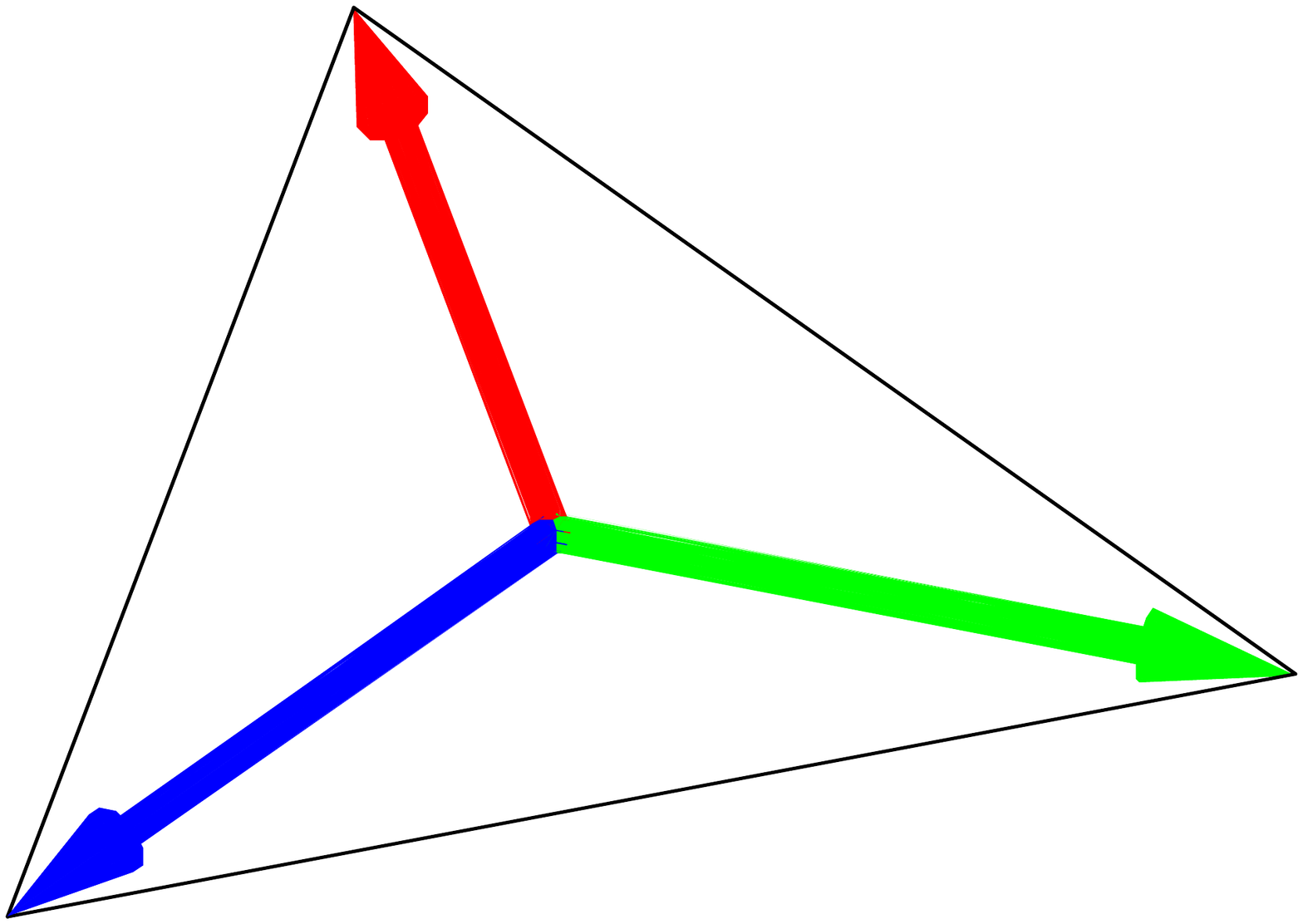}
   \includegraphics[trim = 86mm 63mm 76mm 51mm, clip,width=.14\textwidth]{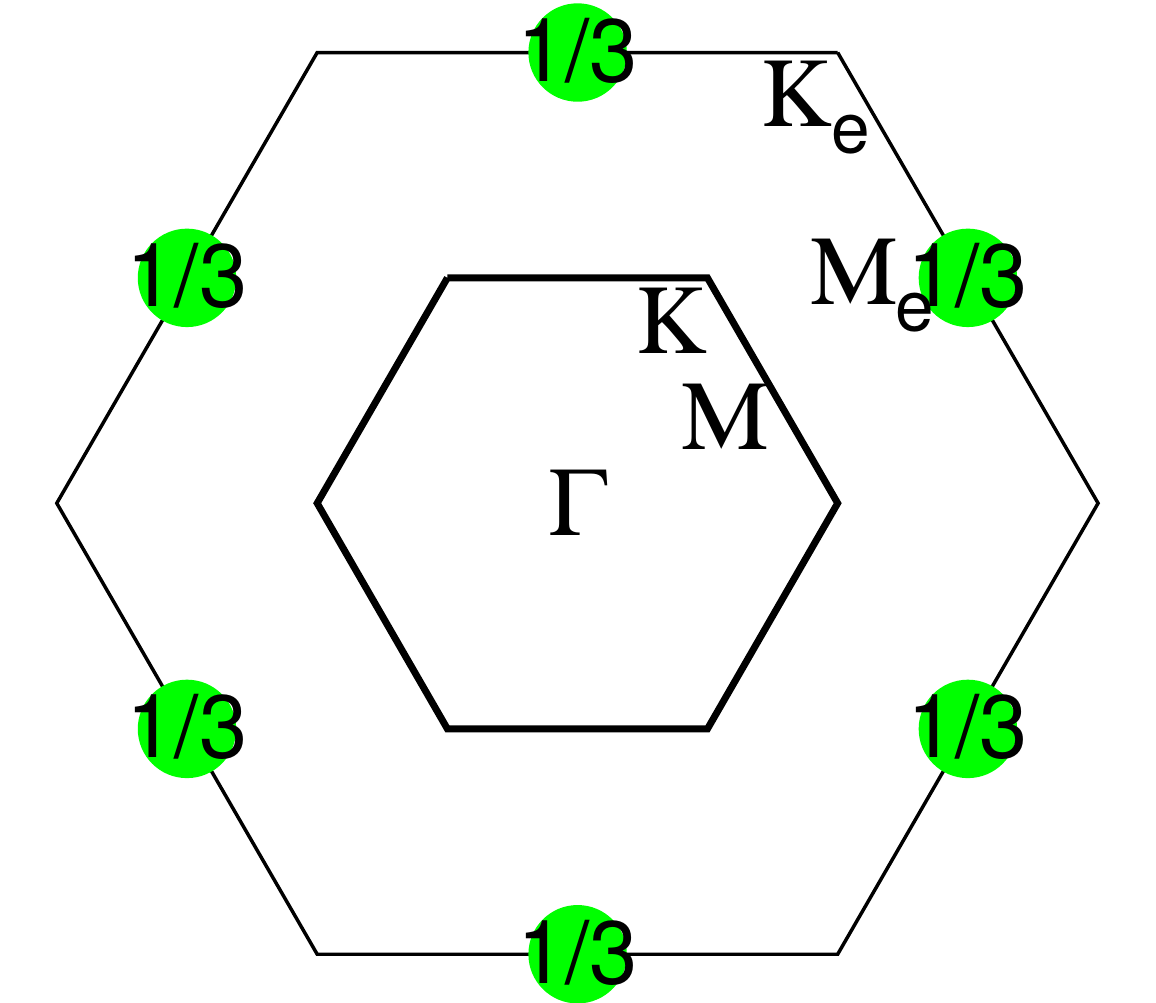}}
 \subfigure[\;$\sqrt{3}\times\sqrt{3}$ state. $E=-2J_1+4J_2-J_3'-2J_3$.\label{fig:reg_kag_c}]{
   \includegraphics[trim = 15mm 81mm 22mm 81mm, clip,width=.185\textwidth]{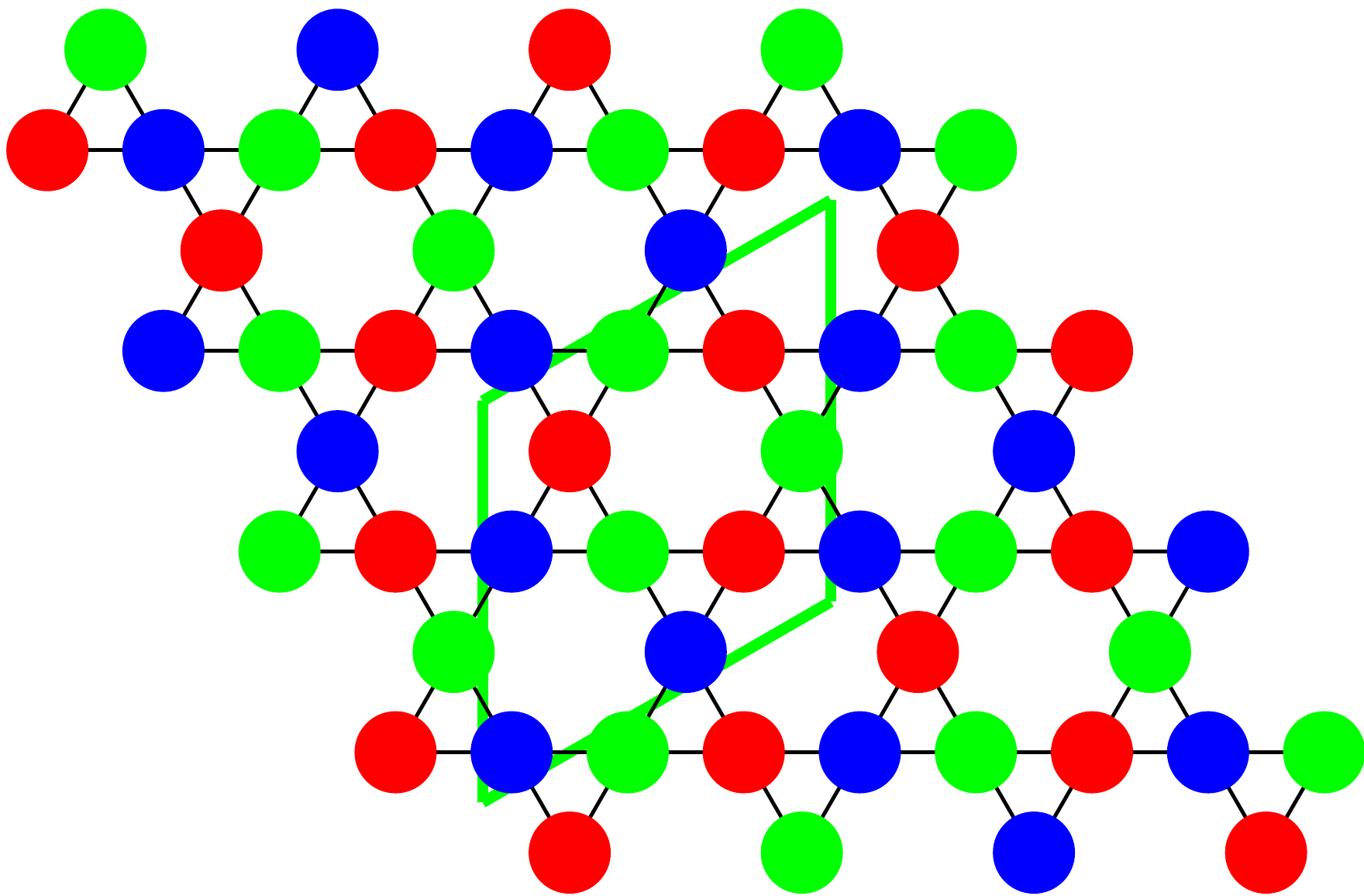}
   \includegraphics[trim = 12mm -50mm 19mm 60mm, clip,width=.08\textwidth]{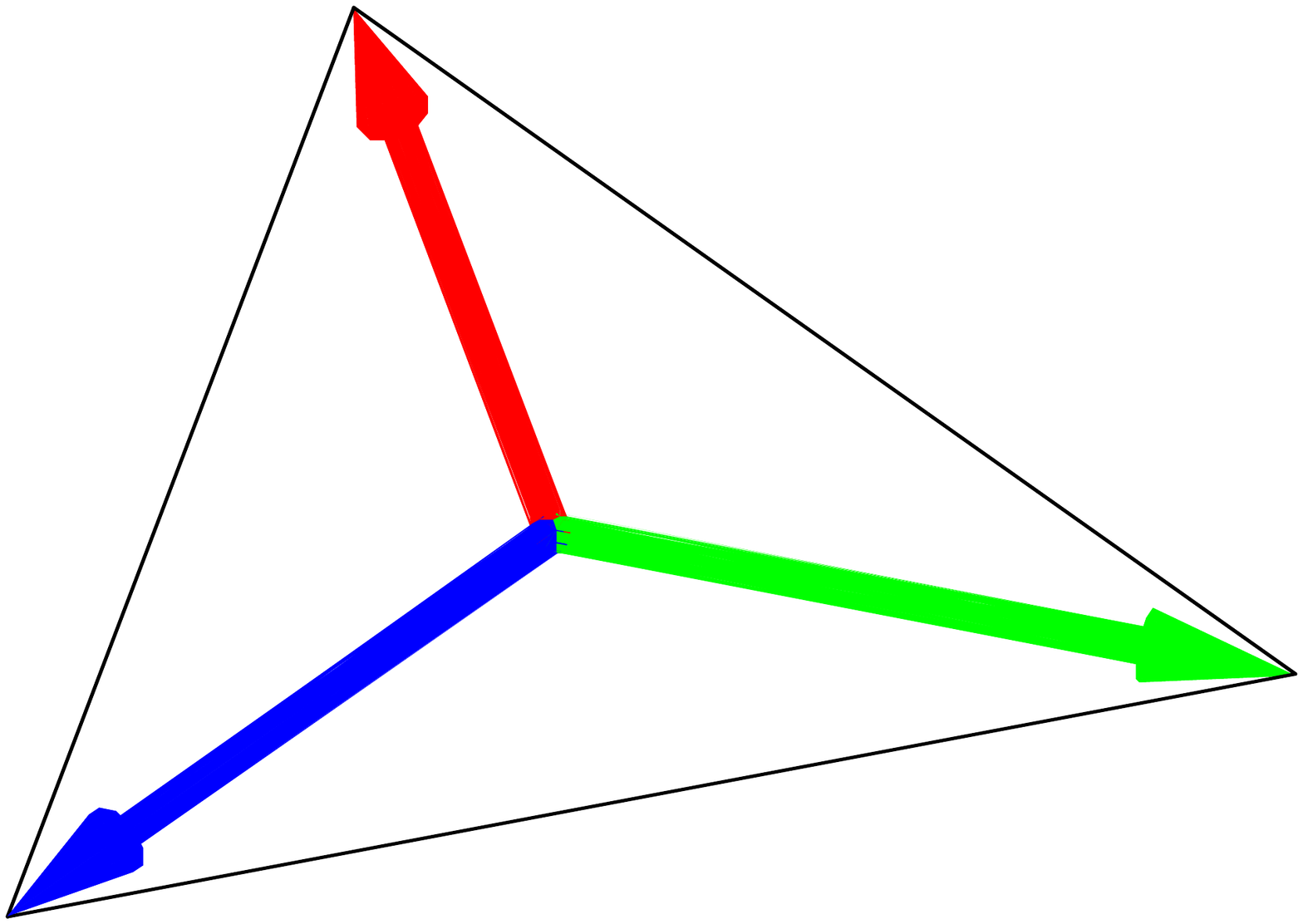}
   \includegraphics[trim = 86mm 63mm 76mm 51mm, clip,width=.14\textwidth]{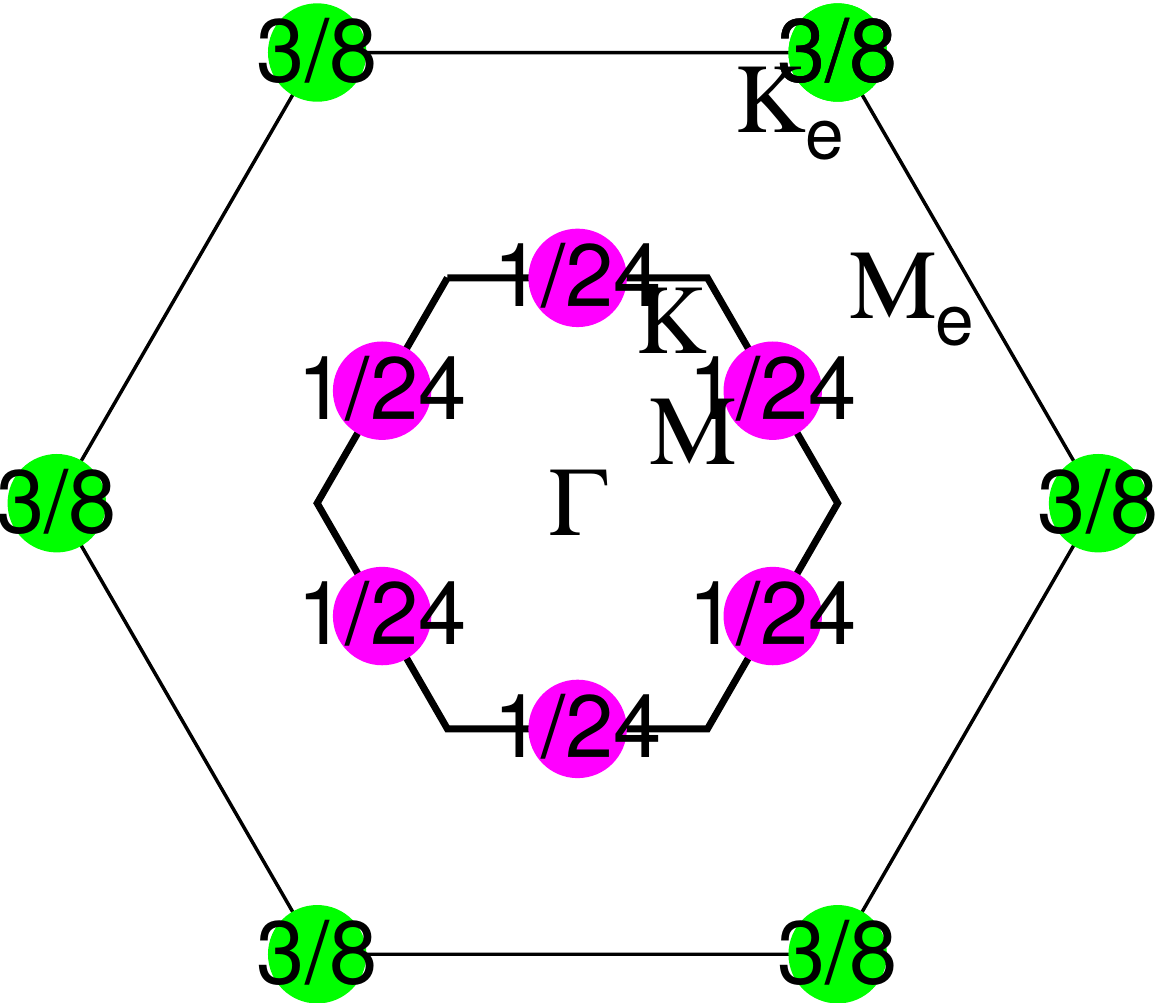}}
 \subfigure[\;Octahedral state. $E=2J_3'-4J_3$.\label{fig:reg_kag_d}]{
   \includegraphics[trim = 15mm 81mm 22mm 81mm, clip,width=.185\textwidth]{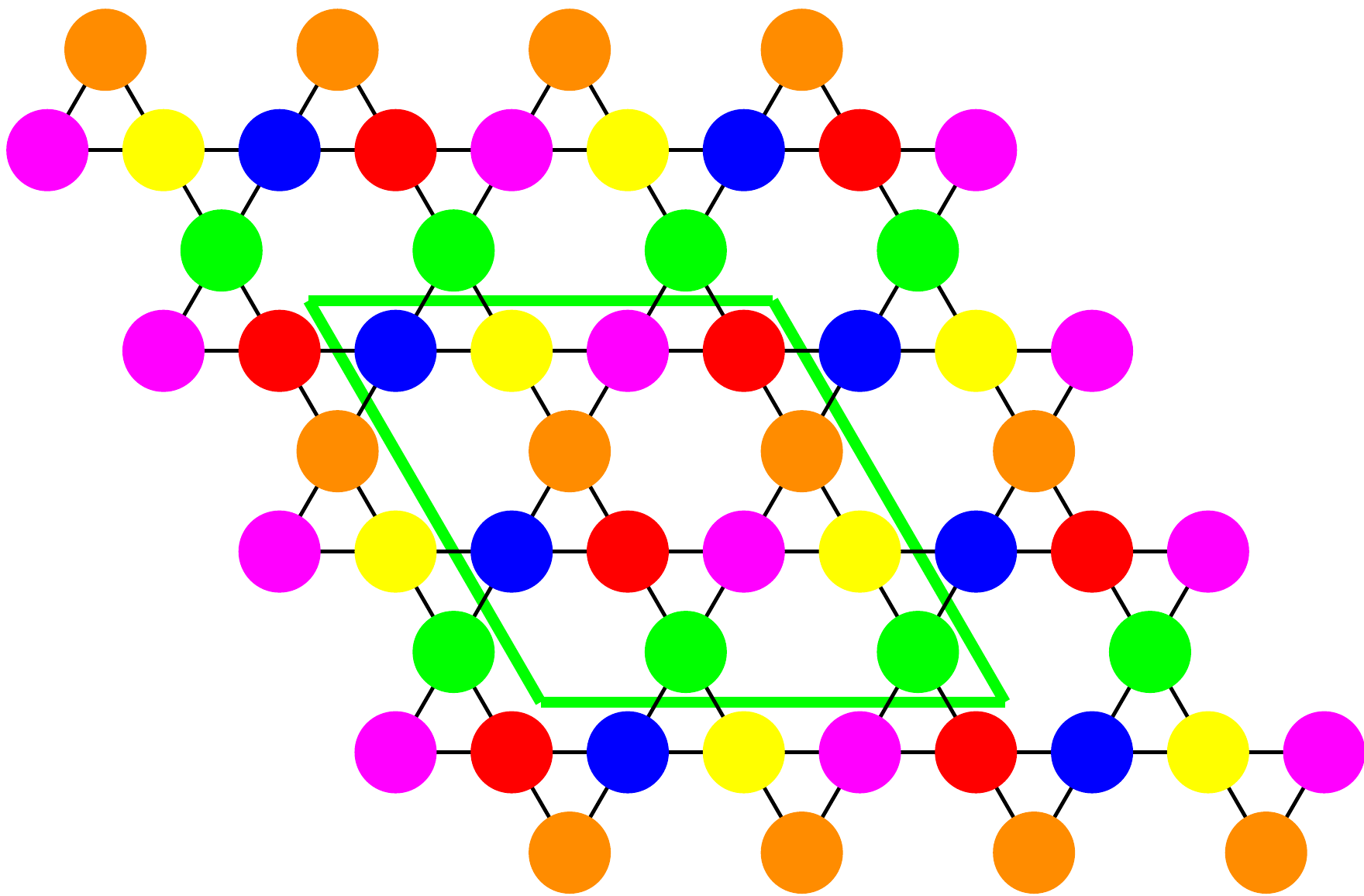}
   \includegraphics[trim = 12mm -50mm 19mm 53mm, clip,width=.08\textwidth]{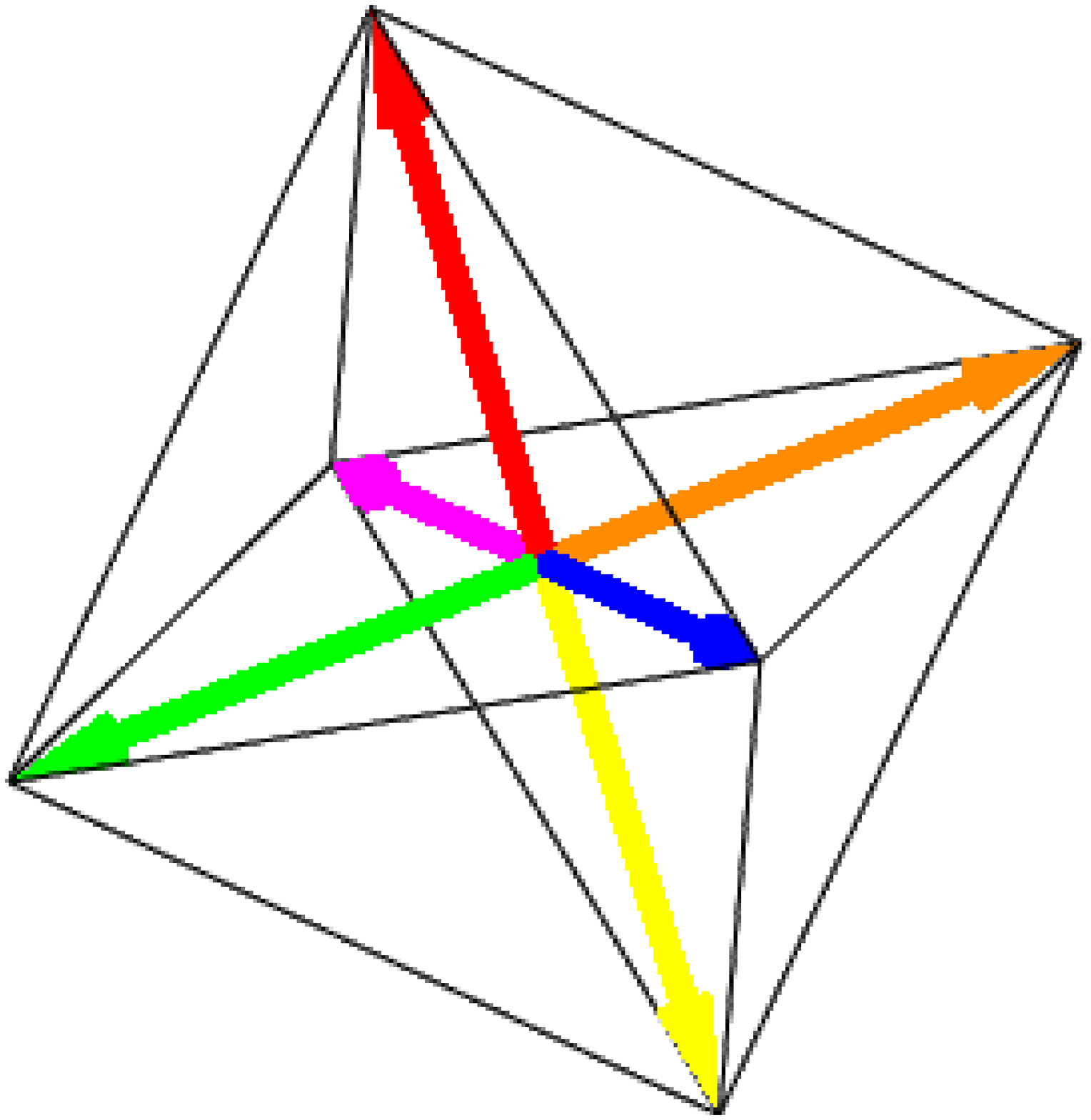}
   \includegraphics[trim = 86mm 63mm 76mm 51mm, clip,width=.14\textwidth]{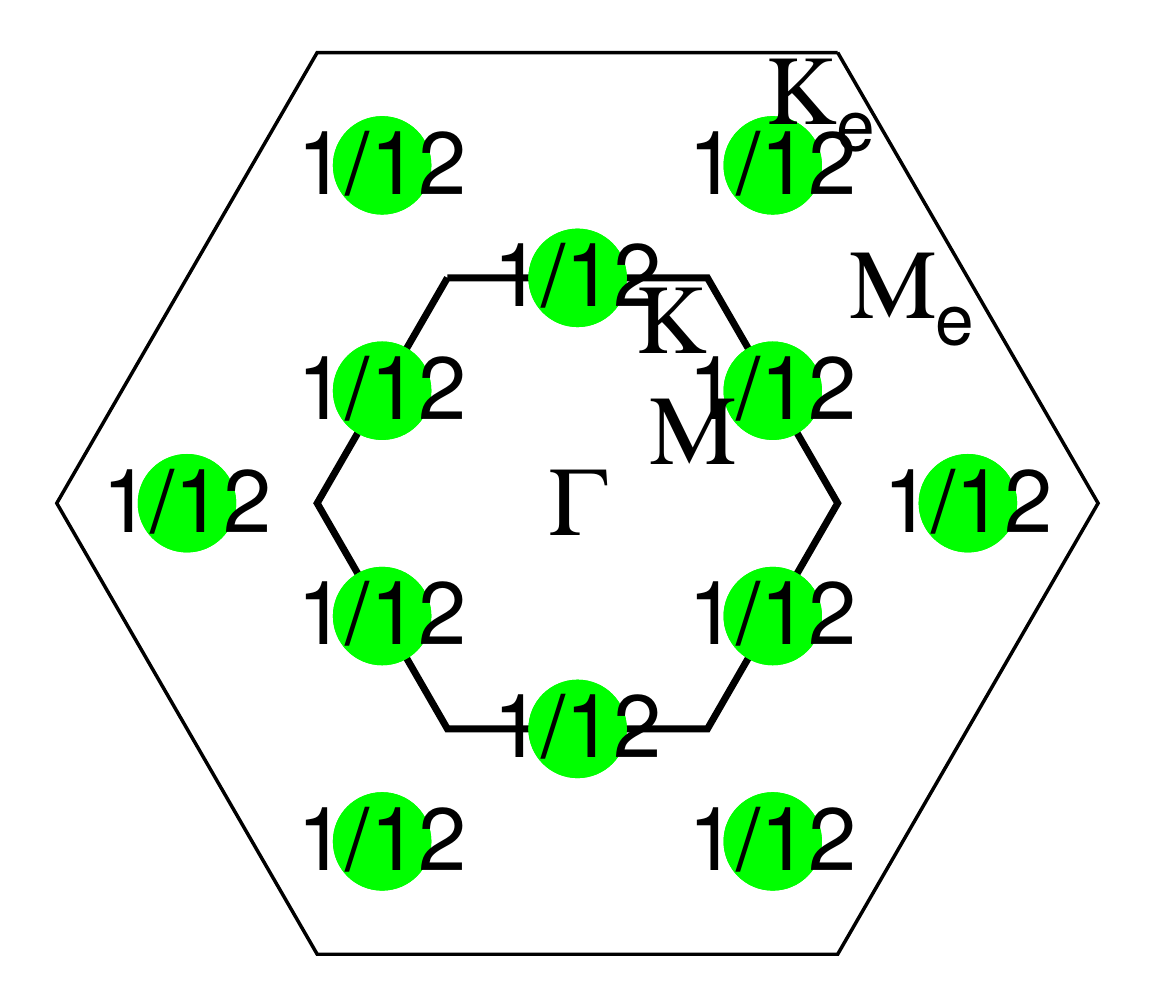}}
 \subfigure[\;Cuboc1 state. $E=-2J_1+2J_2-2J_3'$.\label{fig:reg_kag_e}]{
   \includegraphics[trim = 15mm 81mm 22mm 81mm, clip,width=.185\textwidth]{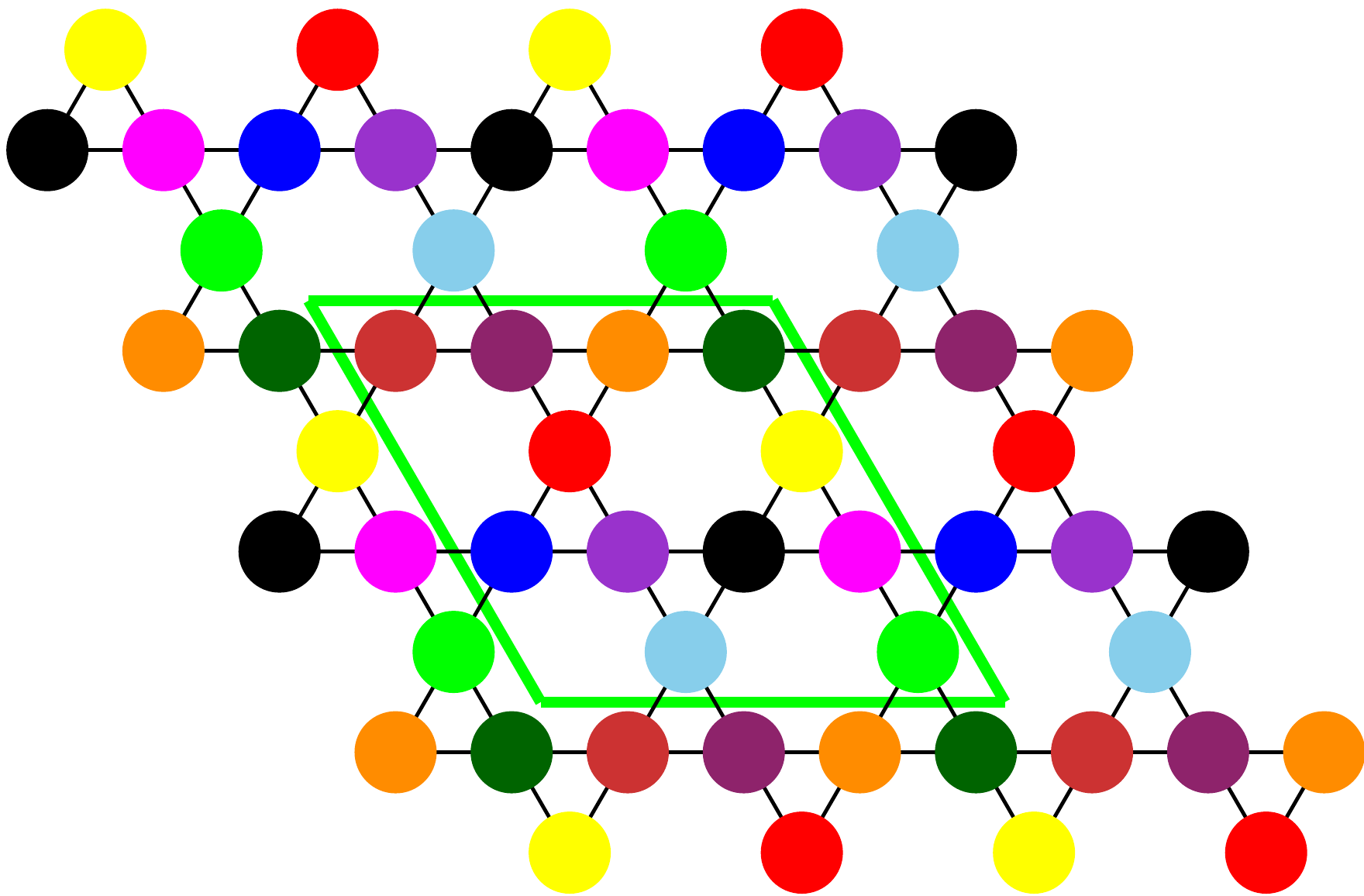}
   \includegraphics[trim = 12mm -50mm 19mm 54mm, clip,width=.08\textwidth]{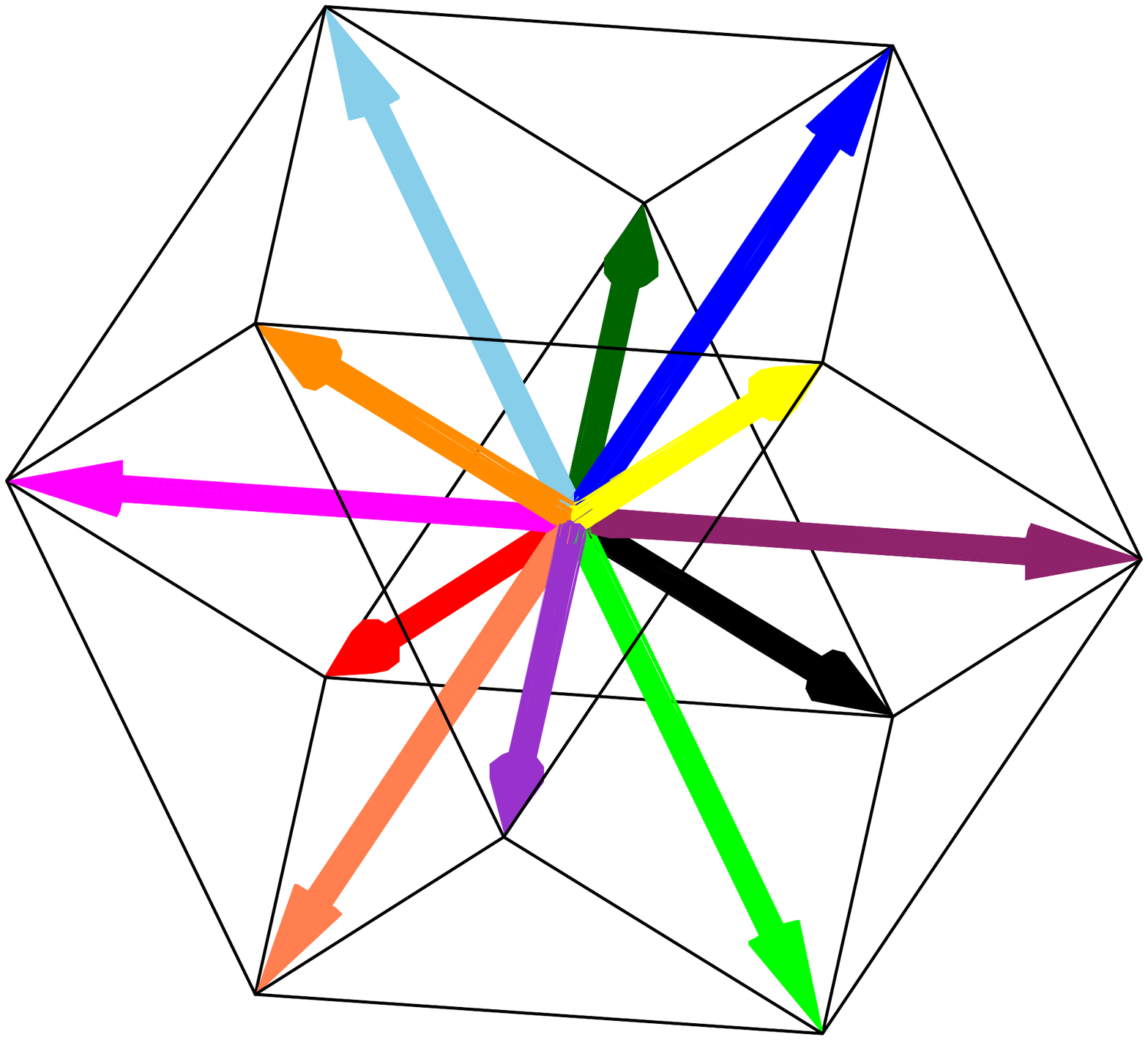}
   \includegraphics[trim = 86mm 63mm 76mm 51mm, clip,width=.14\textwidth]{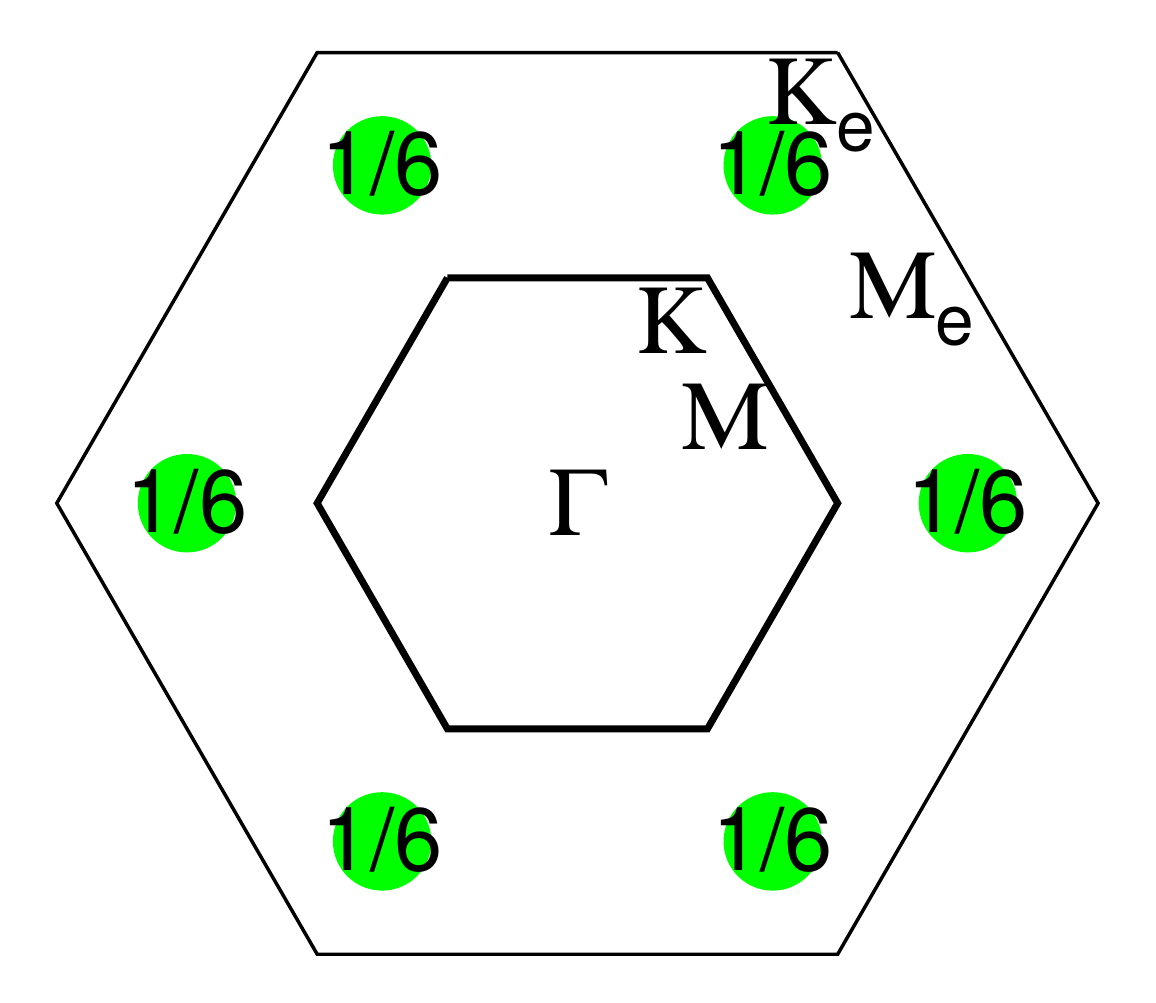}}
 \subfigure[\;Cuboc2 state.  $E=2J_1-2J_2-2J_3'$.\label{fig:reg_kag_f}]{
   \includegraphics[trim = 15mm 81mm 22mm 81mm, clip,width=.185\textwidth]{kag_12ssr.pdf}
   \includegraphics[trim = 12mm -50mm 19mm 54mm, clip,width=.08\textwidth]{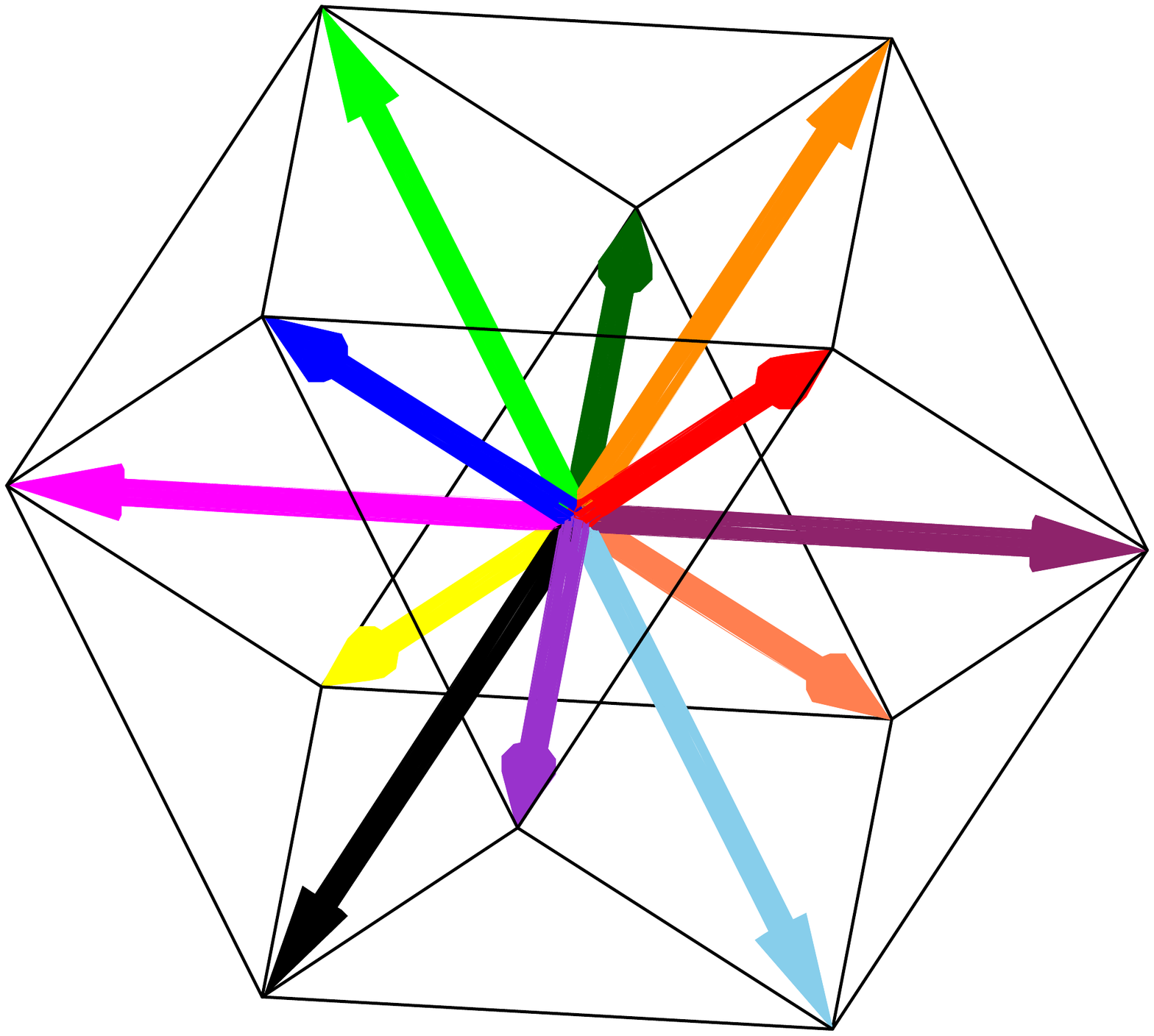}
   \includegraphics[trim = 86mm 63mm 76mm 51mm, clip,width=.14\textwidth]{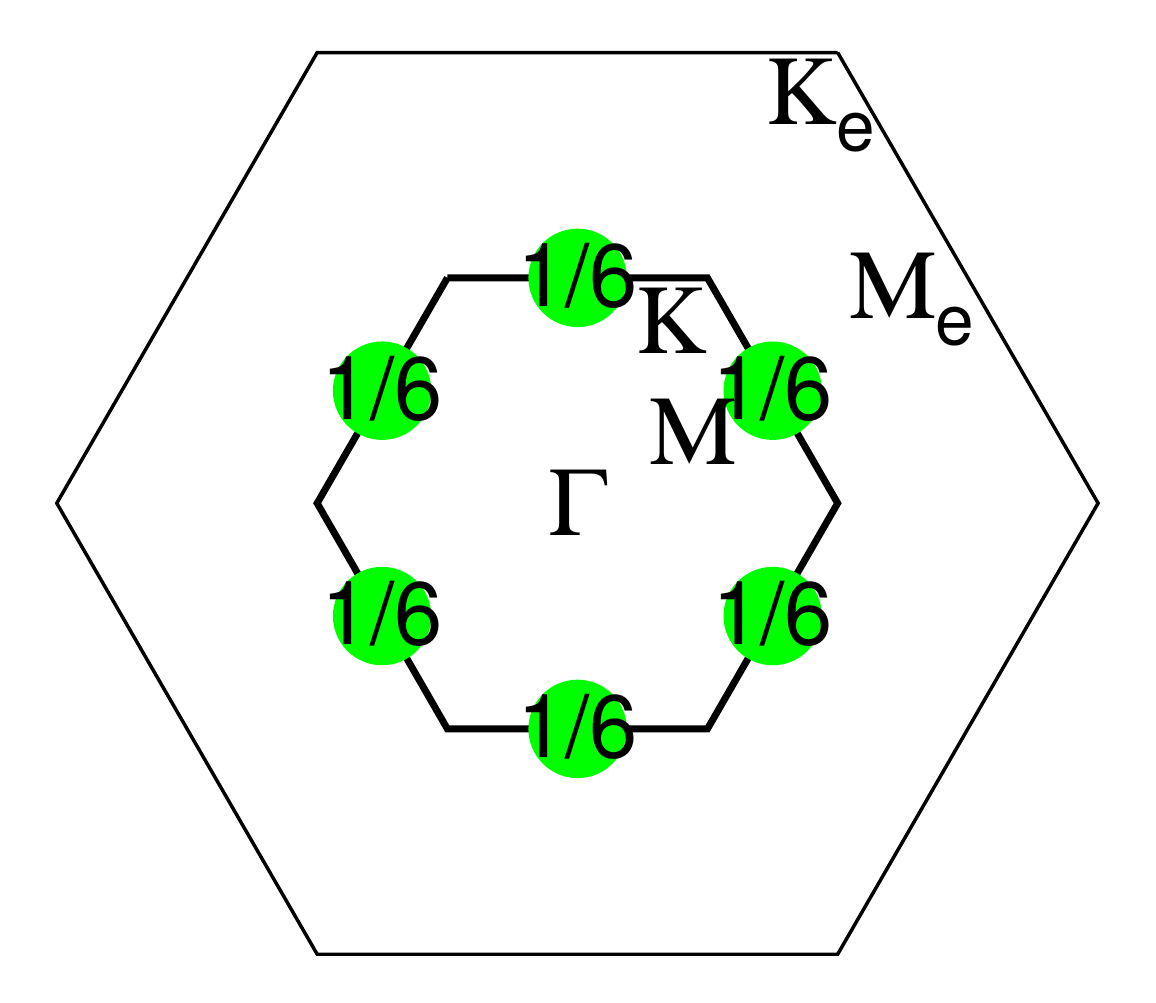}}
 \subfigure[\;$\mathbf q=\mathbf 0$ (left) and $\sqrt{3}\times\sqrt{3}$ (right) umbrella states\label{fig:reg_kag_g}]{
   \includegraphics[trim = 15mm 81mm 22mm 81mm, clip,width=.185\textwidth]{kag_3ssr.pdf}
   \includegraphics[trim = 12mm -50mm 19mm 60mm, clip,width=.08\textwidth]{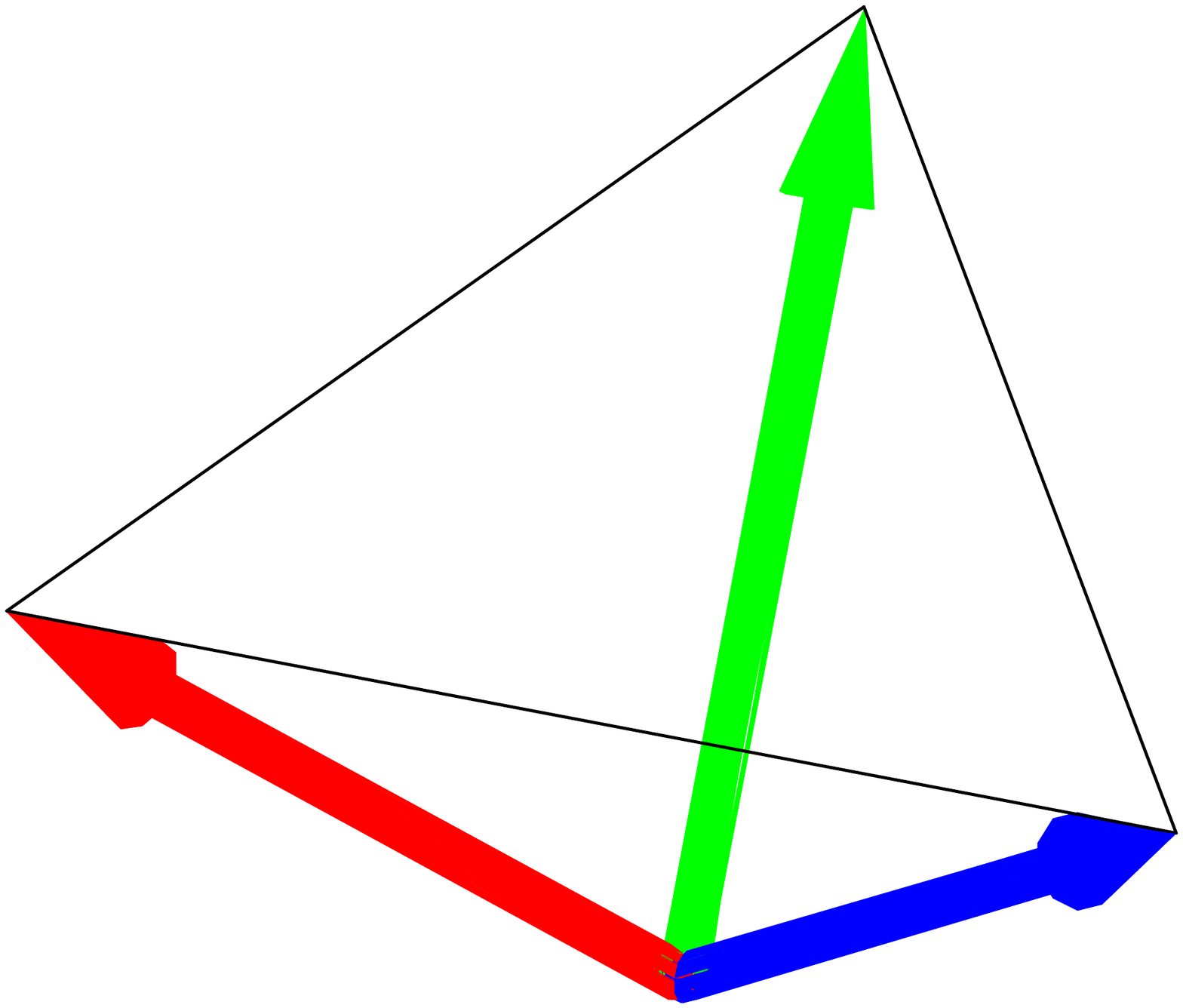}
   \includegraphics[trim = 15mm 81mm 22mm 81mm, clip,width=.185\textwidth]{kag_3ssr_c.pdf}}
  \caption{(Color online) Regular states on the kagome lattice and their equal time structure factors  in the EBZ (see text).
The energies (per site) of these states are given for the $J_1$-$J_2$-$J_3$-$J_3'$ model described in Sec~\ref{sec:energetics}. }
 \label{fig:reg_kag}
\end{center}
\end{figure}

The symmetry group $S_L$ of the kagome lattice is isomorphic to that of the triangular lattice, thus the algebraic solutions Eqs.~(\ref{eq:solution}) remain valid.
Carrying out the approach of Sec.~\ref{ssec:comp_states} for this new lattice, one obtains all the regular states on the kagome lattice.
They are displayed in Fig.~\ref{fig:reg_kag} with the positions and weights of the Bragg peaks.
The equal time structure factor is depicted in the Extended Brillouin Zone (EBZ), drawn with thin lines in Fig.~\ref{fig:reg_kag}: the kagome lattice has 3 spins per unit cell of the underlying triangular lattice and the EBZ has a surface four times larger than the BZ of the underlying triangular Bravais lattice, drawn with dark lines.
Powder-averaged structure factors of the regular states are given in App.~\ref{App:powder}.

One regular state is colinear ($H_c^S=O(2)$):
\begin{itemize}
 \item the ferromagnetic (F) state of Fig.~\ref{fig:reg_kag_a}.
\end{itemize}

Two states with a zero total magnetization are coplanar ($H_c^S=\mathbb Z_2$):
\begin{itemize}
 \item the $\mathbf q=\mathbf 0$ state of Fig.~\ref{fig:reg_kag_b} has 3 sublattices of spins at $120^\circ$ and a 3 sites unit cell,
 \item the $\sqrt{3}\times\sqrt{3}$ state of Fig.~\ref{fig:reg_kag_c} has 3 sublattices of spins at $120^\circ$ and a 9 sites unit cell.
\end{itemize}

Three states with a zero total magnetization completely break $O(3)$ ($H_c^S=\{I\}$):
\begin{itemize}
 \item the octahedral state of Fig.~\ref{fig:reg_kag_d} has 6 sublattices of spins oriented toward the corners of an octahedra and a 12 sites unit cell,
 \item the cuboc1 state of Fig.~\ref{fig:reg_kag_e} has 12 sublattices of spins oriented toward the corners of a cuboctahedron and a 12 sites unit cell,
 \item the cuboc2 state of Fig.~\ref{fig:reg_kag_f} has 12 sublattices of spins oriented toward the corners of an cuboctahedron and a 12 sites unit cell. Note that the first neighbor spins have relative angles of $60^\circ$, in contrast to $120^\circ$ for the cuboc1 state.
\end{itemize}

Two continua of states with a non-zero total magnetization completely break $O(3)$ ($H_c^S=\{I\}$): \begin{itemize}
 \item the $\mathbf q=\mathbf 0$ umbrella states of Fig.~\ref{fig:reg_kag_g}, left,
 \item the $\sqrt{3}\times\sqrt{3}$ umbrella states of Fig.~\ref{fig:reg_kag_g}, right. \end{itemize}
These continua interpolate between the ferromagnetic state and the coplanar states Fig.~\ref{fig:reg_kag_b} and Fig.~\ref{fig:reg_kag_c}.

\subsection{Honeycomb lattice}
\label{ssec:hexa}

\begin{figure}
\begin{center}
 \subfigure[\;Ferromagnetic (F) state. $E=3J_1+6J_2+3J_3$.\label{fig:reg_hexa_a}]{
   \includegraphics[trim = 15mm 81mm 22mm 81mm, clip,width=.2\textwidth]{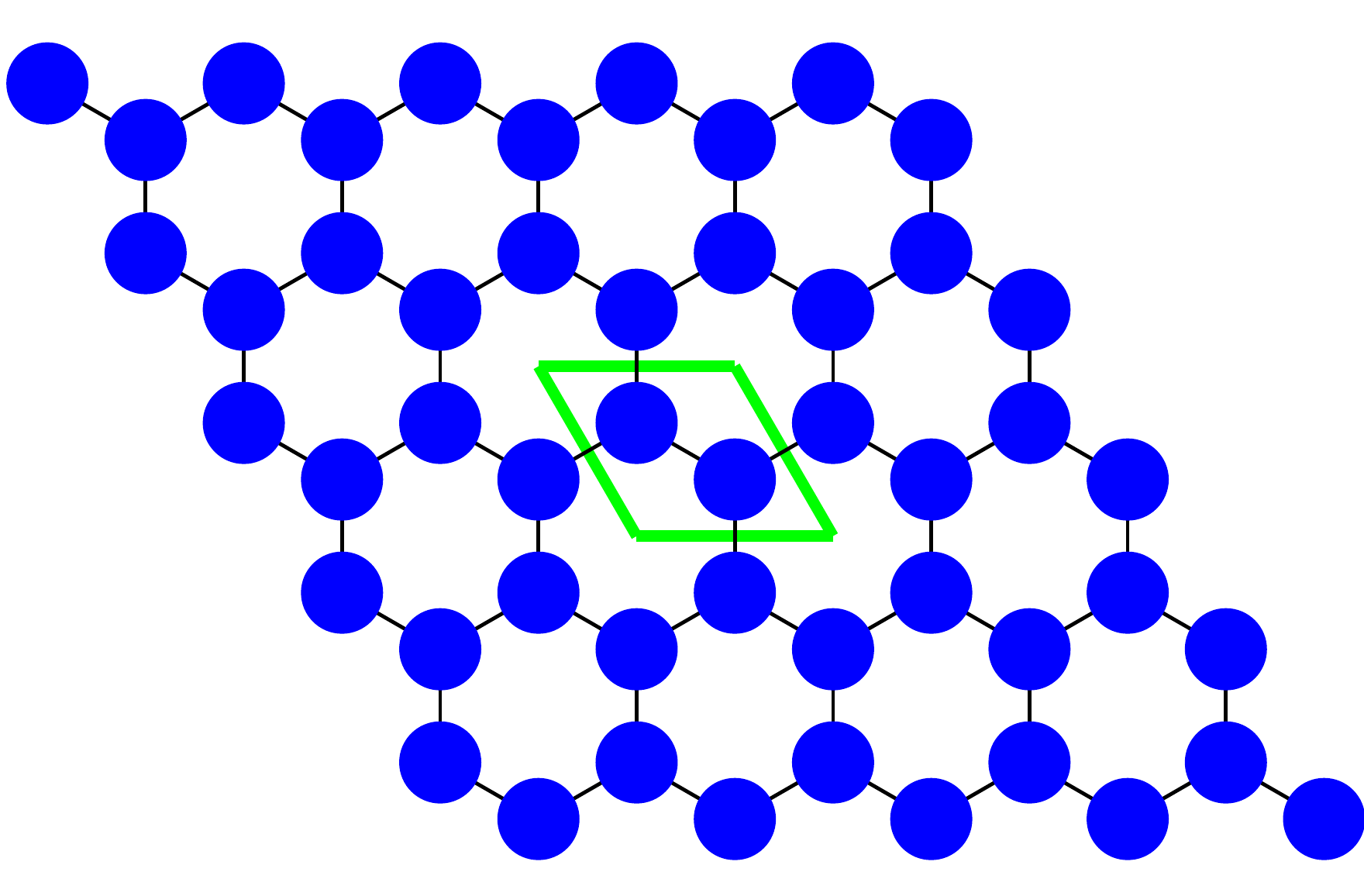}
   \includegraphics[trim = 15mm -50mm 22mm 54mm, clip,width=.08\textwidth]{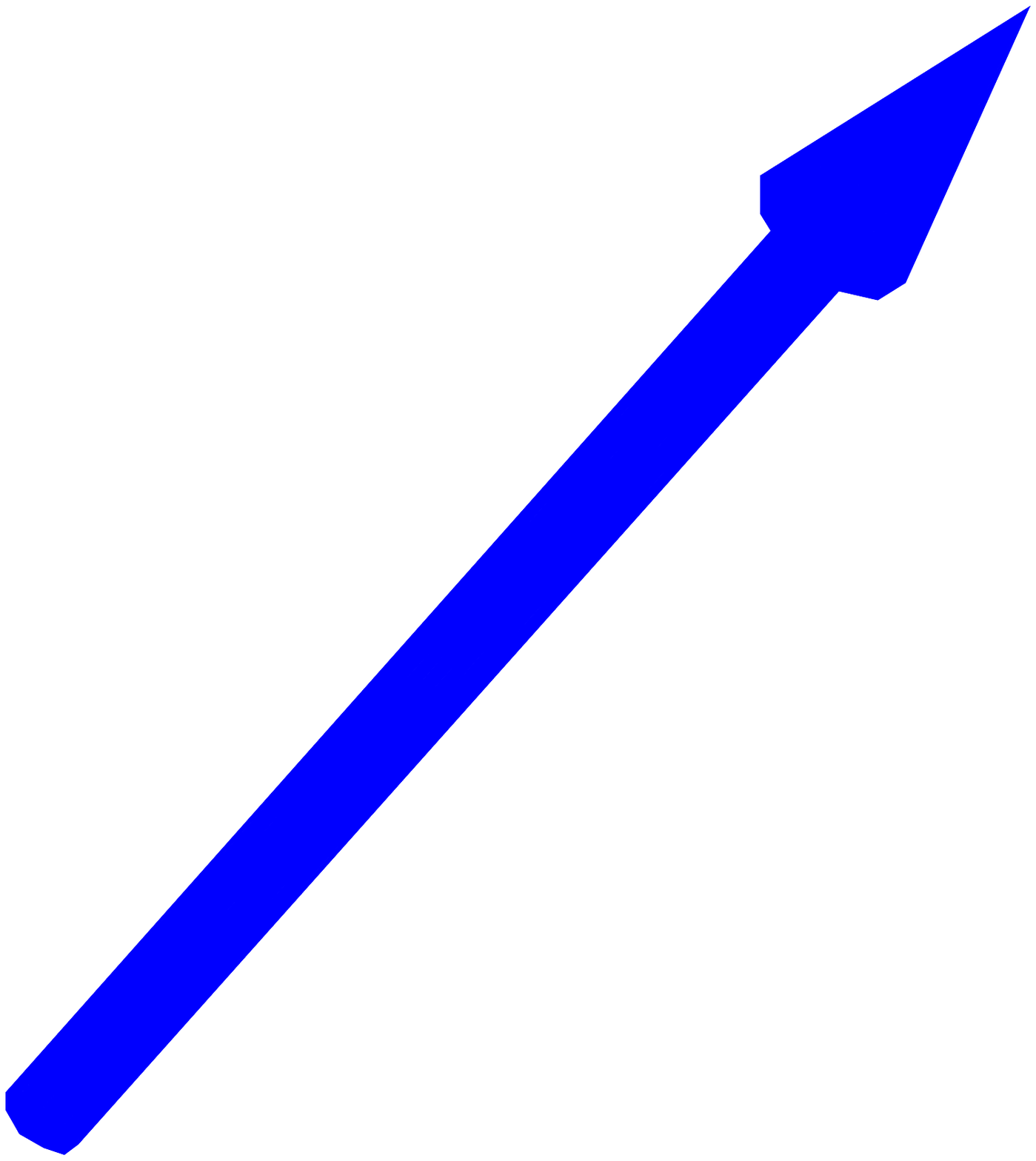}
   \includegraphics[trim = 93mm 56mm 83mm 44mm, clip,width=.14\textwidth]{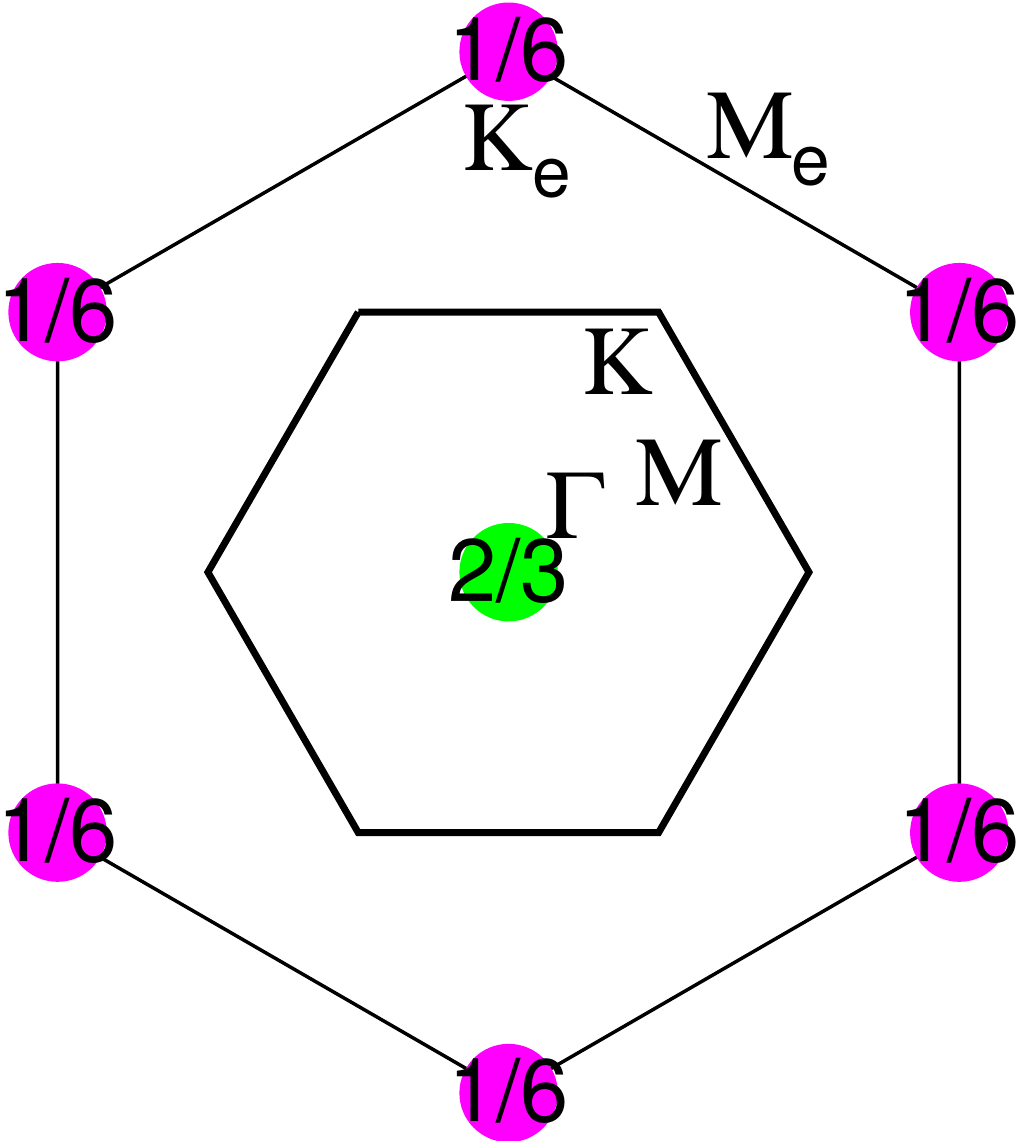}}
 \subfigure[\;Antiferromagnetic (AF) state. $E=-3J_1+6J_2-3J_3$.\label{fig:reg_hexa_b}]{
   \includegraphics[trim = 15mm 81mm 22mm 81mm, clip,width=.2\textwidth]{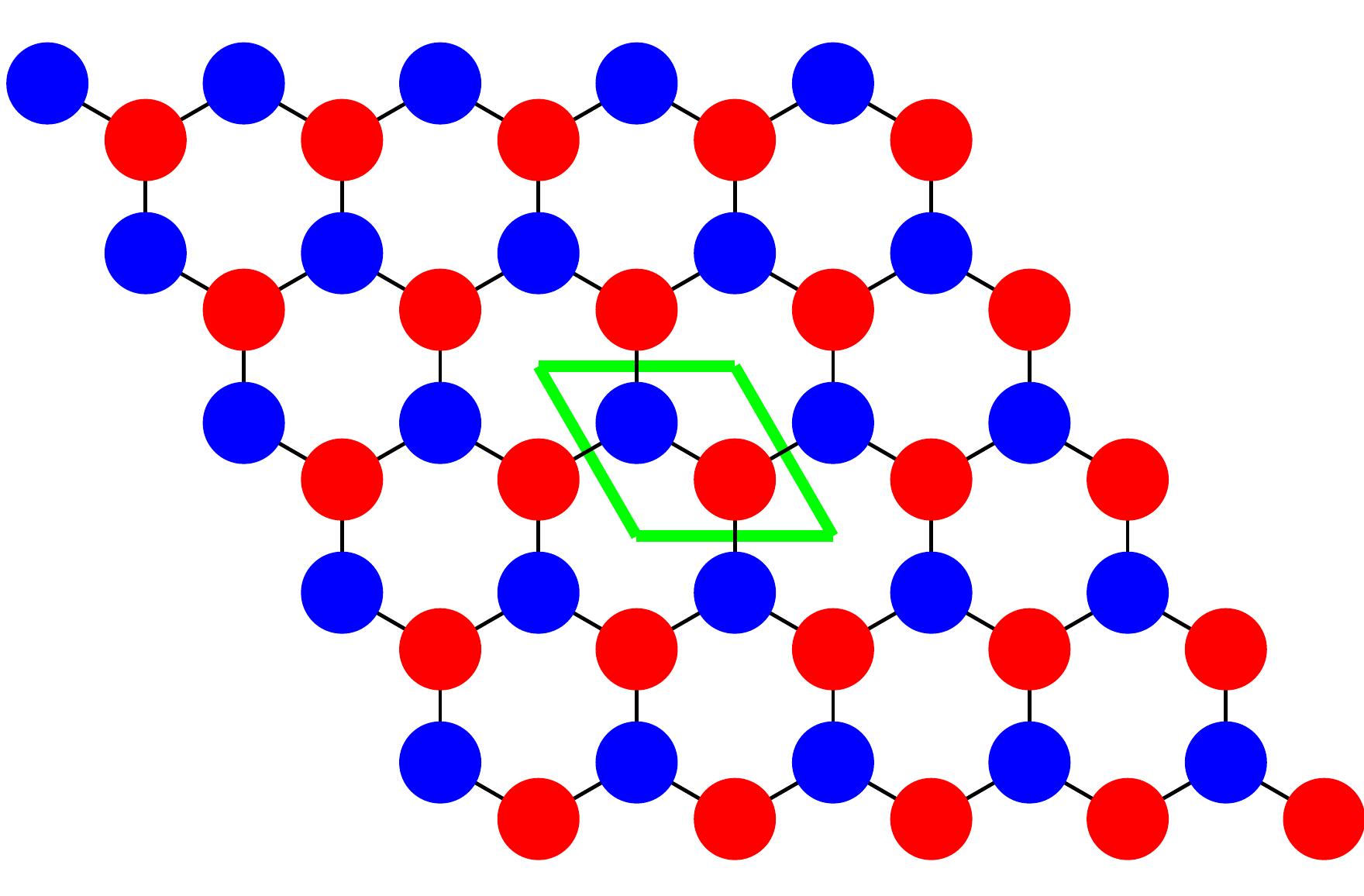}
   \includegraphics[trim = 22mm -50mm 29mm 44mm, clip,width=.08\textwidth]{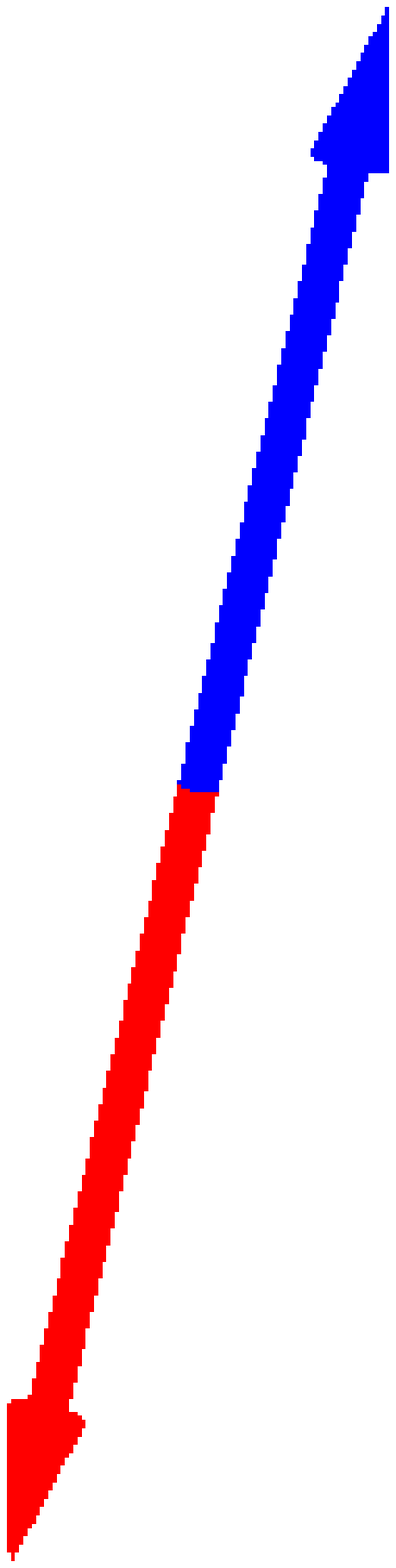}
   \includegraphics[trim = 93mm 56mm 83mm 44mm, clip,width=.14\textwidth]{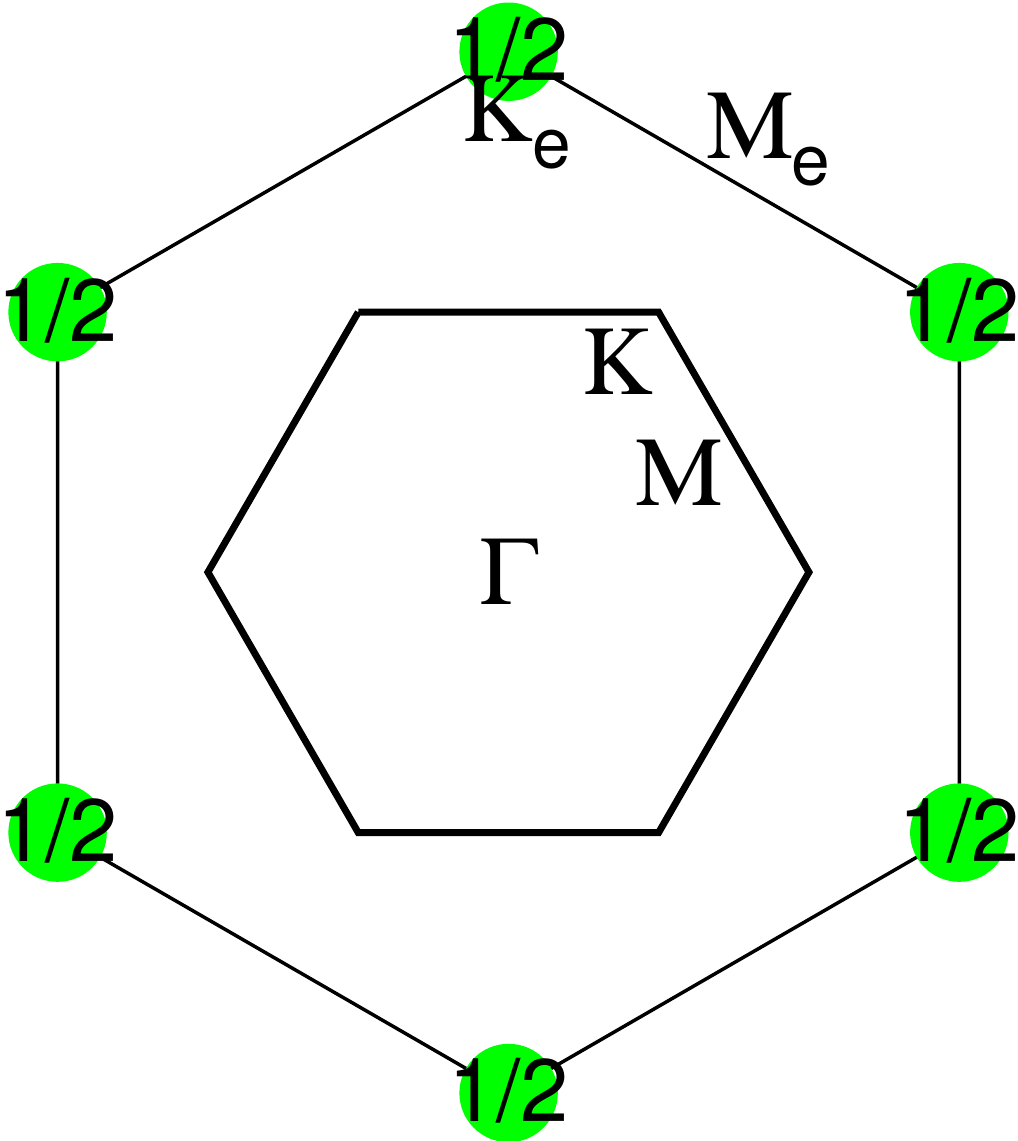}}
 \subfigure[\;Cubic state. $E=J_1-2J_2-3J_3$.\label{fig:reg_hexa_c}]{
   \includegraphics[trim = 15mm 81mm 22mm 81mm, clip,width=.2\textwidth]{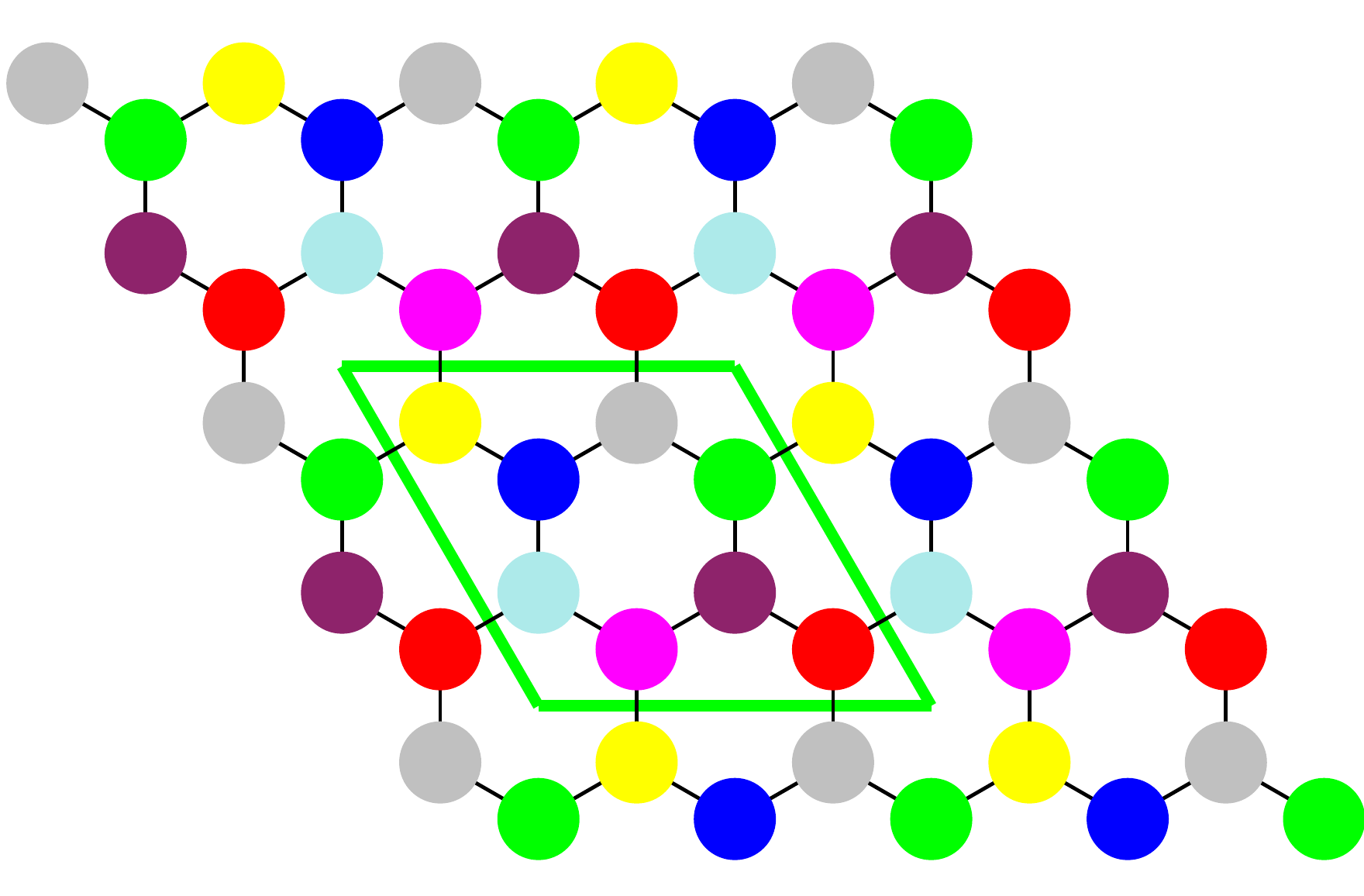}
   \includegraphics[trim = 12mm -40mm 19mm 44mm, clip,width=.08\textwidth]{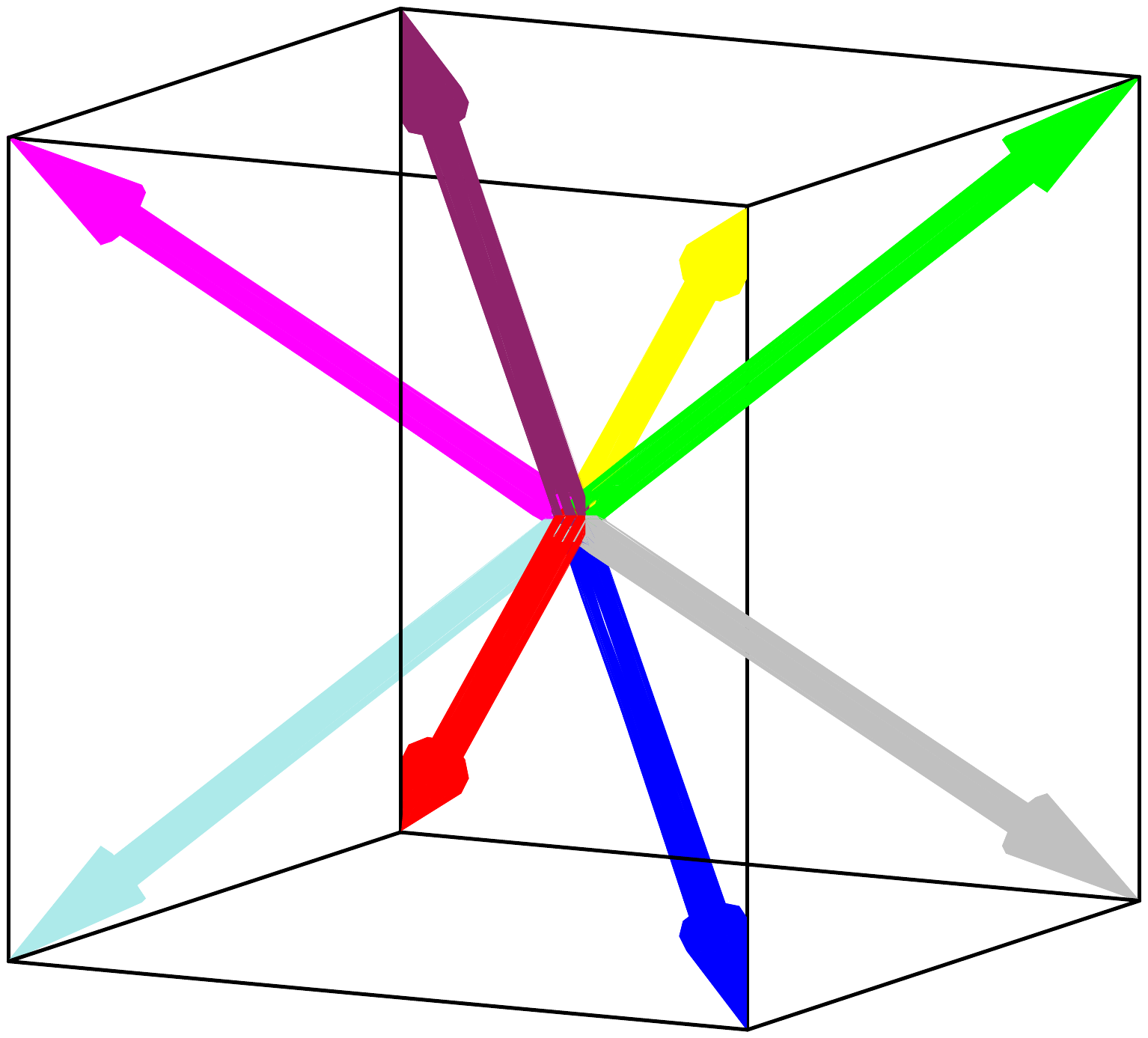}
   \includegraphics[trim = 93mm 56mm 83mm 44mm, clip,width=.14\textwidth]{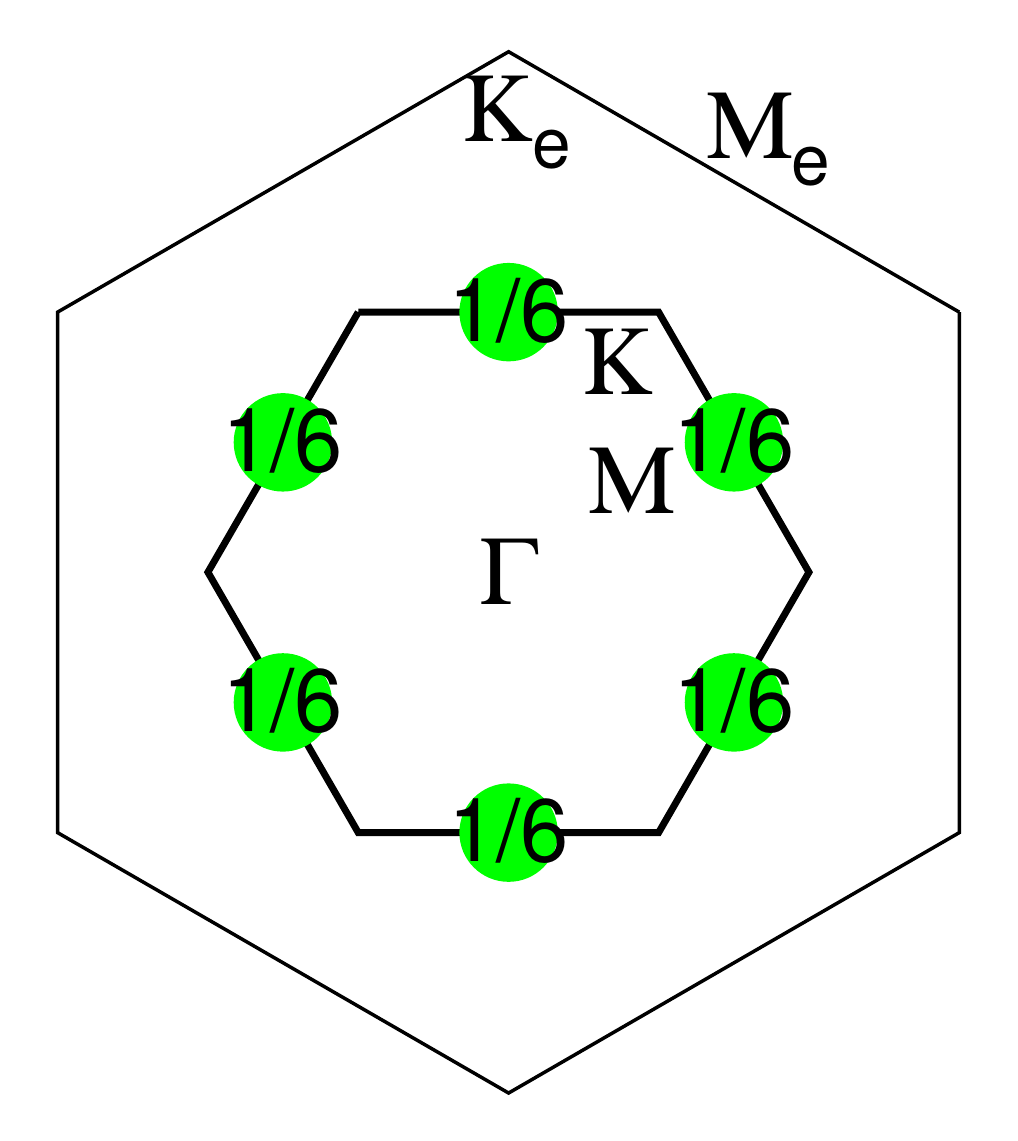}}
 \subfigure[\;Tetrahedral state. $E=-J_1-2J_2+3J_3$.\label{fig:reg_hexa_d}]{
   \includegraphics[trim = 15mm 81mm 22mm 81mm, clip,width=.2\textwidth]{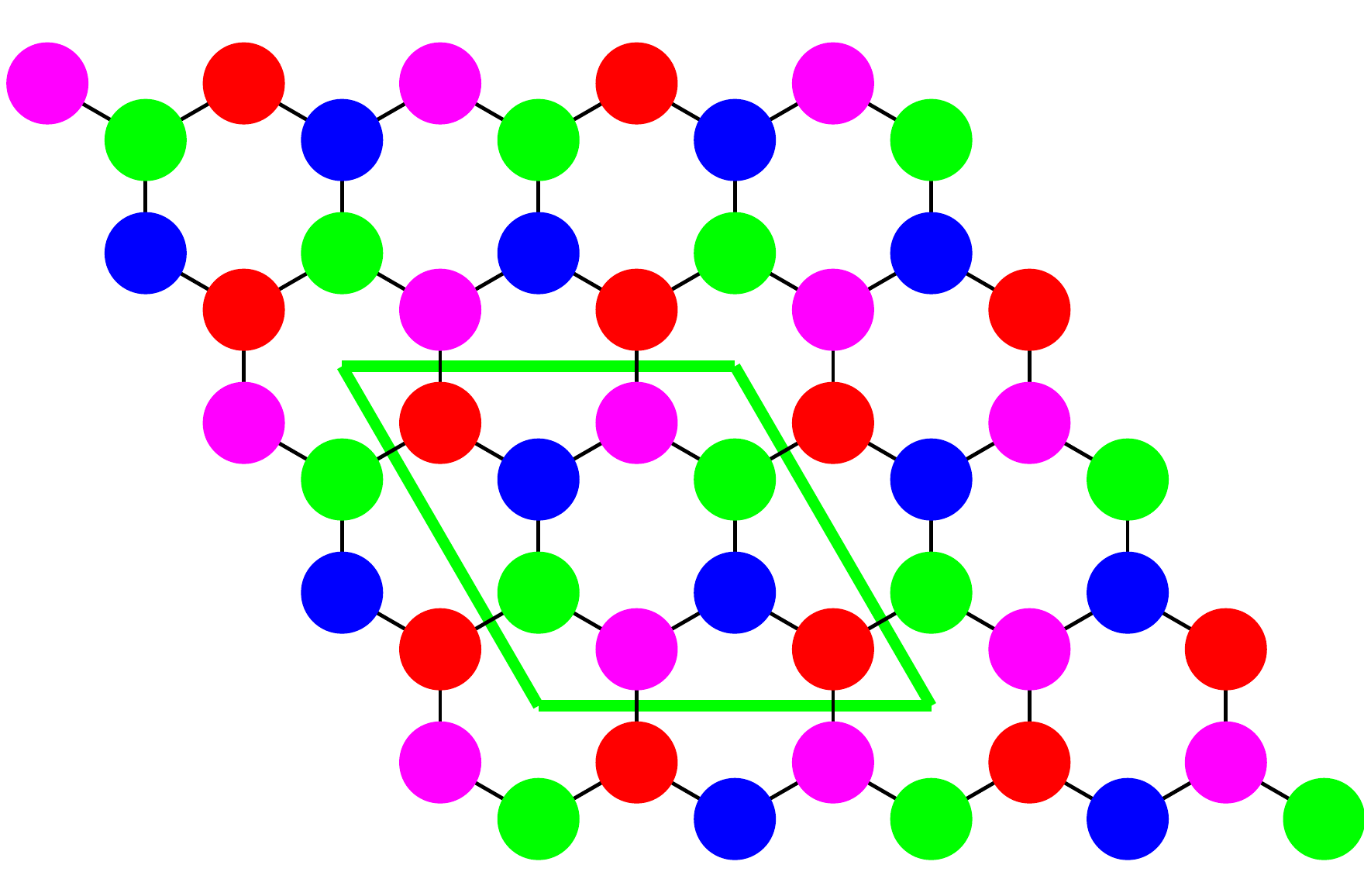}
   \includegraphics[trim = 12mm -30mm 19mm 34mm, clip,width=.089\textwidth]{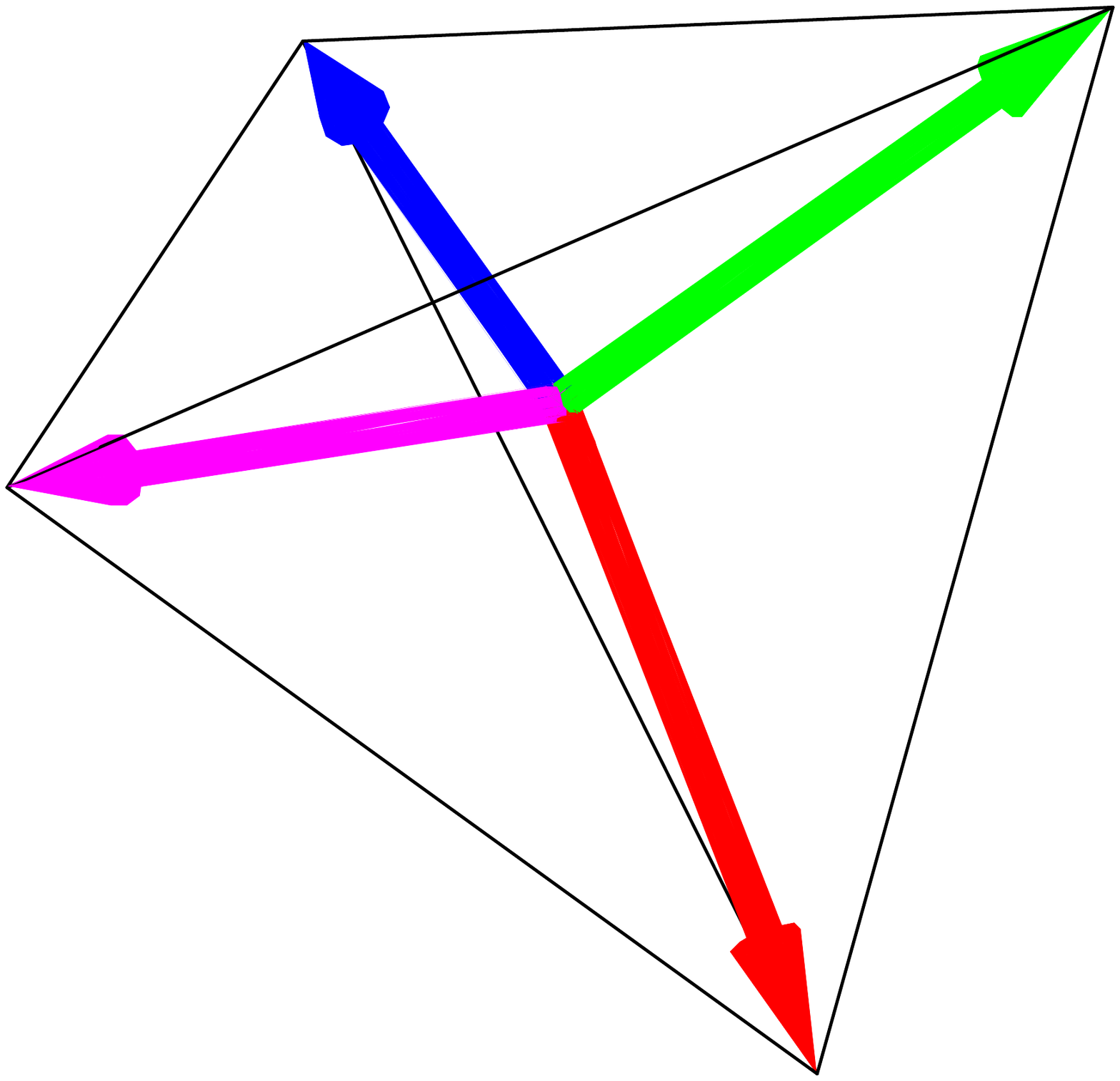}
   \includegraphics[trim = 93mm 56mm 83mm 44mm, clip,width=.14\textwidth]{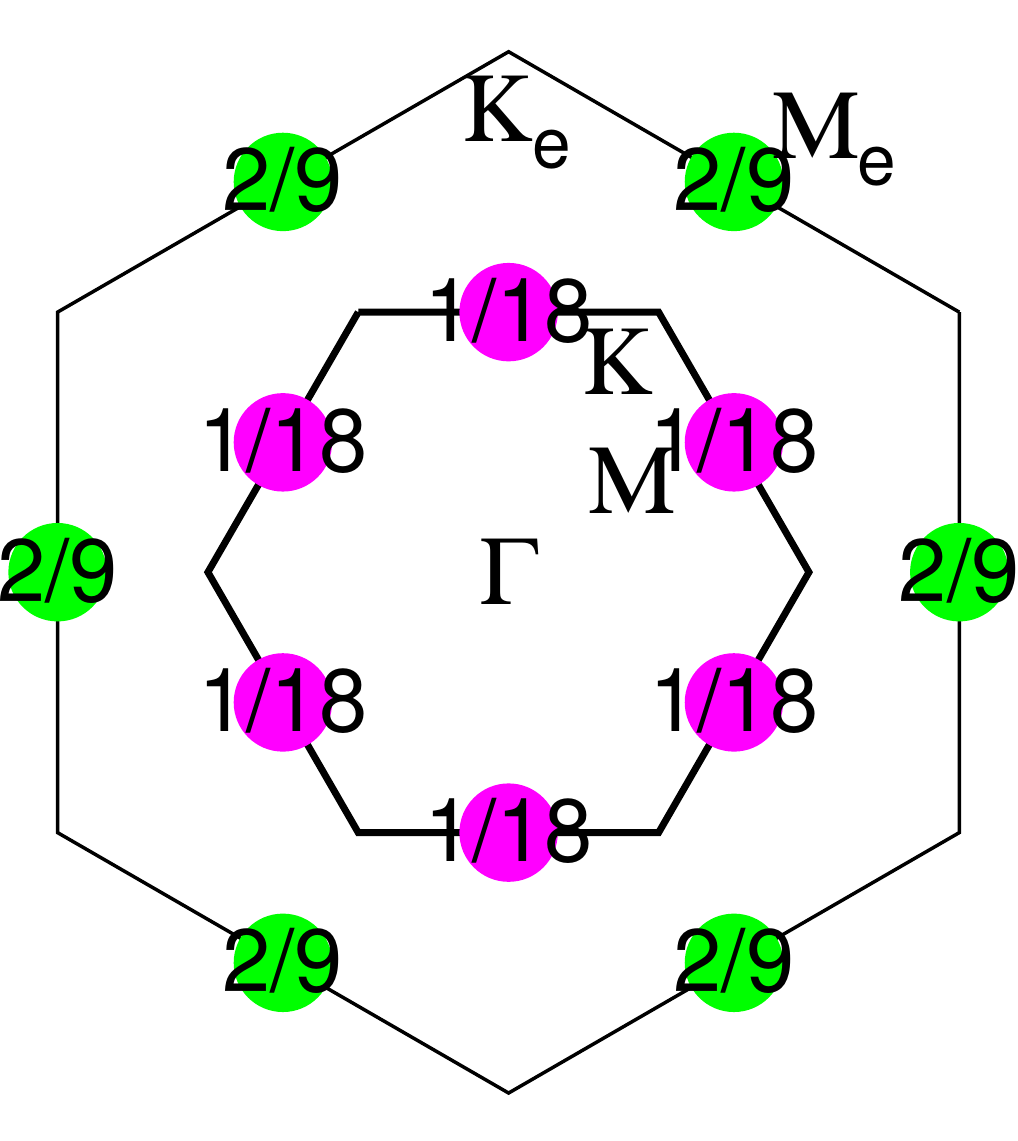}}
 \subfigure[\;V states\label{fig:reg_hexa_e}]{
   \includegraphics[trim = 15mm 81mm 22mm 81mm, clip,width=.2\textwidth]{hexa_2ssr.pdf}
   \includegraphics[trim = 22mm -20mm 29mm 54mm, clip,width=.08\textwidth]{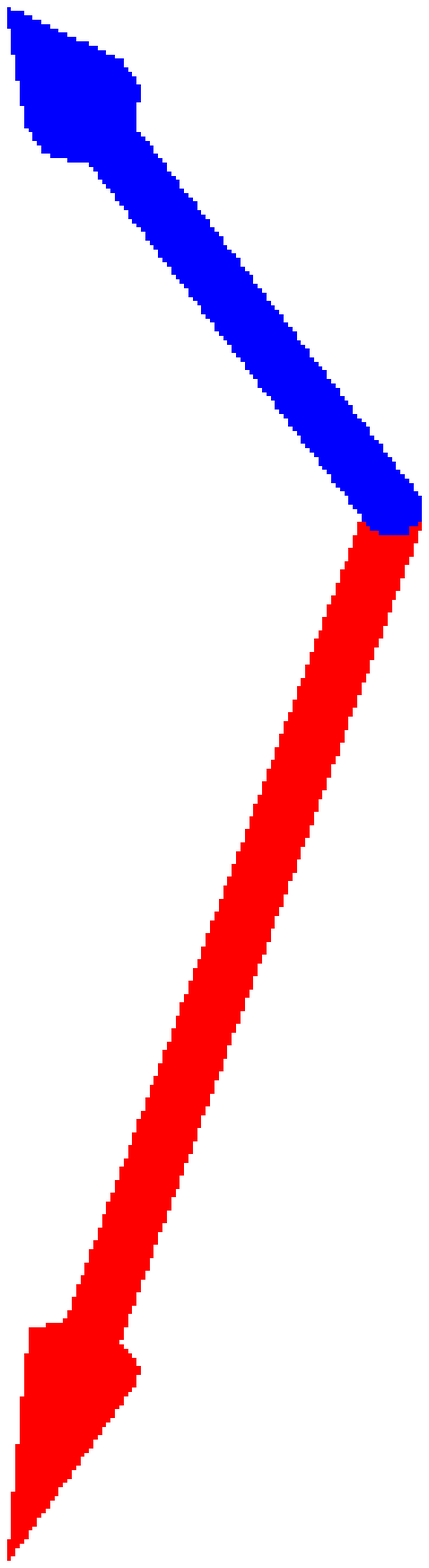}}
\caption{(Color online) Regular states on the honeycomb lattice and their equal time structure factors in the EBZ (see text).
The energies (per site) are given for a $J_1$-$J_2$-$J_3$ Heisenberg model.  }
 \label{fig:reg_hexa}
\end{center}
\end{figure}

All the regular states on the honeycomb lattice are depicted in Fig.~\ref{fig:reg_hexa} and listed below.
The EBZ is drawn with thin lines (its surface is three times larger than that of the BZ).

Two regular states are colinear ($H_c^S=O(2)$):
\begin{itemize}
 \item the ferromagnetic state of Fig.~\ref{fig:reg_hexa_a},
 \item the antiferromagnetic state of Fig.~\ref{fig:reg_hexa_b} has 2 sublattices of spins oriented in opposite directions and a 2 sites unit cell.
\end{itemize}

Two states with a zero total magnetization completely break $O(3)$ ($H_c^S=\{I\}$):
\begin{itemize}
 \item the cubic state of Fig.~\ref{fig:reg_hexa_c} has 8 sublattices of spins oriented toward the corners of a cube and a 8 sites unit cell,
 \item the tetrahedral state of Fig.~\ref{fig:reg_hexa_d} has 4 sublattices of spins oriented toward the corners of a tetrahedron and a 4 sites unit cell.
\end{itemize}

A continuum of states with a non-zero total magnetization partially breaks $O(3)$ ($H_c^S=\mathbb Z_2$):\begin{itemize}
 \item  the V states of Fig.~\ref{fig:reg_hexa_e}, which interpolate between the F and AF states. \end{itemize}

\subsection{Square lattice}
\label{ssec:square}

The symmetry group $S_L$ of the square lattice is distinct from that of the triangular lattice (see Fig.~\ref{fig:sym_latt}) and one has to determine its algebraic symmetry groups.
They are listed below:
\begin{subequations}
\label{eq:solution_carre}
 \begin{gather}
G_{T_1}=G_{T_2}=\varepsilon_1I,
G_\sigma=\varepsilon_\sigma R_{z\pi},
G_{R_4}=\varepsilon_R R_{x\pi},
\nonumber
\\
G_{T_1}=G_{T_2}=\varepsilon_1 R_{\mathbf z\pi\delta_1},
G_\sigma=\varepsilon_\sigma R_{\mathbf x\pi},
G_{R_4}=\varepsilon_R R_{\mathbf z\frac{\pi}{2} },
\nonumber
\\
G_{T_1}=G_{T_2}=\varepsilon_1R_{\mathbf z\pi\delta_1},
G_\sigma=\varepsilon_\sigma R_{\mathbf z\pi\delta_\sigma},
G_{R_4}=\varepsilon_R R_{\mathbf z\pi\delta_R},
\nonumber
\\
G_{T_1}=G_{T_2}=\varepsilon_1 R_{\mathbf z\pi},
G_\sigma=\varepsilon_\sigma R_{\mathbf z\pi\delta_\sigma},
G_{R_4}=\varepsilon_R R_{\mathbf x\pi},
\nonumber
\\
G_{T_1}=G_{T_2}=\varepsilon_1 R_{\mathbf z\pi},
G_\sigma=\varepsilon_\sigma R_{\mathbf x\pi},
G_{R_4}=\varepsilon_R R_{\mathbf z\pi\delta_R},
\nonumber
\\
G_{T_1}=G_{T_2}=\varepsilon_1 R_{\mathbf z\pi },
G_\sigma=\varepsilon_\sigma R_{\mathbf x\pi },
G_{R_4}=\varepsilon_R R_{\mathbf x\pi},
\nonumber
\\
G_{T_1}=G_{T_2}=\varepsilon_1 R_{\mathbf z\pi},
G_\sigma=\varepsilon_\sigma R_{\mathbf x\pi},
G_{R_4}=\varepsilon_R R_{\mathbf y\pi},
\nonumber
\\
\begin{array}{ll}
G_{T_1}=\varepsilon_1R_{\mathbf x\pi},
G_{T_2}=\varepsilon_1R_{\mathbf y\pi},\qquad\qquad\\
\qquad\qquad G_\sigma=\begin{pmatrix}
      0 & 1 & 0\\
      1 & 0 & 0\\
      0 & 0 & -\varepsilon_\sigma
      \end{pmatrix},
G_{R_4}=\begin{pmatrix}
      0 & e_1 & 0\\
      e_2 & 0 & 0\\
      0 & 0 & e_3
      \end{pmatrix},
\end{array}
\nonumber
 \end{gather}
\end{subequations}
where $\mathbf x$, $\mathbf y$, $\mathbf z$ are orthogonal vectors, $e_1$, $e_2$, $e_3$, $\varepsilon_1$, $\varepsilon_\sigma$, $\varepsilon_R=\pm1$ and $\delta_R$, $\delta_\sigma$, $\delta_1=0$ or $1$.

\begin{figure}[top!]
\vspace*{-0.5cm}
\begin{center}
 \subfigure[\;Ferromagnetic (F) state. $E=4J_1+4J_2+4J_3+14K$. \label{fig:reg_carre_a}]{
   \includegraphics[trim = 18mm 30mm 18mm 30mm, clip,width=.13\textwidth]{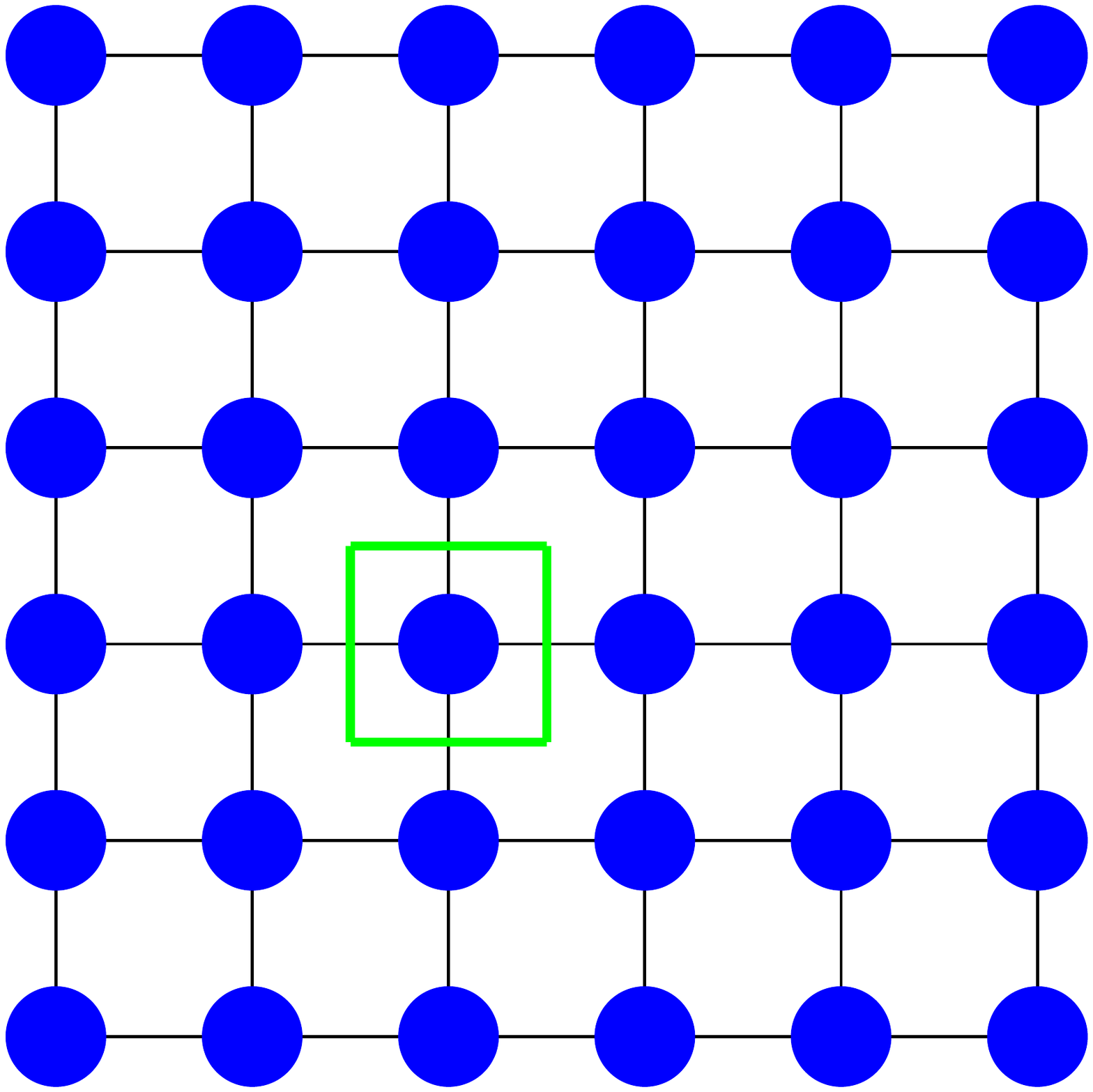}
   \includegraphics[trim = 16mm -50mm 16mm 0mm, clip,width=.08\textwidth]{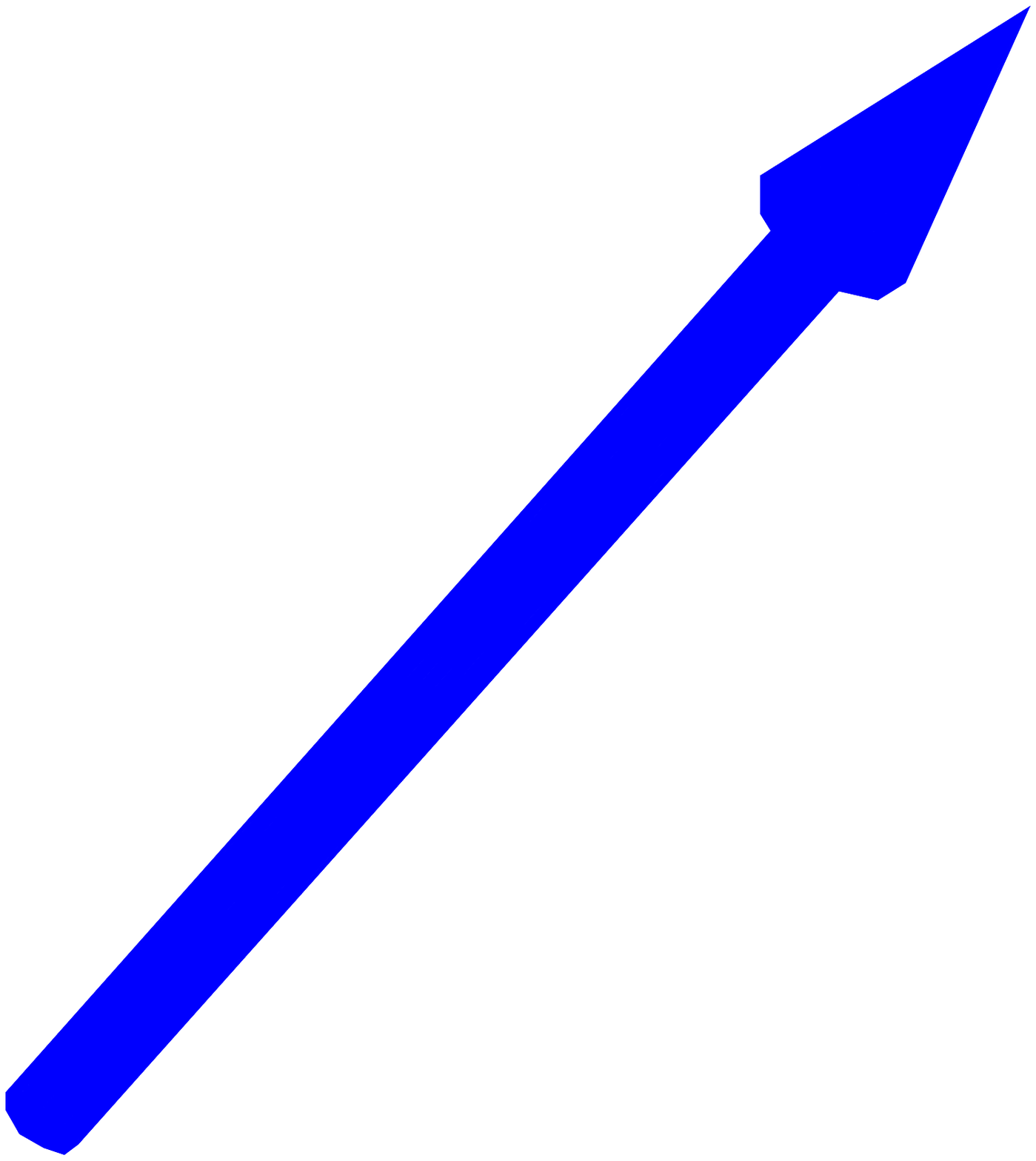}
   \includegraphics[trim = 90mm 51mm 80mm 51mm, clip,width=.14\textwidth]{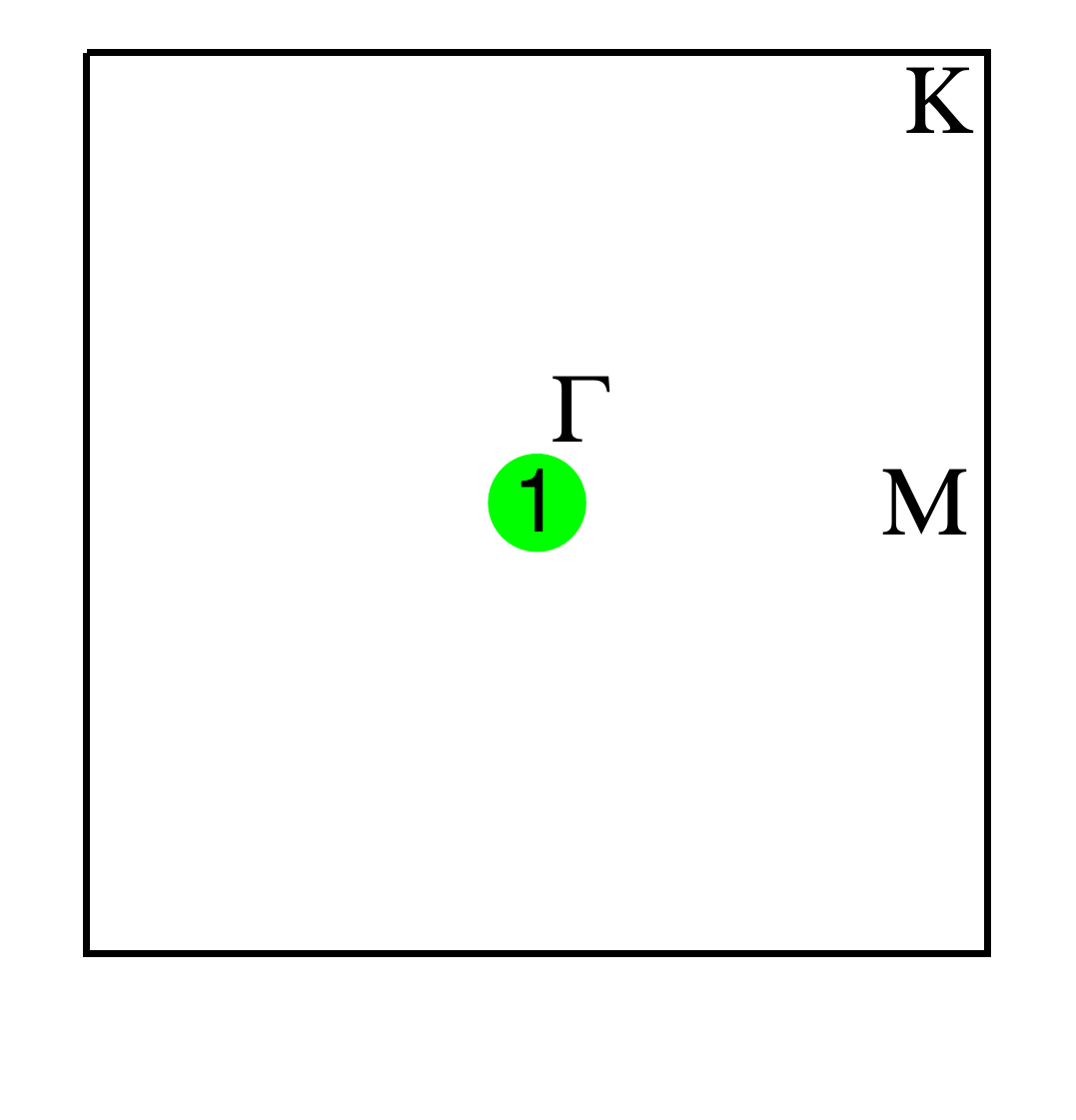}}
 \subfigure[\;$(\pi,\pi)$ N\'eel (AF) state. $E=-4J_1+4J_2+4J_3-2K$. \label{fig:reg_carre_b}]{
   \includegraphics[trim = 18mm 30mm 18mm 30mm, clip,width=.13\textwidth]{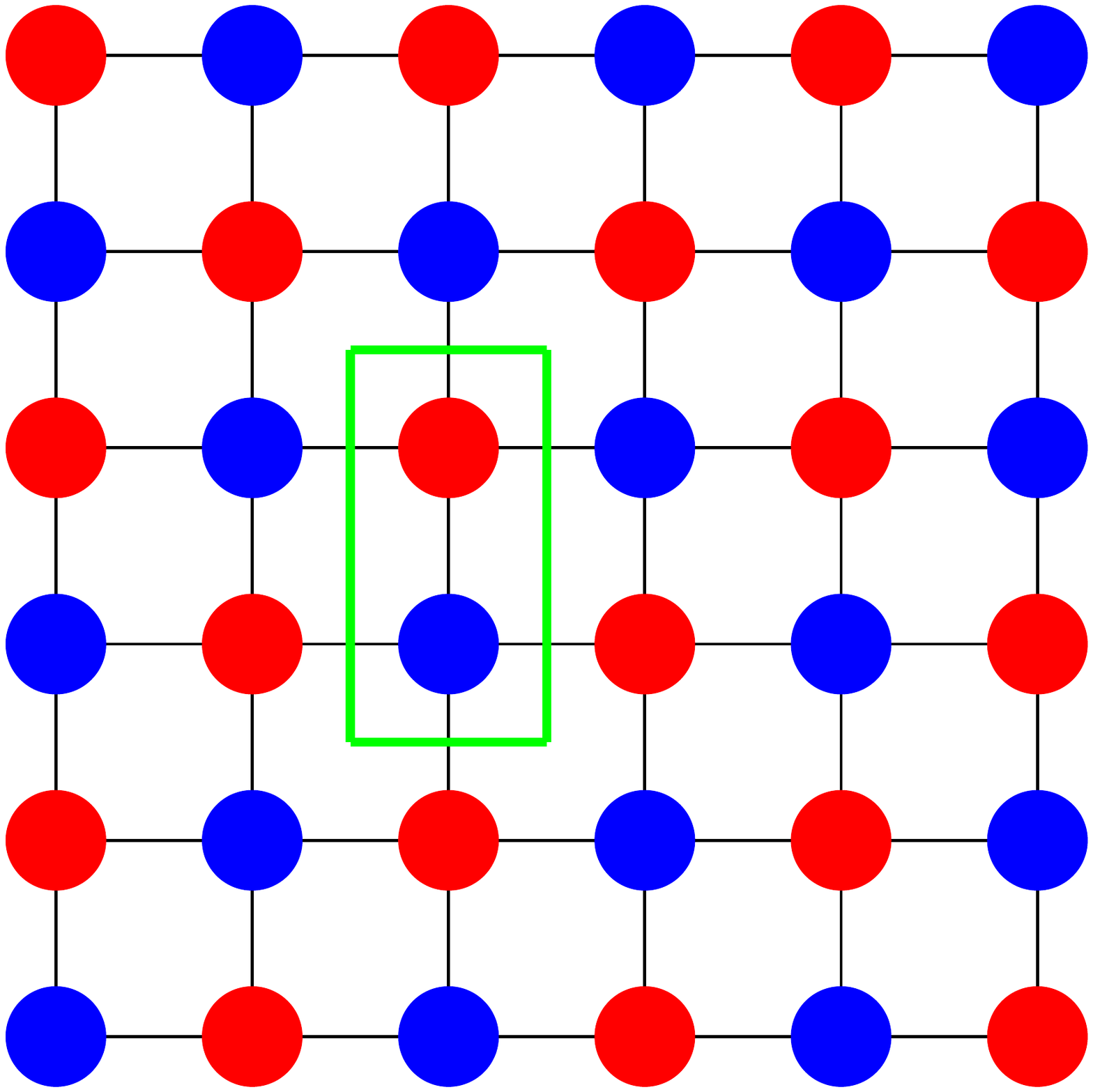}
   \includegraphics[trim = 13mm -50mm 19mm 0mm, clip,width=.08\textwidth]{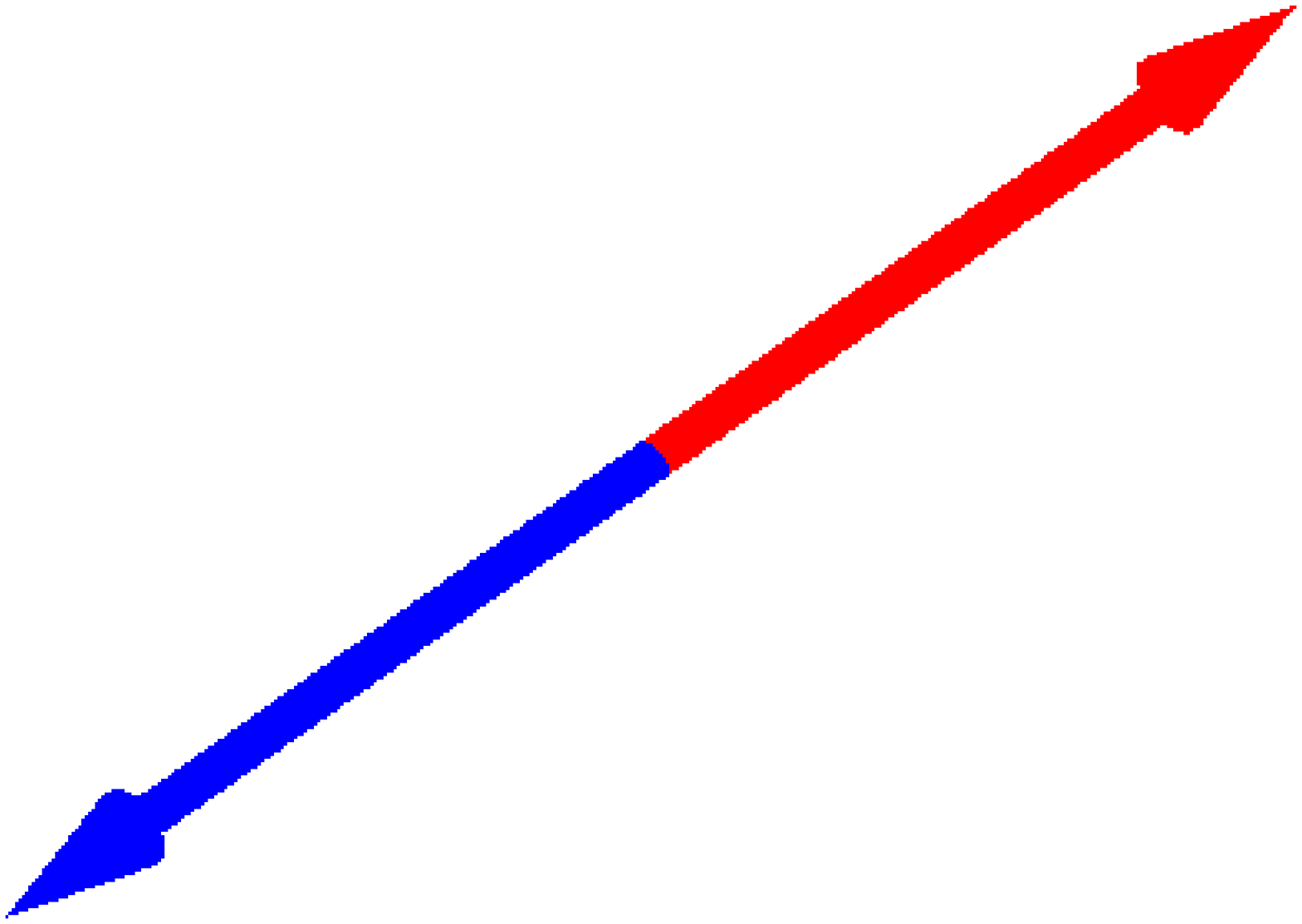}
   \includegraphics[trim = 90mm 51mm 80mm 51mm, clip,width=.14\textwidth]{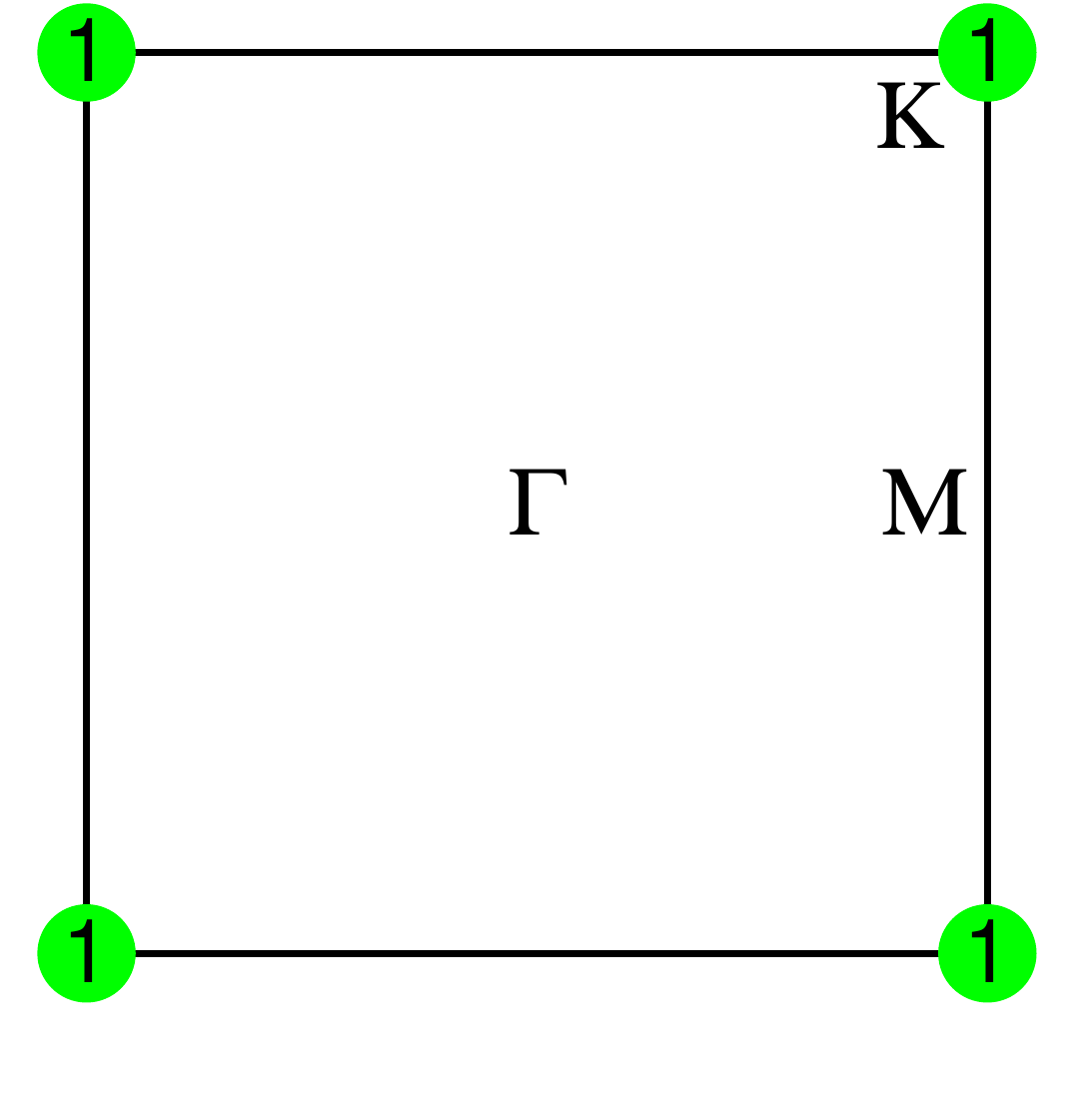}}
 \subfigure[\;Orthogonal coplanar state. $E=-4J_2+4J_3-6K$. \label{fig:reg_carre_c}]{
   \includegraphics[trim = 18mm 30mm 18mm 30mm, clip,width=.13\textwidth]{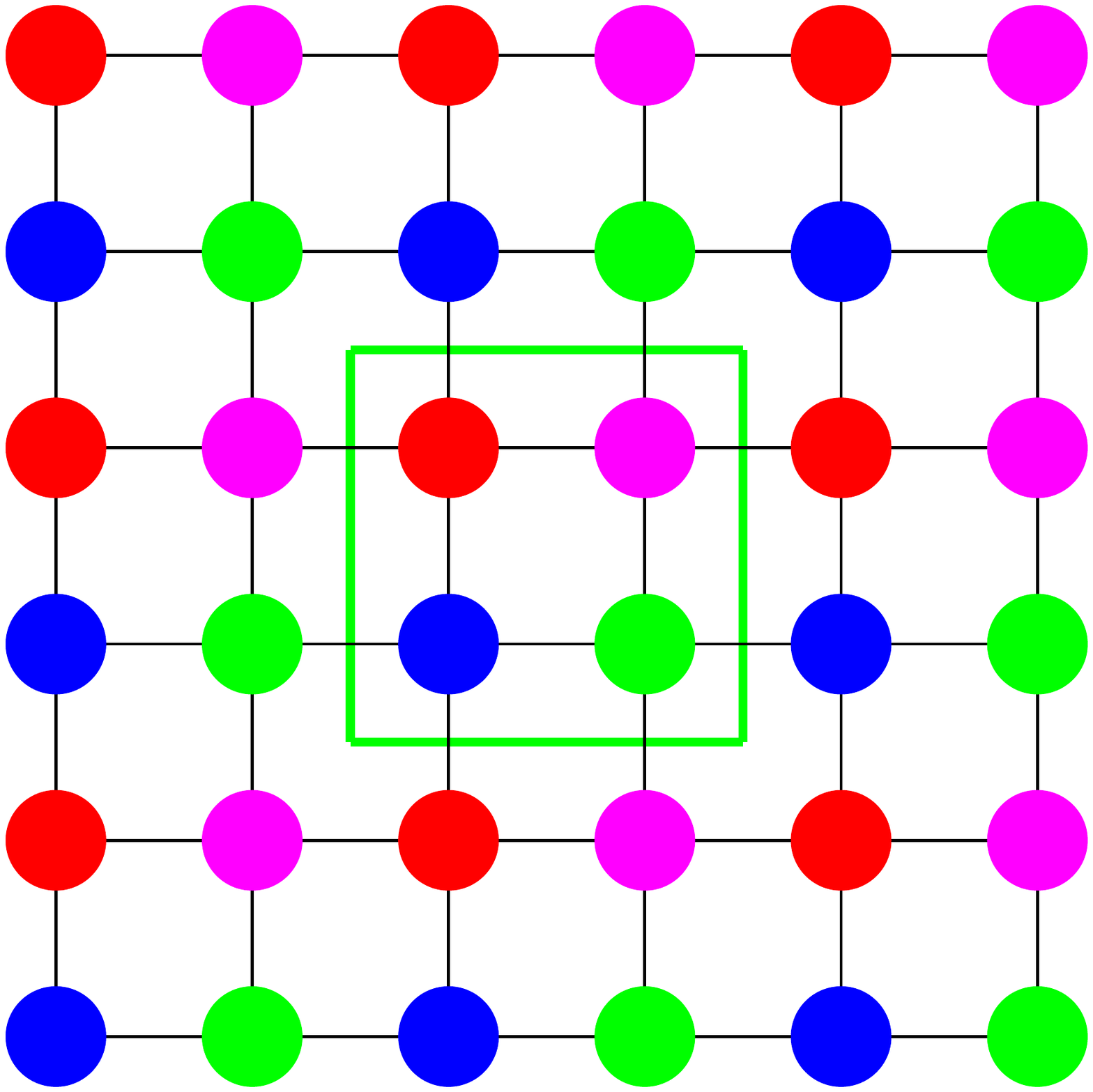}
   \includegraphics[trim = 15mm -50mm 20mm 0mm, clip,width=.08\textwidth]{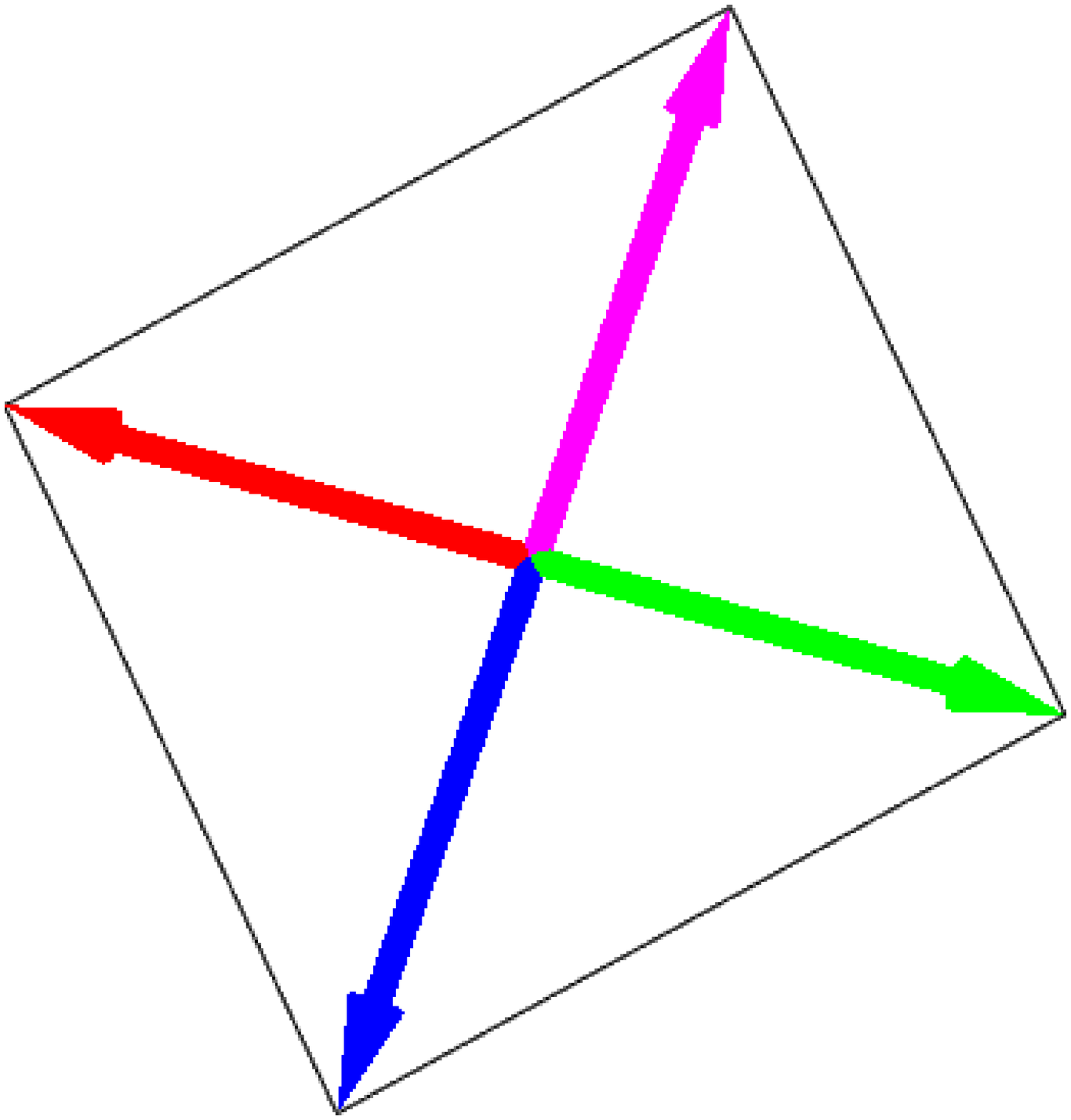}
   \includegraphics[trim = 80mm 41mm 80mm 41mm, clip,width=.14\textwidth]{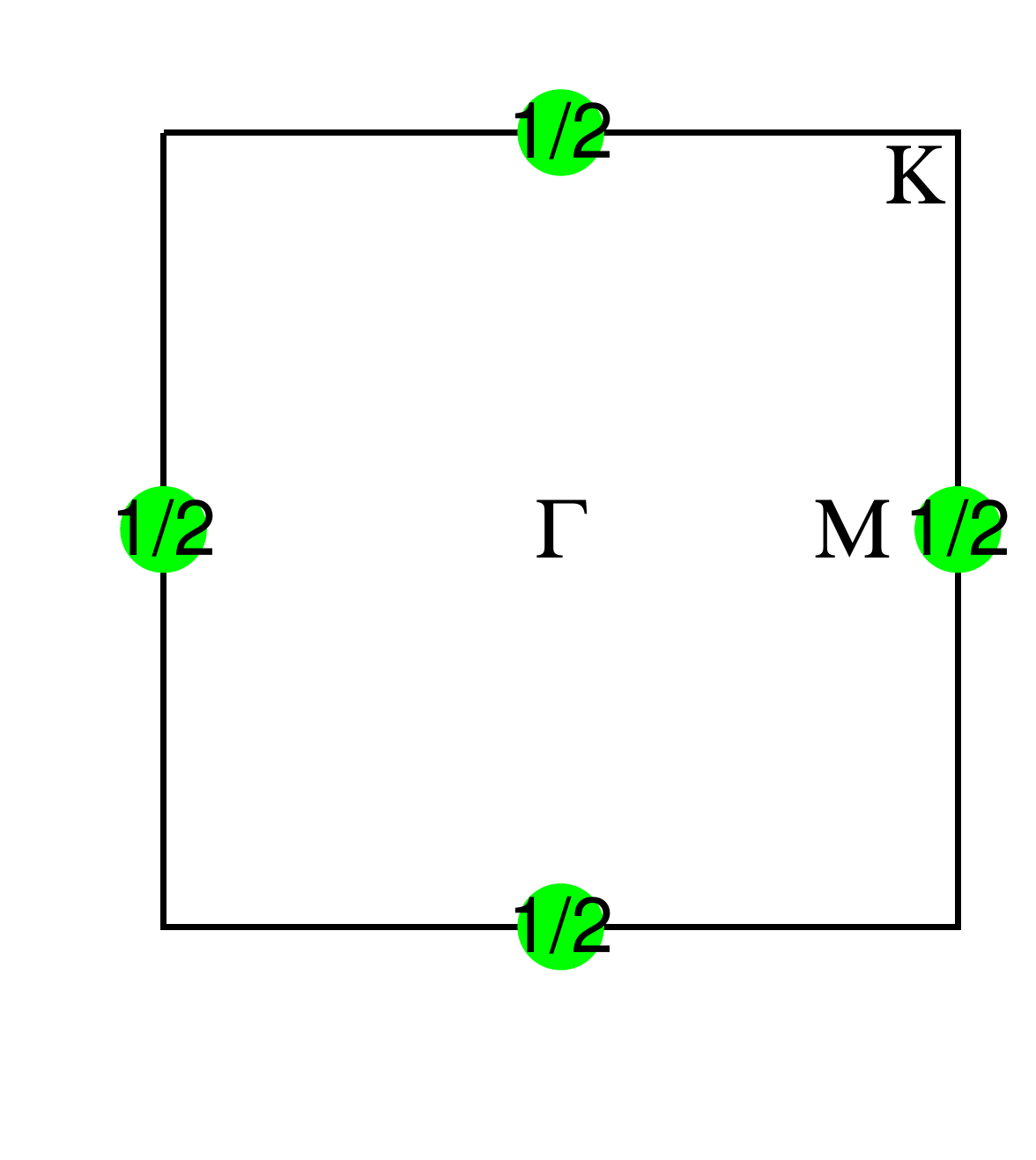}}
 \subfigure[\;V states\label{fig:reg_carre_d}]{
   \includegraphics[trim = 18mm 30mm 18mm 30mm, clip,width=.13\textwidth]{carre_2ssr.pdf}
   \includegraphics[trim = 20mm -50mm 20mm 0mm, clip,width=.08\textwidth]{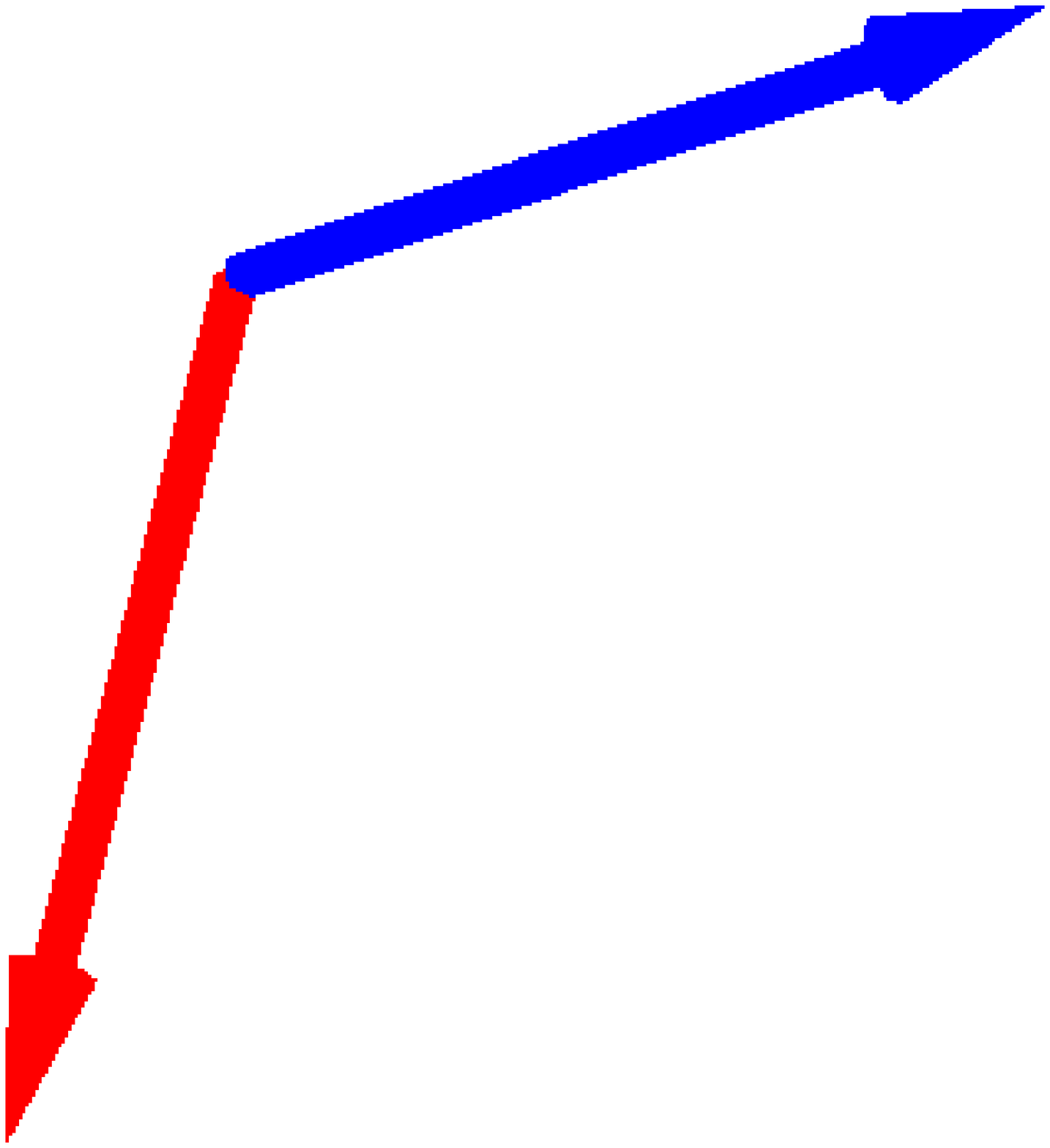}}\\
 \subfigure[\;Tetrahedral umbrella states ($AF~umbrellas$)\label{fig:reg_carre_e}]{
   \includegraphics[trim = 18mm 30mm 18mm 30mm, clip,width=.13\textwidth]{carre_4ssr.pdf}
   \includegraphics[trim = 20mm -50mm 20mm 0mm, clip,width=.08\textwidth]{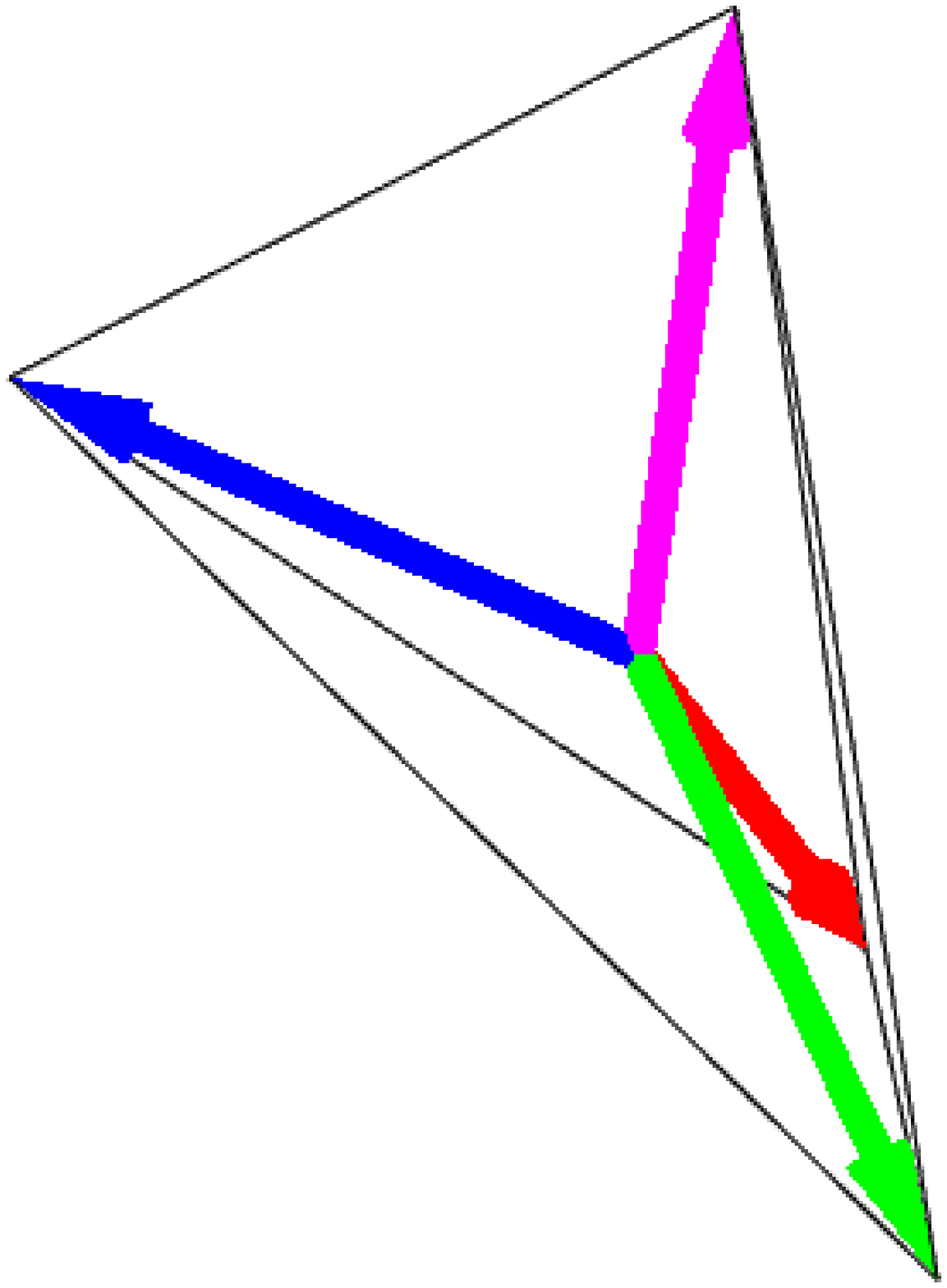}}
 \subfigure[\;Umbrella states ($F~umbrellas$) \label{fig:reg_carre_f}]{
   \includegraphics[trim = 18mm 30mm 18mm 30mm, clip,width=.13\textwidth]{carre_4ssr.pdf}
   \includegraphics[trim = 20mm -50mm 20mm 0mm,width=.08\textwidth]{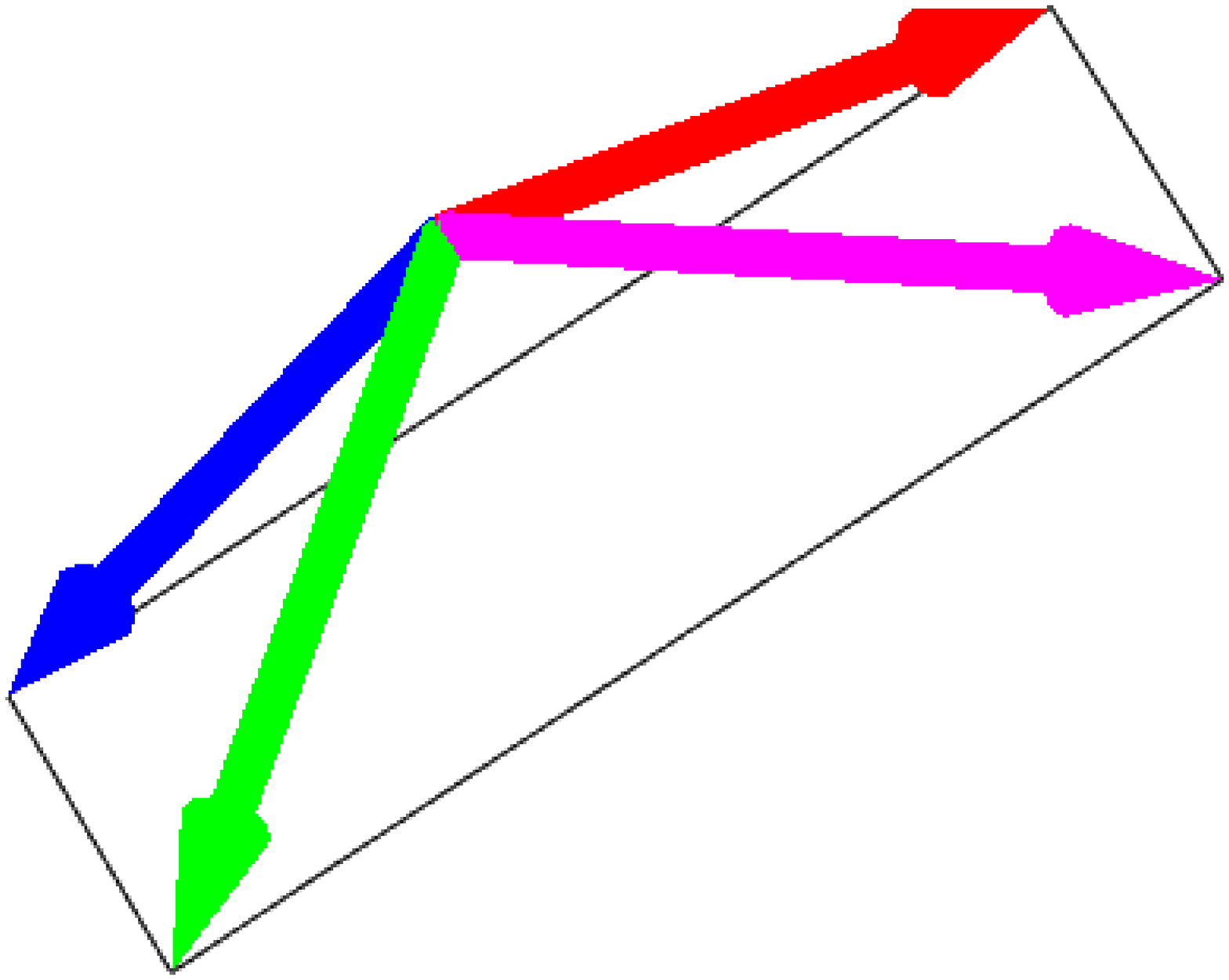}}\\
 \caption{(Color online) Regular states on the square lattice and their equal time structure factors in the square BZ.
The energy  per site of each structure is given as a function of the parameters of the models described in Sec~\ref{sec:energetics}. }
 \label{fig:reg_carre}
\end{center}
\end{figure}

Then, the construction of the compatible states leads to the regular states depicted in Fig.~\ref{fig:reg_carre} and listed below.

Two regular states are colinear ($H_c^S=O(2)$):
\begin{itemize}
 \item the ferromagnetic state of Fig.~\ref{fig:reg_carre_a},
 \item the $(\pi,\pi)$ N\'eel (AF) state of Fig.~\ref{fig:reg_carre_b} has 2 sublattices of spins oriented in opposite directions and a 2 sites unit cell.
\end{itemize}
One state with a zero total magnetization is coplanar ($H_c^S=\mathbb Z_2$):
\begin{itemize}
 \item the orthogonal coplanar state of Fig.~\ref{fig:reg_carre_c} has 4 sublattices of spins with angles of $90^\circ$ and a 4 sites unit cell.
\end{itemize}
Then we have three continua of states with different spin symmetry group $H_c^S$ :
\begin{itemize}
 \item the V states of Fig.~\ref{fig:reg_carre_d} have a non-zero total magnetization and partially break $O(3)$ ($H_c^S=\mathbb Z_2$). They interpolate between the F and the $(\pi,\pi)$ N\'eel states,
 \item the tetrahedral umbrella states of Fig.~\ref{fig:reg_carre_e} have a zero total magnetization and completely
break $O(3)$ ($H_c^S=\{I\}$). They interpolate between the $(\pi,\pi)$ N\'eel and
the orthogonal coplanar state,
 \item the 4-sublattice umbrella states of Fig.~\ref{fig:reg_carre_f} have a non-zero total magnetization and completely break $O(3)$ ($H_c^S=\{I\}$).
They interpolate between the F and the orthogonal coplanar state.
\end{itemize}

\subsection{Regular states with only translations}
\label{ssec:spiral}

When the lattice symmetry group is commutative, the construction of regular states
is particularly simple. This occurs if only translations are considered. In that case, one may chose some arbitrary directions for the spins of the reference unit cell.
Then, one has to chose an $O(3)$ element $G_{T_i}$ associated to each unit lattice translation $T_i$ in direction $i$ (with as many generators as space dimensions).
Assuming that $H_c^S=I$, and using the fact that the translations commute with each other, we find that the  $G_{T_i}$ also commute.
A first family of solutions consists in choosing a set of rotations with the same axis $\mathbf n$, and unconstrained angles.
This gives the conventional spiral states.
Thanks to the arbitrary choice of the spin directions in the reference unit cell, such states are not necessarily planar.

Finally, an other family of solutions can be obtained by choosing the  $G_{T_i}$
among the set of $\pi$-rotations with respect to some orthogonal spin directions, therefore insuring the commutativity.

All these solutions may be generalized by combining one or more $G_{T_i}$ with a spin inversion $-I$.
These generalized spiral states will be noted SS in the following.

\section{Geometrical remarks}
\label{sec:geom}

In this section, we discuss some geometrical properties of regular states.

\subsection{Groups and polyhedra}

From a regular state $c$, one can consider the set $\Sigma\subset \mathcal A$ of  all the different orientations taken by the spins.
We assume that $c$ has a finite number of sublattices/spin directions, so that $\Sigma$ is finite.
For a three component spin system, $\Sigma$ is just a set of points on the unit sphere ${\mathcal S}_2$,
as displayed in Figs.~\ref{fig:reg_tri}, \ref{fig:reg_kag}, \ref{fig:reg_hexa} and \ref{fig:reg_carre}.
$\Sigma$ may be a single site, the ends of a segment, the corners of a polygon or a polyhedra.

The four lattices studied here share some special properties: all the sites and all first-neighbors bonds are equivalent (linked by a $S_L$ transformation).
Due to this equivalence, $\Sigma$ also form a segment/polygon/polyhedron with
equivalent vertices and bonds.\footnote{Notice that nearest neighbors on the lattice do not necessarly correspond to nearest neighbor spin directions in spin space.}
A polyhedron with this property is said to be quasi-regular.
If the elementary plaquettes of the lattices are also equivalent (as in the triangular, square and hexagonal lattices, but not in the kagome lattice where both triangular and hexagonal elementary plaquettes are present) and if $\Sigma$ is a polyhedron, its faces should also be equivalent.
$\Sigma$ must then be one of the five regular convex polyhedra (Platonic solids):\cite{coxeter} tetrahedron, cube, octahedron, dodecahedron or icosahedron.\footnote{Again, the plaquettes of the lattice need not to map to the faces of the polyhedron.}

We now only consider the case where $H_c^S=\{I\}$ (this condition can always be verified by reducing ${\mathcal A}$ to its elements invariant by $H_c^S$ and by consequently modifying $S_S$). 
Clearly, the lattice symmetries constraint the possibilities for the set $\Sigma$, since each lattice symmetry $X$ permutes the sites in $\Sigma$ but leaves it globally unchanged.\footnote{This relation is particularly easy  to visualize in the case of the  tetrahedral state on the triangular lattice, since both the lattice and the polydedron $\Sigma$ have triangular plaquettes/faces: one can put  a tetrahedron with a face posed onto a lattice face.
Then, one {\it roll} the tetrahedron over the lattice to obtain a spin direction at each lattice site. Notice that such a construction would {\it not} work with a cube on the square lattice (and indeed, there is no such eight-sublattice regular state on the square lattice,
see Sec.~\ref{ssec:square}).}
But since the state $c$ is regular, these permutations can also be achieved by a spin symmetry in $S_S$, and the symmetry group $S_\Sigma$ of $\Sigma$ should be viewed as a finite subgroup of $S_S$.

For $S_S=O(3)$, the classification of these subgroups -- called point groups -- is a classical result in geometry,\cite{coxeter} it contains seven groups (related to the three symmetry groups of the five regular polyhedra) and seven infinite series (conventionally noted $C_n$, $C_{nv}$,  $C_{nh}$, $D_n$ $D_{nh}$ $D_{nd}$ and $S_n$ with $n\in \mathbb{N}$. They  are related to the cyclic and dihedral groups).
Of course, the non planar regular states we have discussed so far (Sec.~\ref{ssec:ex_tri} and \ref{sec:reg_states}) fall into this classification.
For instance, the three- and four- sublattice umbrella states of Fig.~\ref{fig:reg_tri_d}, \ref{fig:reg_carre_e} and \ref{fig:reg_carre_f} correspond to $C_{3v}$, $D_{2d}$ and $C_{4v}$ (with respectively 6, 8 and 8 elements).
The cubic, octahedral and cuboctahedronl states correspond to the symmetry group of the cube (48 elements), and the tetrahedral state corresponds (of course) to its own symmetry group.

\subsection{Regular states and  representation of the lattice symmetry group}

We again focus on three-component spin systems with a spin symmetry group $S_S=O(3)$.
In a regular state $c$  each lattice symmetry $X$ can be associated to a matrix $G_X$ in $O(3)$.
Now, as in Sec.~\ref{ssec:asg}, we can compare the actions of two lattice symmetries $X$ and $Y$.
$G_X G_Y G_{XY}^{-1}$ belongs to $H_c^S$.
By an appropriate choice of $G_X$, it is possible to obtain $G_X G_Y=G_{XY}$, which implies that $G$ is a {\it representation} of the lattice symmetry group $S_L$.
Its dimension is 1 for a colinear state, 2 for planar states, and 3 for the others.
Is this representation reducible ?
If yes, it must contain at least one representation of dimension 1 (because the maximal dimension considered here is 3), thus there exist at least one spin direction which is stable under all the spin symmetry operations spanned by $G_X$ with $X\in S_L$.
Except in the trivial colinear case, one can easily check that it is the case only for the states belonging to a continuum.
For the V-states, $G$ is the tensor product of a trivial and a non trivial 1d representation of $S_L$ (and of any 1d third one).
For the umbrella states, $G$ is the tensor product of a trivial 1d and a 2d irreducible representation (IR).
For the tetrahedral state of Fig.~\ref{fig:reg_carre_e}, $G$ is the tensor product of a non-trivial 1d and a 2d IR representation.
For the other cases, the associated representation is irreducible.

There is another context where antiferromagnetic Néel states are known to be related to irreducible representations.
If a quantum antiferromagnet has a GS with  long-range Néel order, its spectrum displays a special structure, called ``tower of states''.\cite{bllp94,lblp95} It reflects the fact that a symmetry breaking Néel state is a linear combination of specific eigenstates with different quantum numbers
describing the spatial symmetry breaking, and with different values of the total spin $S$,
describing the $SU(2)$ symmetry breaking.
If such a quantum system has a GS with a regular Néel order, its tower of state
should have an $S=1$ state with the same quantum numbers as those of the irreducible representation $X\mapsto G_X$ discussed above.
The reason why this representation shows up in the $S=1$ sector of the tower of state is because
$S=1$ corresponds to the action of the lattice symmetries onto a three-dimensional vector, as the classical spin directions.

\section{Energetics}
\label{sec:energetics}

As discussed in the introduction, there is no simple way to find the GS
of a classical spin model if the lattice is not a Bravais lattice, and/or if spin-spin  interactions are not simply quadratic in the spin components.
So far, we have discussed regular states from pure symmetry considerations, but in Sec.~\ref{ssec:stat} we show that, under some rather general conditions, a regular state is a {\it stationary} point for the energy, whatever the Hamiltonian (provided it commutes with the lattice symmetries).

In addition, we argue that regular states  are good candidates to be {\it global energy minima}.
To justify this, we first discuss a rigorous energy lower bound (Sec.~\ref{ssec:Espiral}) for Heisenberg like Hamiltonians and investigate in Sec.~\ref{ssec:j1j2j3} several Heisenberg models with further neighbor interactions ($J_1$, $J_2$, $J_3$, etc.) on non-Bravais lattices such as the hexagonal and kagome lattices.
In large regions of the phase diagrams, one regular state energy reaches the lower bound and is one (may be not unique) exact GS.

\subsection{A condition for a regular state to be ``stationary'' with respect to
small spin deviations}
\label{ssec:stat}

To  address the question of energetic stability of regular states, we
give some conditions under which an infinitesimal variation of the spin directions would not change the energy  (necessary condition to have a GS).
To simplify the notations we consider an Heisenberg model with some competing interactions (such as in Eq.(~\ref{eq:Heisenberg})), but the arguments easily generalize to
multi-spin interactions of the form  $(\mathbf S_i\cdot \mathbf S_j) (\mathbf S_k\cdot \mathbf S_l)\dots$ (respecting the lattice symmetries).

We assume that there is a non trivial lattice symmetry $X$ which leaves one site $i$ unchanged: $X(i)=i$ (existence of a non-trivial point group).
In addition, we assume that a spin rotation $R_s$ of axis $\mathbf n$ and angle $\theta\ne 0$ can be associated to $X$ in order to have $R_sX c=c$.
These conditions insure that the invariant direction of $R_s$ is $\mathbf n=\pm\mathbf S_i$.
Excepted states belonging to a continuum, all regular states verify these conditions on the lattices we have studied .

With these conditions, the derivatives of the energy with respect to the spin directions vanish.
The proof is as follows.
One considers the local field $\mathbf h_i=\frac{\partial E}{\partial \mathbf S_i}$ which is experienced by the spin $i$.
$\mathbf h_i$ is a linear combination of the $\mathbf S_j$ where $j$ runs over the sites which interact with the site $i$:
\begin{equation}
\mathbf h_i=\sum_{d} J_d \sum_{j\in N_d(i)} \mathbf S_j,
\end{equation}
where $N_d(i)$ is the set of the neighbors of $i$ at distance $d$ on the lattice.
Since the  configuration $c$ is invariant under $R_sX$, one may also compute $\mathbf h_i$ as
\begin{equation}
  \mathbf h_i=\sum_{d} J_d \sum_{j\in X(N_d(i))} R_s(\mathbf S_j).
  \label{eq:h_i}
\end{equation}
$X$ reshuffles the neighbors of $i$ (at any fixed distance) but since $X(i)=i$, $N_d(i)$ is globally stable:  $N_d(i)=X(N_d(i))$. So, from Eq.~(\ref{eq:h_i}), we have
\begin{equation}
\mathbf h_i=R_s (\mathbf h_i).
\end{equation}
We therefore conclude that $\mathbf h_i$ is colinear with $\mathbf n$ and thus colinear with $\mathbf S_i$.
This shows that the energy derivative $\frac{\partial E}{\partial \mathbf S_i}$ vanishes for spin variations orthogonal to $\mathbf S_i$ (longitudinal spin variations are not allowed as $(\mathbf S_i)^2$ must be kept fixed).

\textit{All regular states studied in the previous examples that do not belong to a continuum are thus energetically stationnary with respect to small spin deviations}.
They are thus interesting candidates for global energy minima.

\subsection{Lower bound on the energy of Heisenberg models}
\label{ssec:Espiral}

The Fourier transform $\mathbf S_{\mathbf qi}$ of the local spin on a periodic lattice of $N$ unit cells is defined by
\begin{equation}
  \mathbf S_{\mathbf q i}=\frac{1}{\sqrt N} \sum_\mathbf{x} \mathbf S_{\mathbf x i}e^{-i\mathbf{q}\mathbf{x}}.\nonumber
\end{equation}
where each site is labeled by an index $i=1\dots m$ ($m$ is the number of sites per unit cell), $\bf x$ is the position of its unit cell,
and $\bf q$ is a wave vector in the first Brillouin zone.
For an Hamiltonian in the form of Eq.~(\ref{eq:Heisenberg}), the energy can be written as:
\begin{eqnarray}
E&=&\sum_{\substack{\mathbf v,\mathbf x\\i, j=1,\dots, m}}J_{ij}(\mathbf v)\;\mathbf S_{\mathbf xi}\cdot\mathbf S_{\mathbf x+\mathbf v j} \\
&=&\sum_{\substack{\mathbf q\in {\rm BZ}\\i, j=1,\dots, m}}J_{ij}(\mathbf q)\;\mathbf S_{-\mathbf qi}\cdot\mathbf S_{\mathbf qj}\;\;,
\end{eqnarray}
with
\begin{equation}
J_{ij}(\mathbf q)=\sum_{\mathbf v} J_{ij}(\mathbf v)e^{i\mathbf q \mathbf v}.
\end{equation}

Since $(\mathbf S_{i\mathbf{x}})^2=1$ for all $i$ and ${\mathbf x}$, $\sum_{i\mathbf x} \mathbf S_{i\mathbf{x}}^2=\sum_{i\mathbf q} \mathbf S_{q\mathbf{x}}^2=mN$,
we see that a lower bound on the energy (per site) is obtained from the lowest eigenvalue  of the matrices $J(\mathbf q)$:\cite{luttingertisza}
\begin{equation}
\label{eq:minE}
 \frac{E}{m N} \geq  \min_{\{\mathbf q\}}\left( J_{\mathbf q}^{\rm min} \right)
\end{equation}
where $J_{\mathbf q}^{\rm min}$ is the lowest eigenvalue of the matrix $J(\mathbf q)$.

If the lattice has a single site per unit cell ($m=1$) this lower bound is reached by a planar
spiral of the form:\cite{villain77}
\begin{equation}
\mathbf S_{\mathbf x 1}=\mathbf u\cos({\mathbf Q}\cdot{\mathbf x})
		    + \mathbf v\cos({\mathbf Q}\cdot{\mathbf x})
\end{equation}
where ${\bf Q}$ is the  propagation vector (pitch) of the spiral, and corresponds to a minimum of $J_{\mathbf q}^{\rm min}$.
In spin space, the  plane of the spiral is fixed by two orthonormal vectors $\mathbf u$ and $\mathbf v$. When $m=1$, it is only
when $J_{\mathbf q}^{\rm min}$ admits several degenerate minima in the Brillouin zone that additional non-spiral (and possibly non-planar) GS may be constructed.
If the lattice has more than one site per unit cell, an attempt to construct a spiral with a pitch corresponding to the smallest eigenvalue $J(\mathbf Q)^{\rm min}$ will generally {\it not} lead to a physical spin configuration with fixed spin length $\mathbf S_{i\mathbf{x}}^2=1$ at every site.
We will however see in the next section that for some models, a non-planar regular state
may reach the lower bound, whereas all the spiral states are energetically higher.

\subsection{Variational phase diagrams of Heisenberg models on the kagome and hexagonal lattices}
\label{ssec:j1j2j3}

In this section we comment the phase diagrams of  $J_1$-$J_2$-$J_3$(-$J'_3$) Heisenberg models on the kagome and hexagonal lattices. $J_n$ is the interaction between $n^{th}$ neighbors.
On the kagome lattice, there are two types of third neighbors depicted in Fig.~\ref{fig:third_neigh}, and thus two coupling constants $J_3$ and $J_3'$.

\begin{figure}
\begin{center}
\subfigure[\;\label{fig:hexa_J1AF}$J_1=1$ (AF)]
  {\includegraphics[trim= 1cm 4.5cm 1.5cm 4.5cm,clip,width=.2\textwidth]{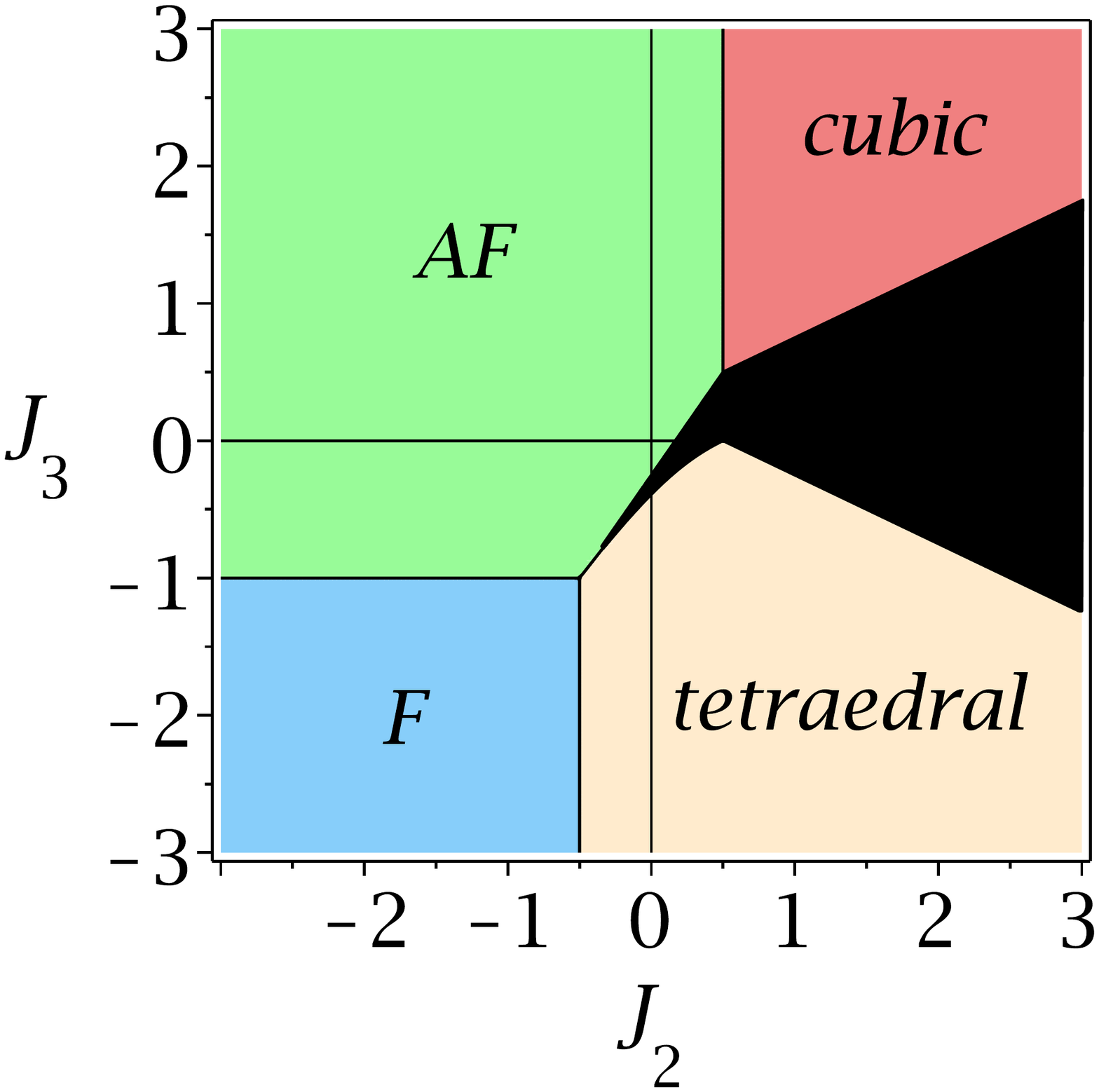}}
  \subfigure[\;\label{fig:hexa_J1F}$J_1=-1$ (F)]
  {\includegraphics[trim= 1cm 4.5cm 1.5cm 4.5cm,clip,width=.2\textwidth]{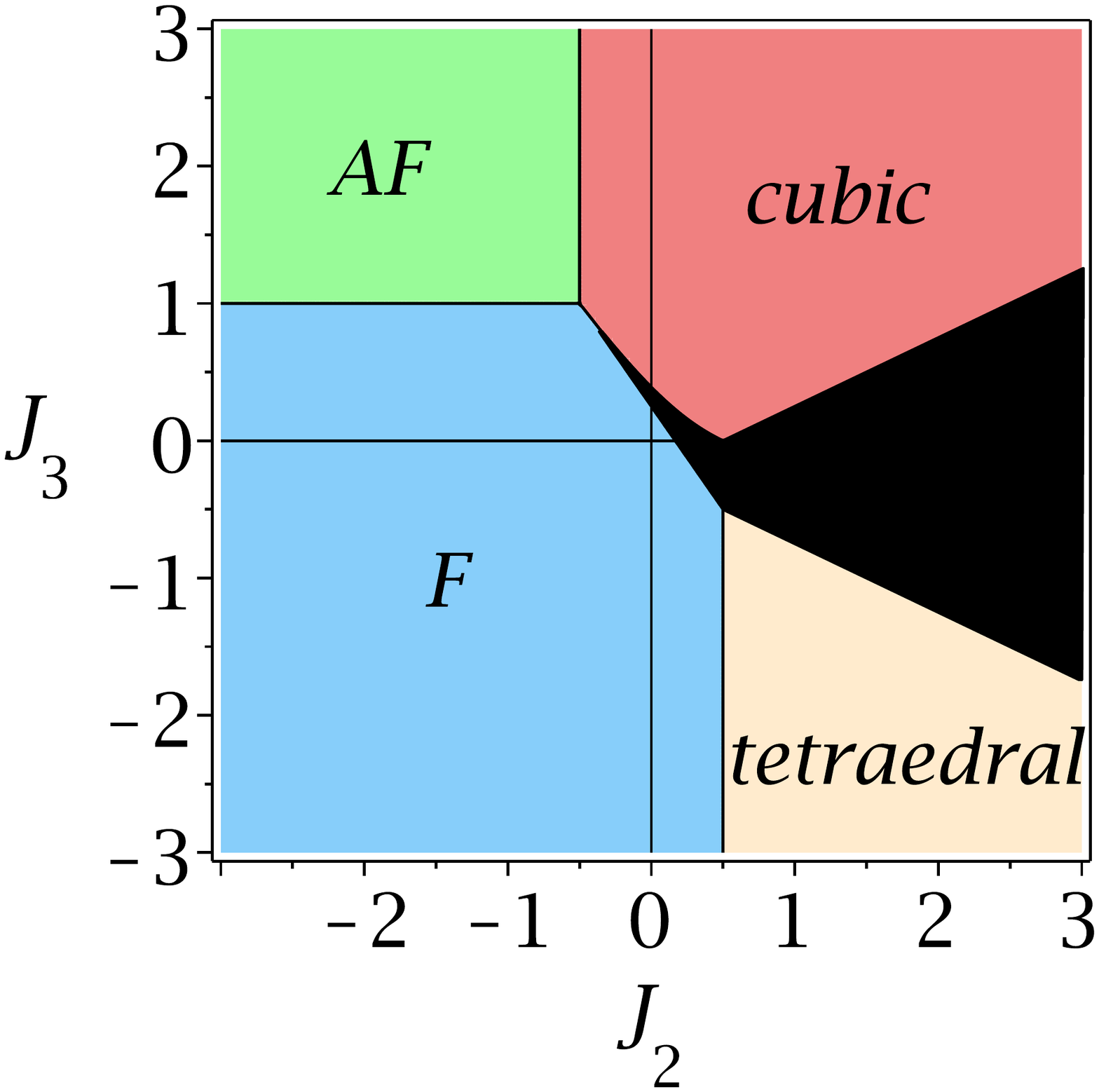}}
\caption{(Color online) Phase diagram of the $J_1$-$J_2$-$J_3$ Heisenberg model on the honeycomb lattice.
 Labels  refer to the regular states described in Fig.\ref{fig:reg_hexa}.
 In each colored region (black excluded), the regular state is an exact GS.
 In the black region a generalized spiral state (SS) has an energy stricly lower than the regular states, but the actual GS energy might still be lower. }
 \label{fig:diag_hexa}
\end{center}
\end{figure}

\begin{figure}
 \begin{center}
  \subfigure[\;\label{fig:third_neigh}Definition of the coupling constants of the model. ]
  {\includegraphics[trim= 1.5cm 11.4cm 2cm 11.4cm,clip,width=.45\textwidth]{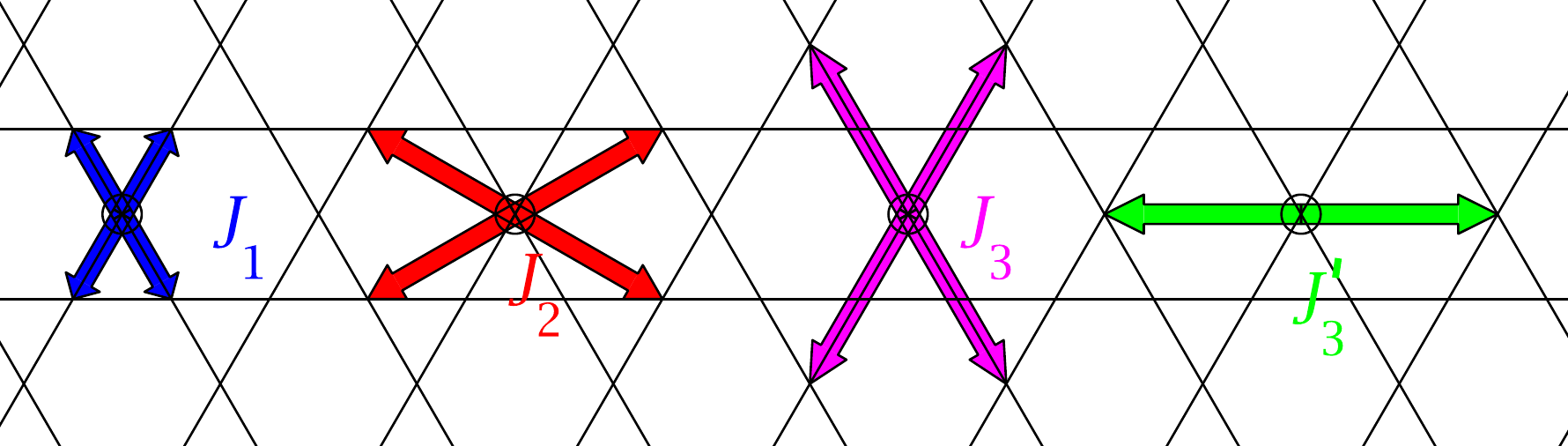}}
  \subfigure[\;\label{fig:diagJ1_1_J3_02a}$J_1=1$, $J_3'=0.2$]
  {\includegraphics[trim= 1cm 4.5cm 1.5cm 4.5cm,clip,width=.2\textwidth]{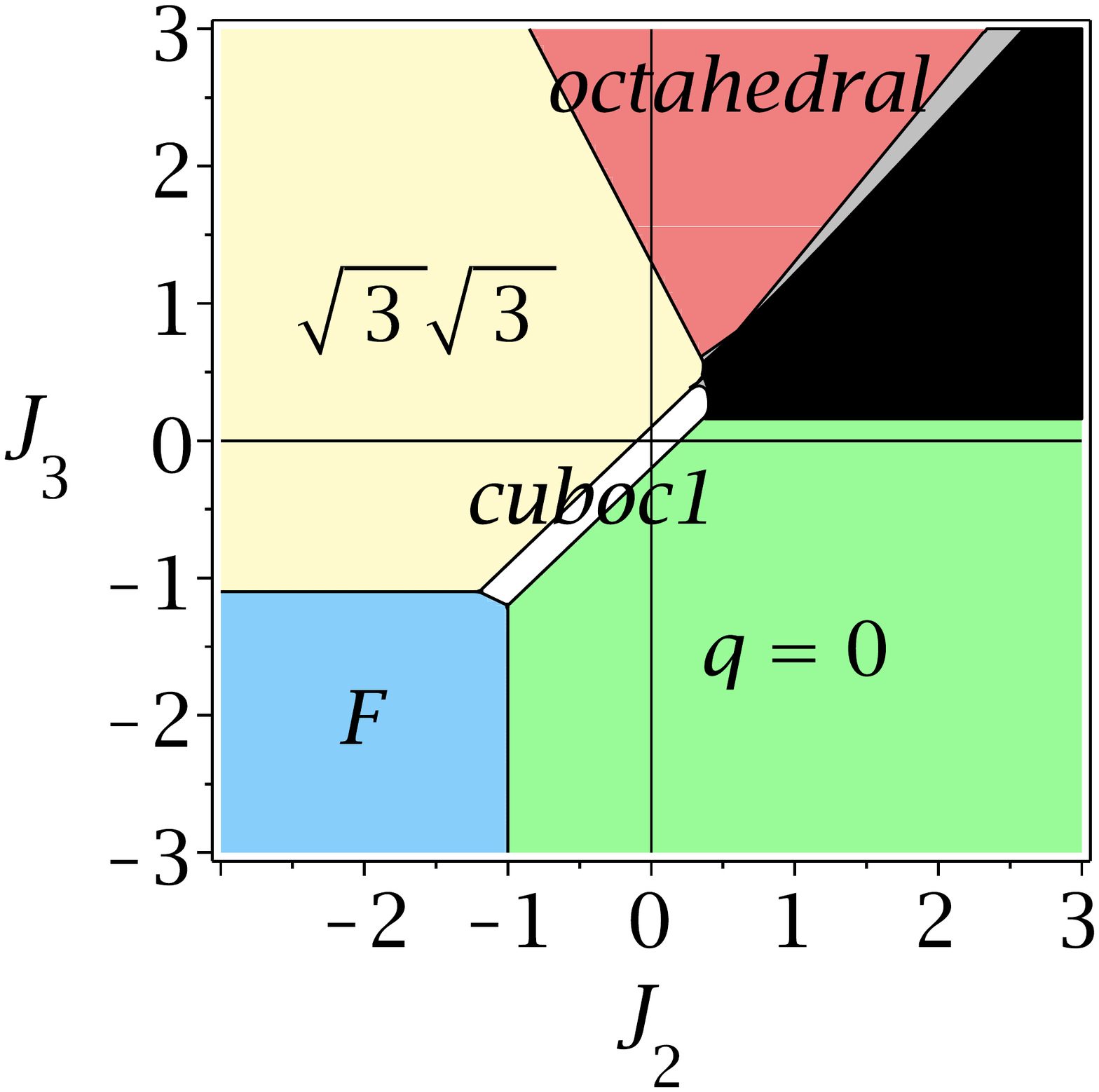}}
  \subfigure[\;\label{fig:diagJ1_1_J3_02b}$J_1=1$, $J_3'=-0.2$]
  {\includegraphics[trim= 1cm 4.5cm 1.5cm 4.5cm,clip,width=.2\textwidth]{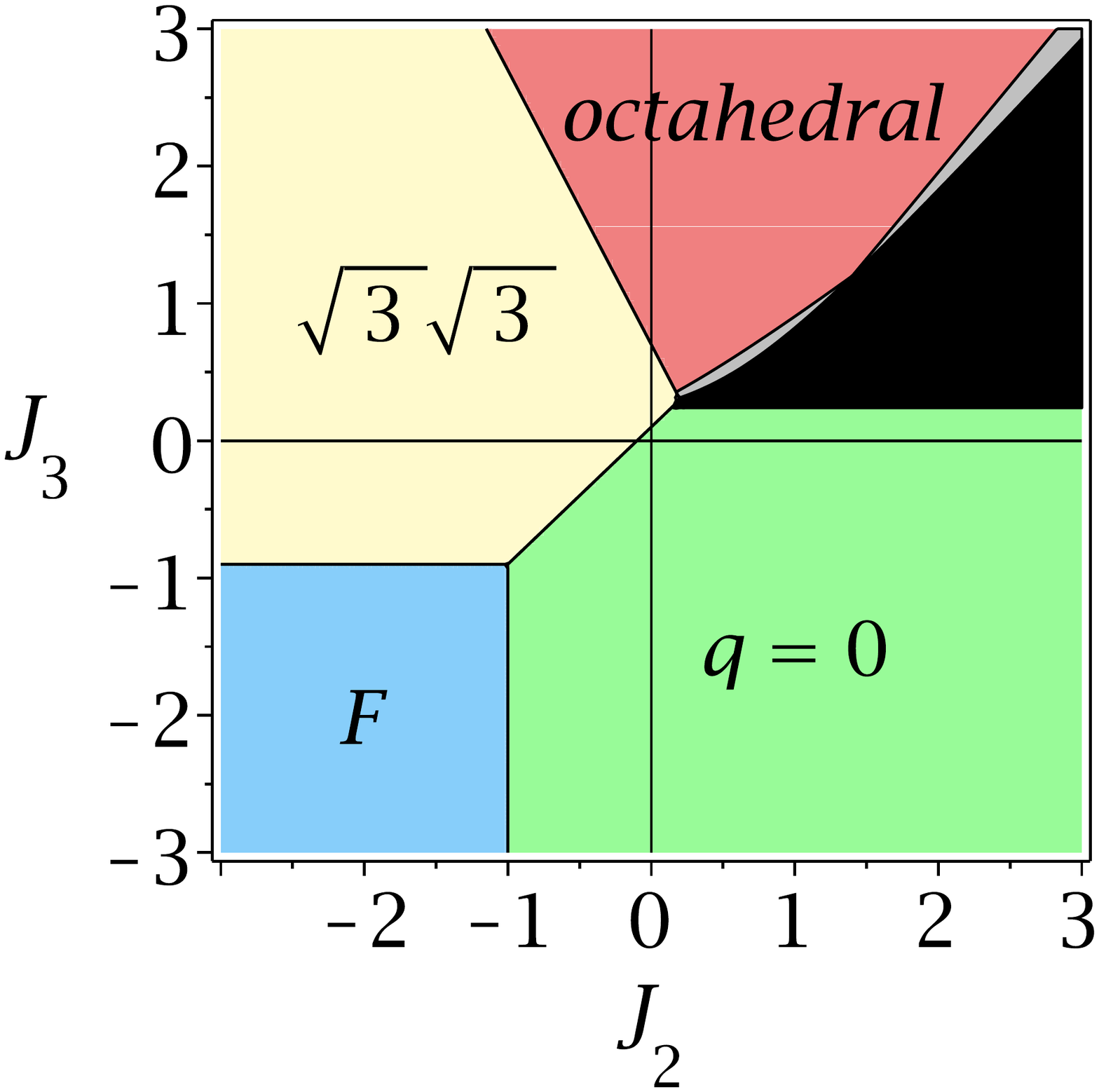}}
  \subfigure[\;\label{fig:diagJ1_1_J3_02c}$J_1=-1$, $J_3'=0.2$]
  {\includegraphics[trim= 1cm 4.5cm 1.5cm 4.5cm,clip,width=.2\textwidth]{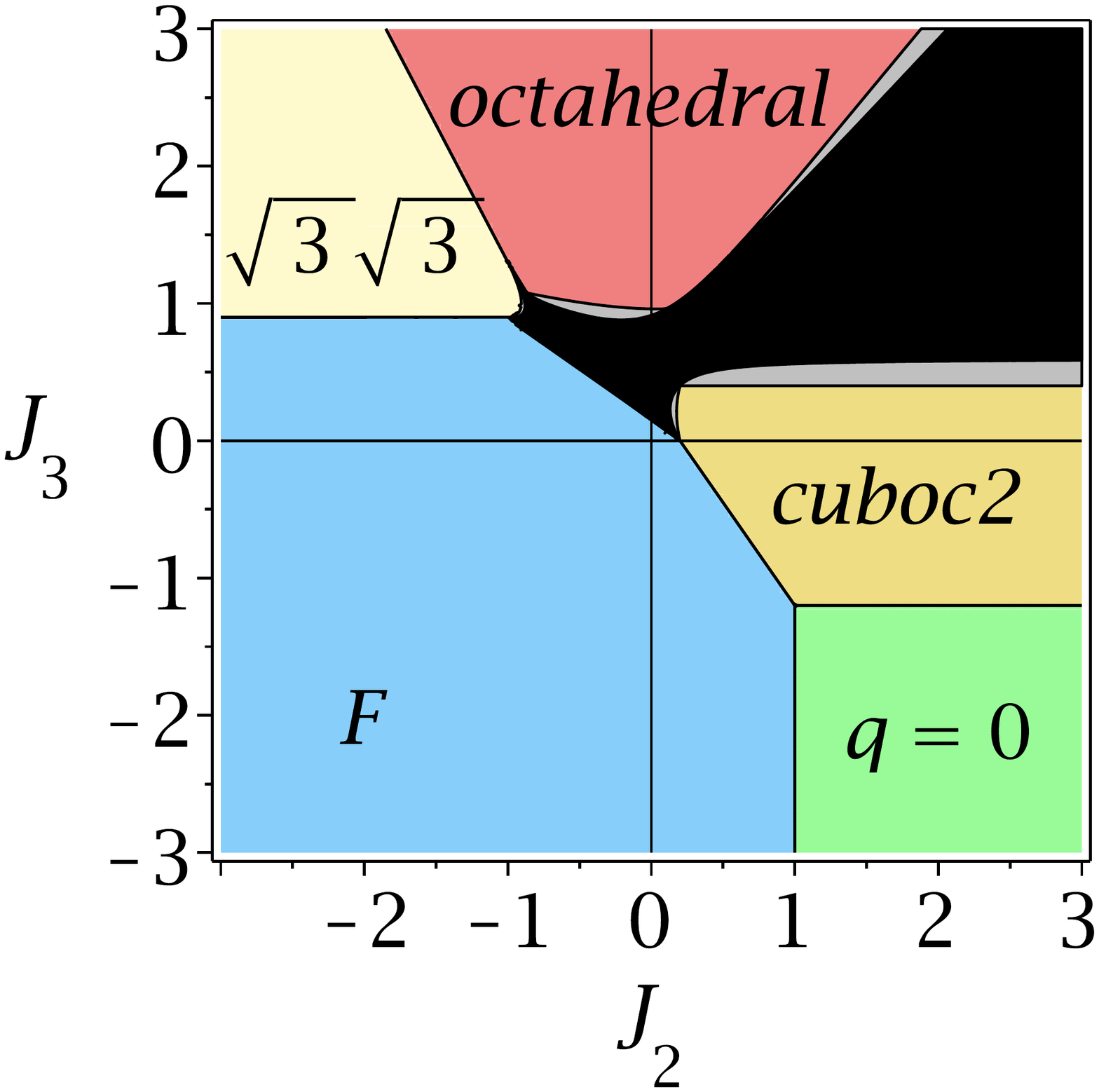}}
  \subfigure[\;\label{fig:diagJ1_1_J3_02d}$J_1=-1$, $J_3'=-0.2$]
  {\includegraphics[trim= 1cm 4.5cm 1.5cm 4.5cm,clip,width=.2\textwidth]{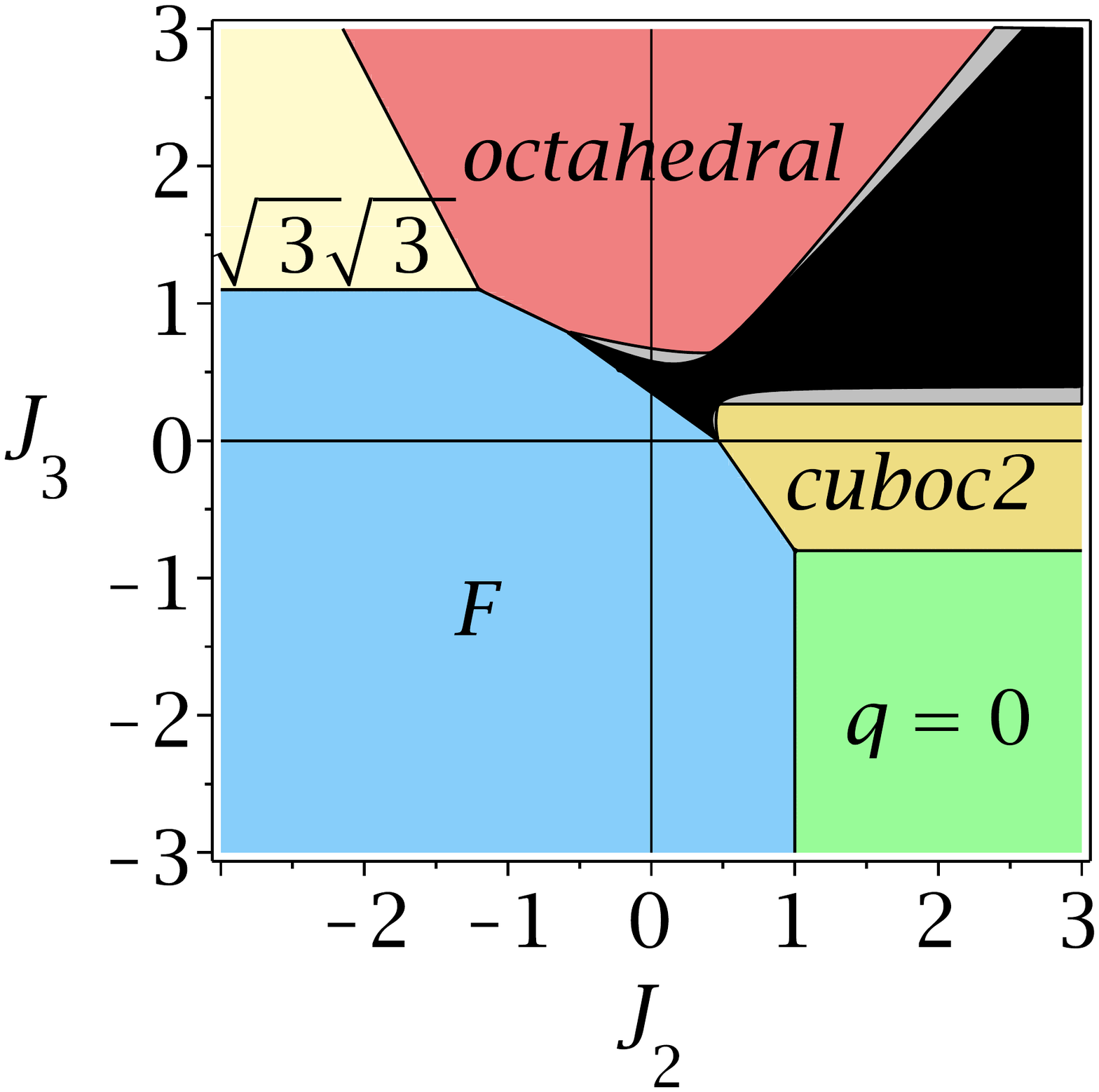}}
  \caption{(Color online) Phase diagram  of the $J_1$-$J_2$-$J_3$-$J_3'$ model on the kagome lattice.
In each colored region (white included,  grey and black excluded), the regular state is an exact GS.
Labels  refer to the regular states described in Fig.\ref{fig:reg_kag}.
In the grey regions, the  near-by regular state does not reach the lower bound of Sec.~\ref{ssec:Espiral} but no SS state is energetically lower.
In the black regions, a SS has a lower energy than the regular states, but the actual GS might yet be lower.
 }
  \label{fig:j1j2j3_kag}
 \end{center}
\end{figure}

For each set of parameters, we determined the regular state with the lowest energy, the SS of Sec.~\ref{ssec:spiral} with the lowest energy, and the value of the lower bound on the energy (the energies of regular states are given in Figs.~\ref{fig:reg_kag} and \ref{fig:reg_hexa}).
The results on these two lattices are described in Figs.~\ref{fig:diag_hexa} and \ref{fig:j1j2j3_kag}.
Such phase diagrams are {\it a priori} variational.
However, it turns out than in all the colored (white included,  grey and black excluded)
regions of Figs.~\ref{fig:diag_hexa} and \ref{fig:j1j2j3_kag}, the regular state with the lowest energy reaches the rigorous lower energy bound of Eq.~(\ref{eq:minE}).
\textit{This demonstrates that (at least) one GS is regular in these regions of the parameter space. }
In the grey areas, the  energy lower bound is not reached, but the regular near-by state could be a GS as no SS has a lower energy.
In the black areas, the GS is not regular: some SS is energetically lower (but sometimes still higher than the lower bound).
All regular states (excepted those from continua) appear in some area of the presented phase diagrams.
This shows that these states are good candidates as variational GSs.
The absence of regular states of a continuum in an Heisenberg model is easily understood.
The energy $E$ of any regular state $c$ belonging to a continuum cannot be lower than the energies $E_1$ and $E_2$ of the two states $c_1$ and $c_2$ between which it interpolates.
One (at least) of the two states, say $c_1$, is colinear along a direction $\mathbf n$.
The $c_2$ spins are then perpendicular to $\mathbf n$.
Let $\theta$ be the angle between the spins of the continuum state and $\mathbf n$.
Then $\mathbf S_i=\mathbf S_i^{c_1}\cos\theta + \mathbf S_i^{c_2}\sin\theta$ and the energy reads $E=E_2+(E_1-E_2)\cos^2\theta$.
Thus, $E$ is in between $E_1$ and $E_2$ and is never strictly the lowest energy.\footnote{In the presence of an external magnetic field $\mathbf h$ and if a one-dimensional representation included in $G$ is ferromagnetic (as it is the case for some umbrella's and for the V-states), $\mathbf n$ aligns on $\mathbf h$.
The energy then reads $E=E_2+(E_1-E_2)\cos^2\theta-\mathbf h \cos\theta$ and an umbrella state becomes stationnary.
It is well known that such structure can be the GSs in presence of a magnetic field.\cite{Roger90,cg91}}

We will now address the possible degeneracies of regular tridimensionnal spin states in these models.
On the hexagonal lattice, our phase diagram is in agreement with  Ref.~\onlinecite{fsl01}.
One should nevertheless notice that the regular tridimensionnal orders (tetrahedral and cubic states) are degenerate with colinear non regular states.
These last states have a higher density of soft excitations (larger energy wells in the phase  space landscape) and will always win as soon as (thermal or quantum) fluctuations are  introduced (order by disorder mechanism\cite{vbcc80,s82,h89,cj92,k93}).
However the non planar configurations could  be stabilized by quartic or ring-exchange interactions.

On the kagome lattice (Fig.~\ref{fig:j1j2j3_kag}) the occurrence of  the cuboc2 (Fig.~\ref{fig:reg_kag_f}) for  $J_1$-$J_2$ interactions\cite{domenge2005} and  of the cuboc1  (Fig.~\ref{fig:reg_kag_e}) for $J_1$-$J_3'$ interactions\cite{Janson_2008} has already been reported.
These two states are not degenerated with SS and are to our knowledge unique and stable GS of the model.
To our knowledge, the octahedral state has not been found before, but this state has the same energy as a continuum of non SS states including colinear states, and it will be destabilized by any fluctuation.

\subsection{Square and triangular lattices: Phase diagrams of Heisenberg versus ring-exchange models}
\label{ssec:MSE}

\begin{figure}[here!]
 \begin{center}
  \subfigure[\;\label{fig:carre_J1AF}$J_1=1$ (AF)]
  {\includegraphics[trim= 1cm 4.5cm 1.5cm 4.5cm,clip,width=.2\textwidth]{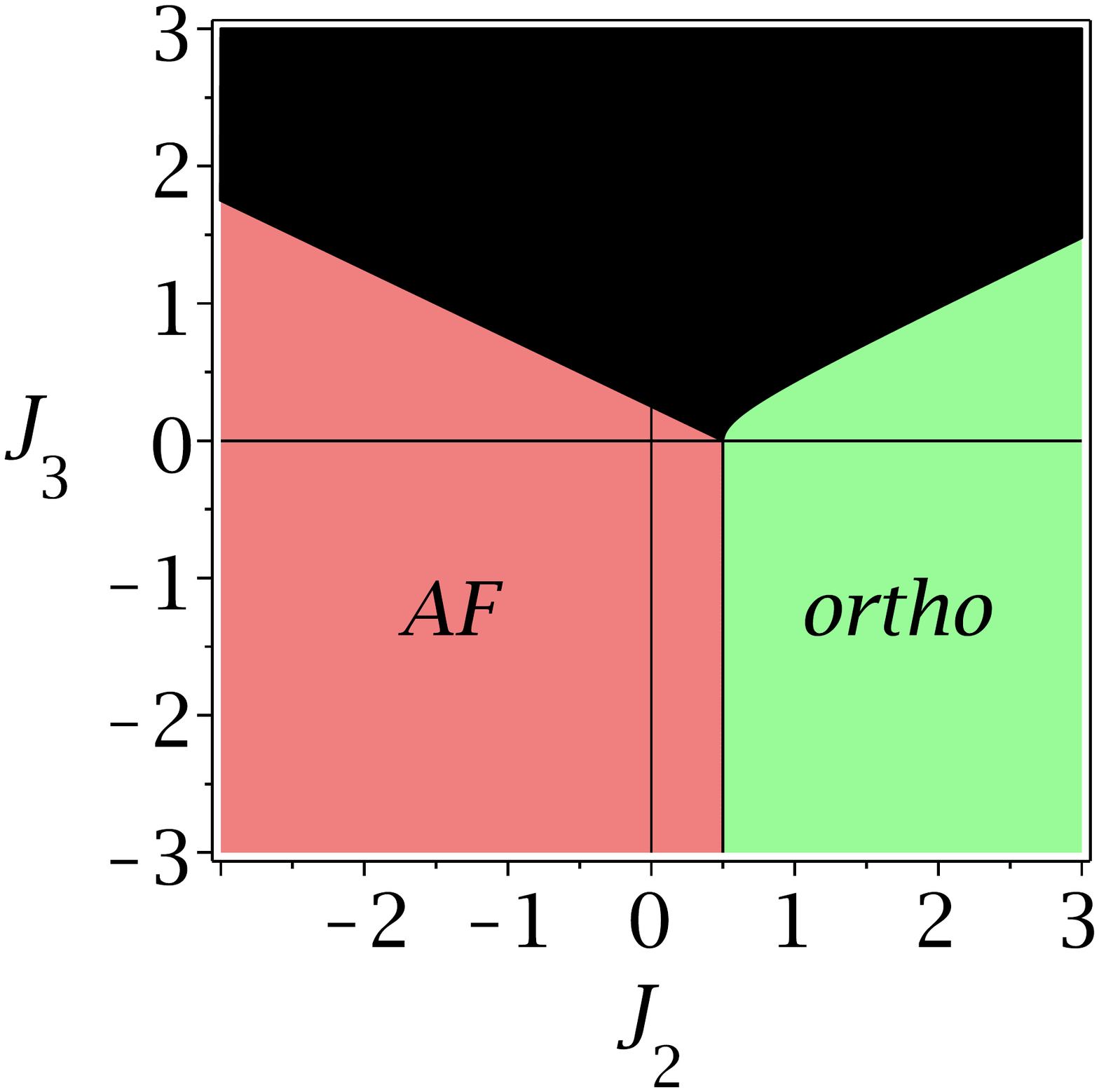}}
  \subfigure[\;\label{fig:carre_J1F}$J_1=-1$ (F)]
  {\includegraphics[trim= 1cm 4.5cm 1.5cm 4.5cm,clip,width=.2\textwidth]{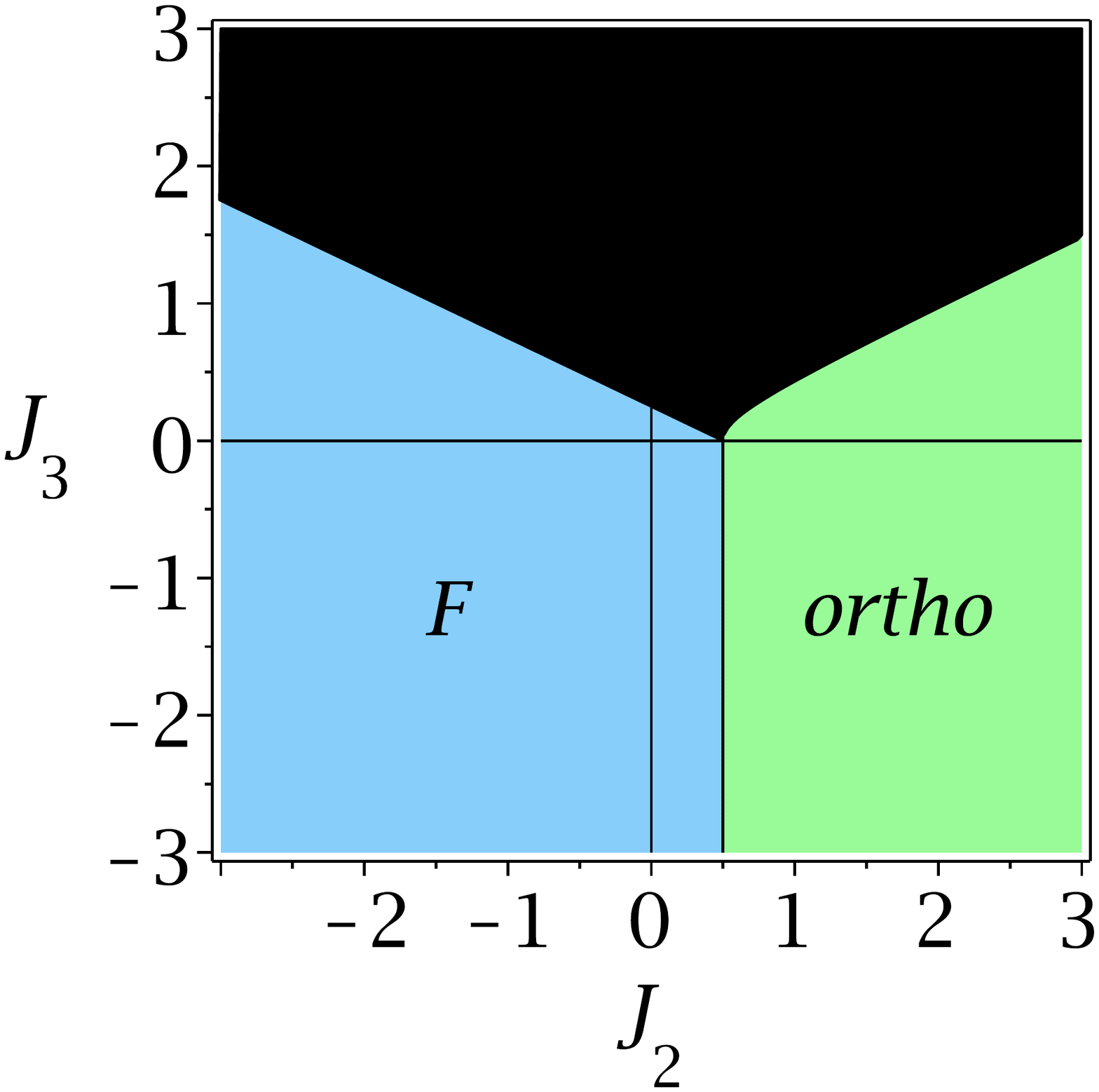}}
  \subfigure[\;\label{fig:carre_MSEJ1AF}$J_1=1$ (AF)]
  {\includegraphics[trim= 1cm 4.5cm 1.5cm 4.5cm,clip,width=.2\textwidth]{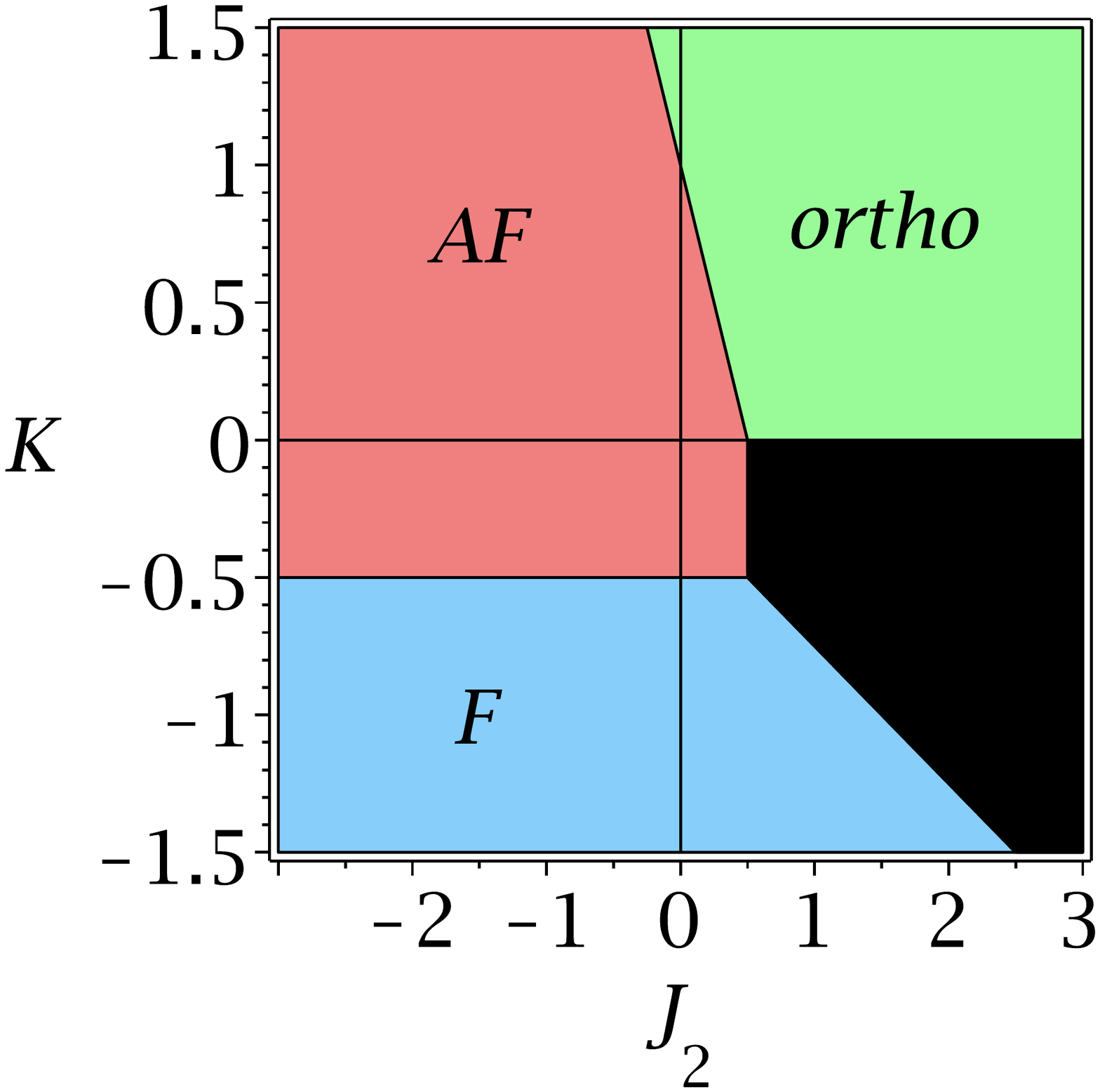}}
  \subfigure[\;\label{fig:carre_MSEJ1F}$J_1=-1$ (F)]
  {\includegraphics[trim= 1cm 4.5cm 1.5cm 4.5cm,clip,width=.2\textwidth]{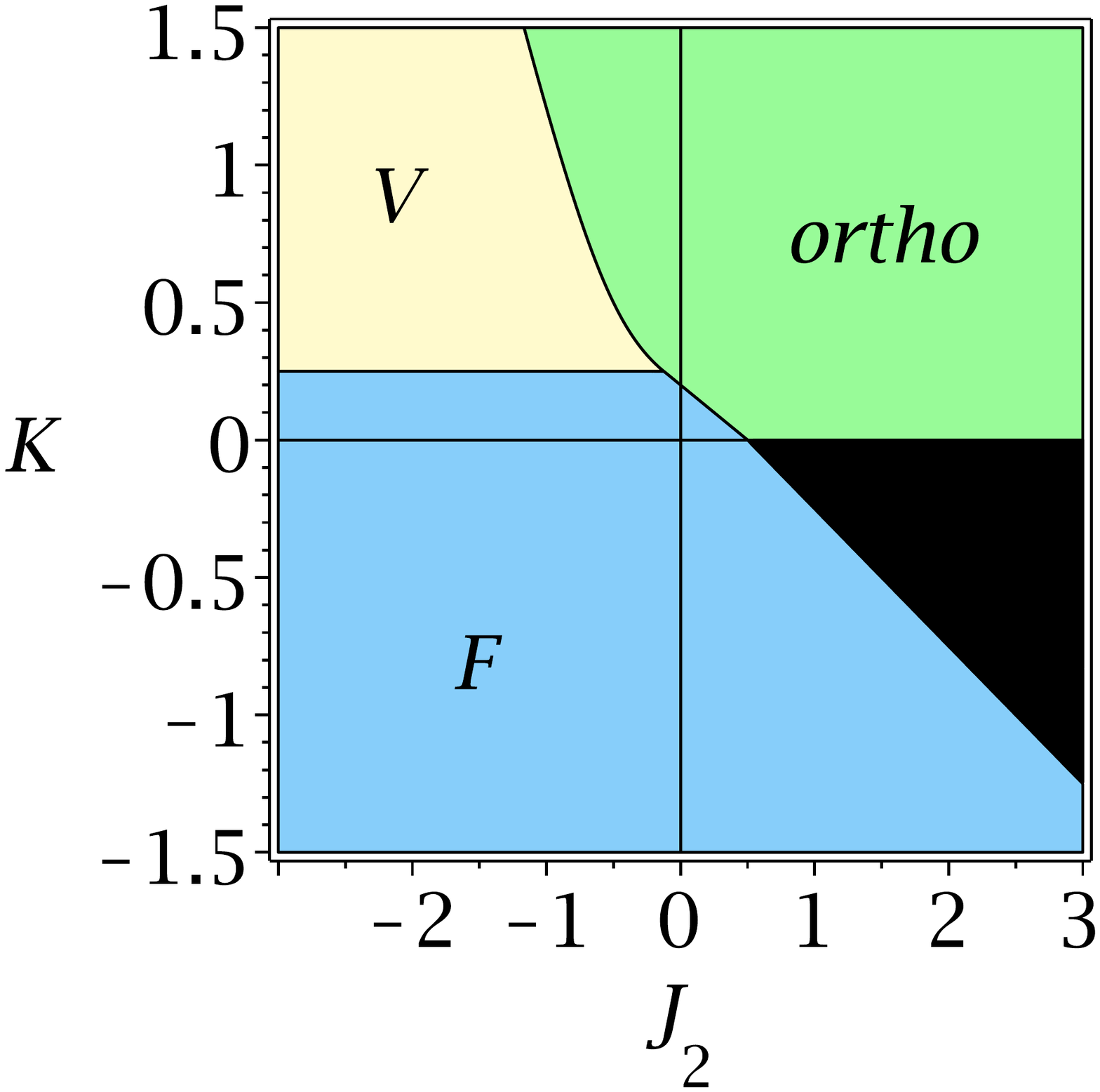}}
   \caption{(Color online) Phase diagrams on the square lattice with $J_1$-$J_2$-$J_3$ Heisenberg interactions (top line) and $J_1$-$J_2$-$K$ model (bottom line).
Labels  refer  to regular states defined  in Fig. \ref{fig:reg_carre}.
In each colored region (black excepted), the regular state has the lowest energy of the set of all regular and SS states.
In the black regions, a SS has a lower energy than the regular states.
For pure Heisenberg interactions, we know that we obtain the GS energy, but for non Heisenberg interactions the actual GS might be lower.
In the $J_1$-$J_2$-$J_3$ model the coplanar (orthogonal  4-sublattice)  phase is degenerate with non regular colinear states, which will win upon introductions of fluctuations.
A contrario the coplanar (orthogonal 4-sublattice)  phase is stable in a large range of parameters in the $J_1$-$J_2$-$K$ model.
 }
   \label{fig:phasediagram_carre}
 \end{center}
\end{figure}

\begin{figure}[here!]
 \begin{center}
  \subfigure[\;\label{fig:tri_J1AF}$J_1=1$ (AF)]
 {\includegraphics[trim= 1cm 4.5cm 1.5cm 4.5cm,clip,width=.2\textwidth]{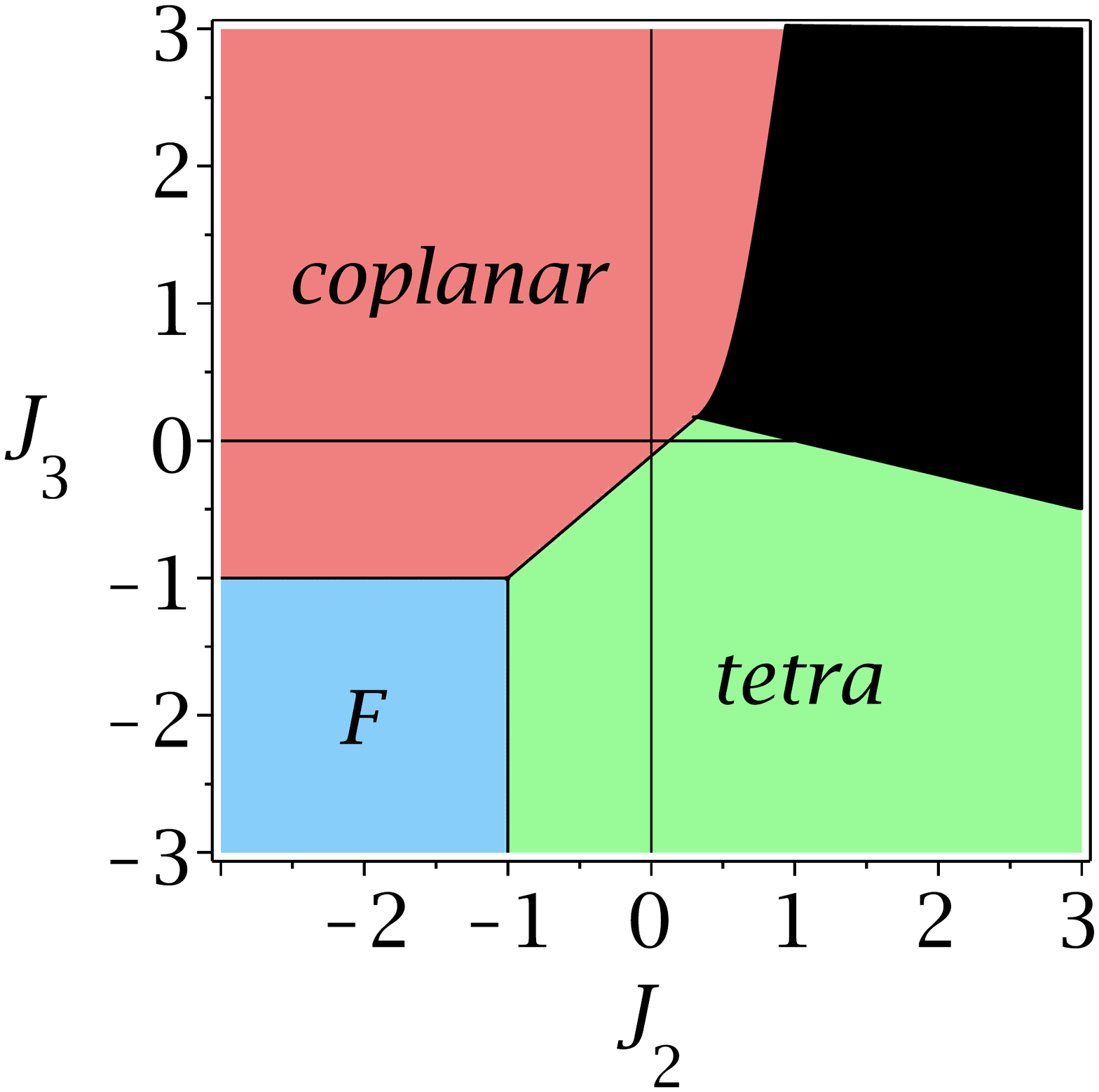}}
  \subfigure[\;\label{fig:tri_J1F}$J_1=-1$ (F)]
  {\includegraphics[trim= 1cm 4.5cm 1.5cm 4.5cm,clip,width=.2\textwidth]{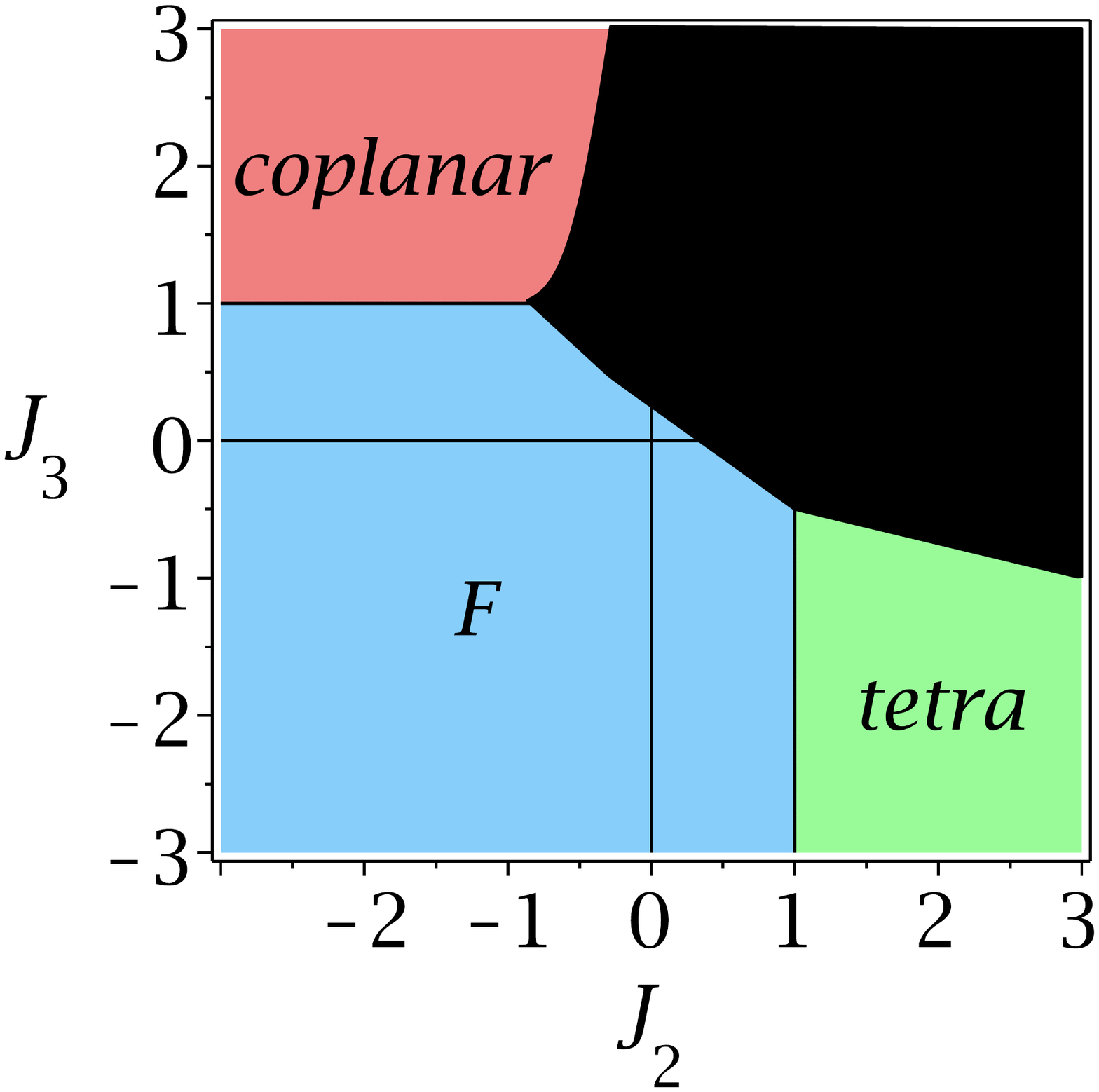}}
  \subfigure[\;\label{fig:tri_JKAF}$J_1=1$ (AF)]
  {\includegraphics[trim= 1cm 4.5cm 1.5cm 4.5cm,clip,width=.2\textwidth]{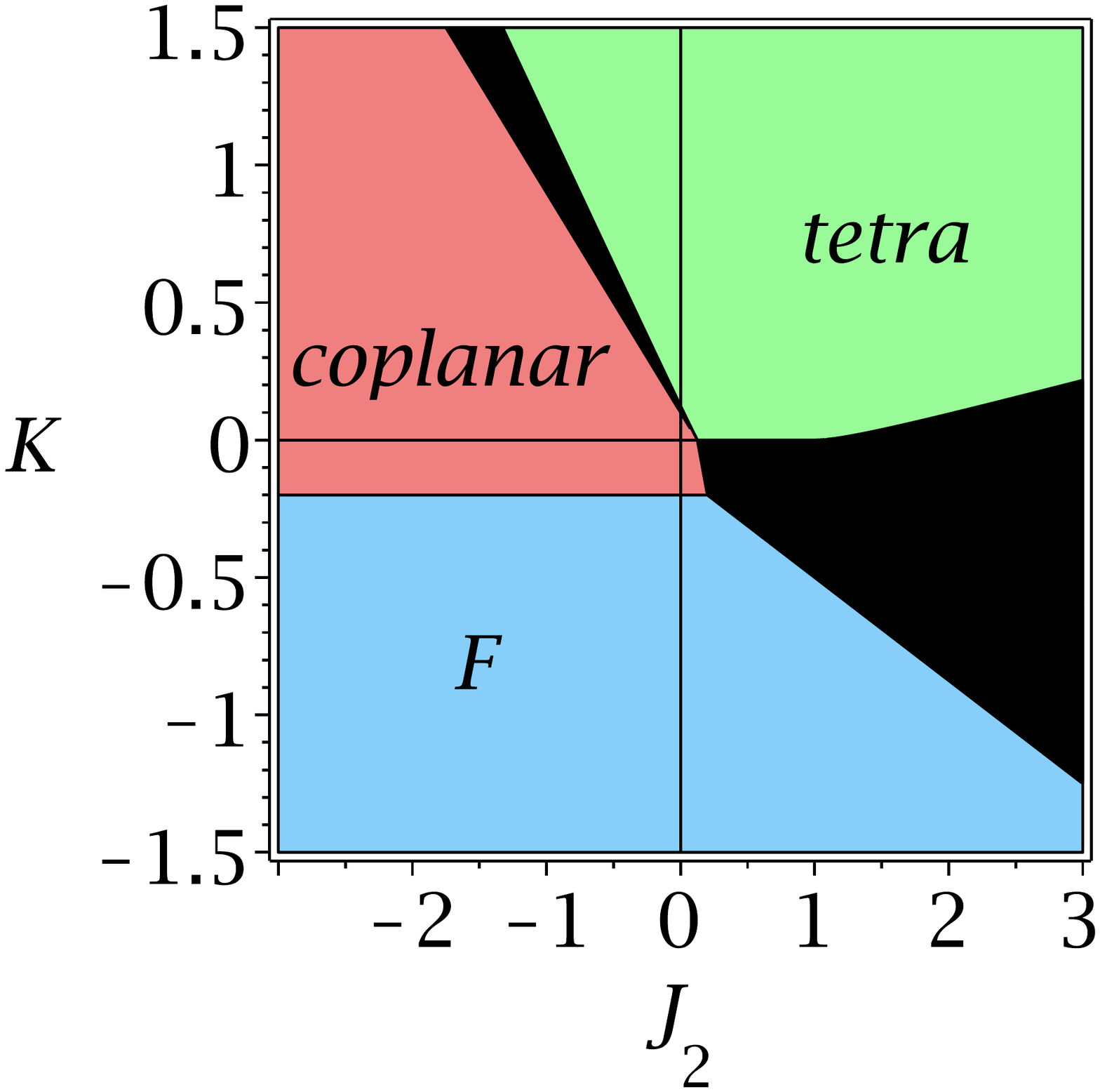}}
  \subfigure[\;\label{fig:tri_JKF}$J_1=-1$ (F)]
  {\includegraphics[trim= 1cm 4.5cm 1.5cm 4.5cm,clip,width=.2\textwidth]{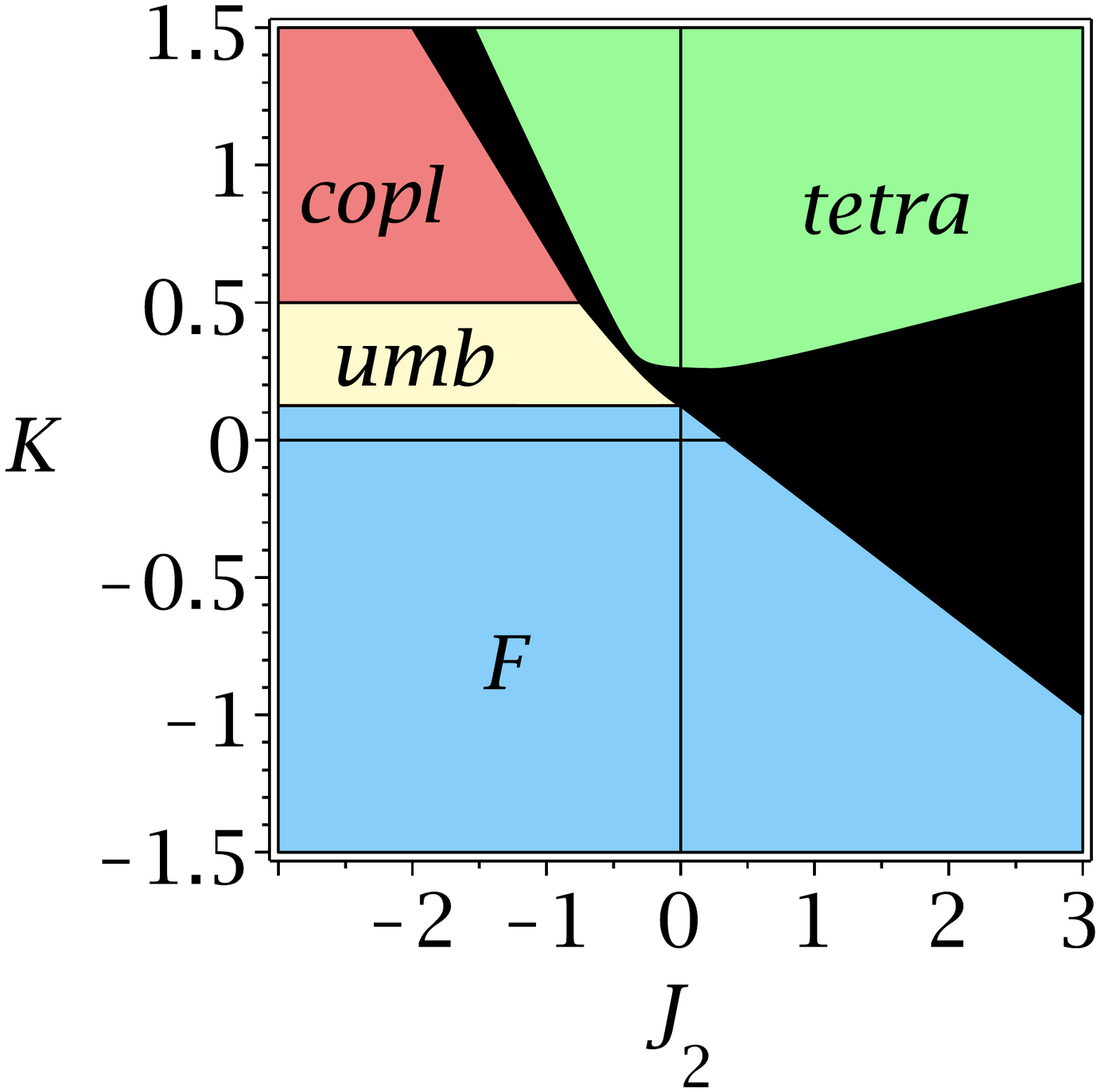}}
   \caption{(Color online) Phase diagrams on the triangular lattice with $J_1$-$J_2$-$J_3$ Heisenberg interactions (top line) and $J_1$-$J_2$-$K$ model (bottom line).
Labels refer to regular states defined  in Fig. \ref{fig:reg_tri}.
In each colored region (black excepted), the regular state has the lowest energy of the set of all regular and SS states.
In the black regions, a SS has a lower energy than the regular states.
For pure Heisenberg interactions, we know that we obtain the GS energy, but for non Heisenberg interactions the actual GS might be lower. }
   \label{fig:phasediagram_tri}
 \end{center}
\end{figure}

In this section we will comment the phase diagram of the Heisenberg models  (Eq.~(\ref{eq:Heisenberg})) on the square and triangular lattices and display the effect
of 4-spin ring-exchange ($J_1$-$J_2$-$K$) on these two lattices.
The $J_1$-$J_2$-$K$ model is defined as:
\begin{eqnarray}
E&=& \sum_{i,j} J(|{\bf x}_i-{\bf x}_j|) \mathbf S_i \cdot \mathbf S_j  + K \sum_{i,j,k,l}\left((\mathbf S_i \cdot \mathbf S_j)(\mathbf S_k\cdot\mathbf S_l)
\right.\nonumber\\
&&
+(\mathbf S_i \cdot \mathbf S_l)(\mathbf S_j \cdot \mathbf S_k)
-(\mathbf S_i \cdot \mathbf S_k)(\mathbf S_j \cdot \mathbf S_l)
+\mathbf S_i \cdot\mathbf S_j \nonumber\\
&&\left.
+\mathbf S_j \cdot \mathbf S_k
+ \mathbf S_k \cdot \mathbf S_l
+ \mathbf S_l \cdot \mathbf S_i
+\mathbf S_i \cdot \mathbf S_k
+ \mathbf S_j \cdot \mathbf S_l \right)
\end{eqnarray}
where the sum in the $K$ term runs on rhombi $i,j,k,l$.\cite{km97}
This model encompasses first and second neighbor $J_1$ and $J_2$ couplings and a $K$ ring-exchange term which introduces quartic interactions as well as modifications of first and
second neighbor Heisenberg interactions.\cite{km97}
The phase diagramms are displayed in Figs:~\ref{fig:phasediagram_carre} and \ref{fig:phasediagram_tri}.

In the $J_1$-$J_2$-$J_3$ Heisenberg phase diagrams on the square and triangular lattice, all regular states that do not belong to continua do appear as exact GSs in some parts of the phase diagrams (colored regions - black excepted of Figs.~\ref{fig:carre_J1AF}, \ref{fig:carre_J1F}, \ref{fig:tri_J1AF} and \ref{fig:tri_J1F}).
In black regions, SS are more stable than regular states.
As these lattices are Bravais lattices, we know how to reach the lower bond of Sec.~\ref{ssec:Espiral} thanks to a spiral state.
The orthogonal state on the square lattice and the tetrahedral state on the triangular lattice (Fig.~\ref{fig:reg_tri_b} and \ref{fig:reg_carre_c}) are degenerate with spiral states including colinear states with 2 spins (up, down) in the magnetic unit cell, which will win upon introduction of fluctuations.
On a large part of the phase diagram on the square lattice (spirals excepted) the spins are thus colinear.

The presence of a 4-spin ring exchange on the square lattice gives richer phase diagramms (Figs.~\ref{fig:carre_MSEJ1AF} and \ref{fig:carre_MSEJ1F}) with the appearance of states from continua.
We recall that these phase diagrams are variational and give the minimal energy state among the regular and the SS states.
A dominant 4-spin ring exchange stabilizes the orthogonal 4-sublattice coplanar antiferromagnet, which is known to be robust to large quantum fluctuations.\cite{laeuchli2005}
One of these phases belongs to a continuum: the $V$ states (Fig:~\ref{fig:reg_carre_d}).
Part of this phase diagram on the square lattice has been known for a long time for the $J_1$-$K$ model,\cite{cgb92} but the effect of a second neighbor interaction leads to new phases that might be interesting in various respects.

The $J_1$-$J_2$-$K$ phase diagramm on the triangular lattice (Figs.~\ref{fig:tri_JKAF} and \ref{fig:tri_JKF}) exhibits all the regular phases that can be constructed on this lattice.
In that model, large ring-exchange stabilizes the tetrahedral chiral phase studied by Momoi and co-workers.\cite{Momoi1997,km97}
The presence of large parts of the phase diagramms with planar or 3-dimensional order parameter at $T=0$, and of points where a large number of classical phases are in competition, could give interesting hints in the quest of exotic quantum phases.\cite{liMing2000,motrunich2005,Grover2010}

\subsection{Finite temperature phase transitions in two-dimensions}
\label{ssec:transition}

In two-dimensions, the Mermin-Wagner\cite{MerminWagner} theorem insures that continuous symmetries cannot be spontaneously broken at finite temperature.
It does however not prevent discrete symmetries to be broken.
Indeed, some finite temperature phase transitions associated to discrete symmetries have been found in classical $O(3)$ models: lattice symmetry breaking in the $J_1$-$J_2$ and $J_1$-$J_3$ models on the square lattice,\cite{chandra90,weber03,Capriotti2004} chiral symmetry breaking in a ring-exchange model on the triangular lattice\cite{k93,Momoi1997} and in a $J_1$-$J_2$ model on the kagome lattice.\cite{domenge2005,domenge2008}

What should be expected in a system where the GS is a regular state $c$ ?
Let us first consider the case where $c$ is not chiral, that is when the spin inversion $\mathbf S\to -\mathbf S$ gives a state $c'$ which can also be obtained from $c$ by a rotation in $SO(3)$. 
At an infinitesimally small temperature, the rotational symmetry is restored and the statistical ensemble is that of all the (regular) states obtained from $c$ by $SO(3)$ rotations.
The  thermal average of an observable is therefore also an average over $SO(3)$ rotations.
Now, if we compare an observable $O$ and the same observable after a lattice symmetry $X$, we will get the same
 average (for regular states, the effect of $X$ can be absorbed by a rotation).
So not only the rotational symmetries, but all the lattice symmetries are  restored at $T=0^+$. The simplest scenario is therefore a complete absence of symmetry-breaking phase transition from $T=0^+$ up to $T=\infty$.
Now, for a chiral state, the thermal fluctuations will only partially restore the $O(3)$ symmetry of the model, and a chiral phase transition -- possibly accompanied by some lattice symmetry breaking -- should be expected.
From this point of view, a classical system in  two-dimensions with no finite-temperature phase transition is likely to have a regular and non-chiral GS.

\section{Conclusion}
\label{sec:CCL}

Based on symmetry considerations (and on an analogy with Wen's\cite{Wen_PSG} classification of quantum spin liquids using the concept of PSG), we introduced a family of classical (antiferro-)magnetic structures, dubbed ``regular'' states.
They can be constructed in a systematic way for any lattice, based on the method explained in Sec.~\ref{sec:reg_construction}.
We found that these states are often good variational states to study the zero-temperature phase diagram of ``complex'' problems (non-Bravais lattice and/or multiple spin interactions for instance).
In many cases, one of the regular state is found to reach a lower energy bound, allowing to show that it is a GS.

We note that, although one can always find a planar GS in Heisenberg models on a Bravais
lattice,  non-planar antiferromagnetic spin structures with many sublattices are rather
common in presence of competing interactions, non quadratic spin interactions and non-Bravais lattices.
As mentioned in the introduction, we believe this approach may find an application in the study of real magnetic compounds where the (equal time) spin-spin correlations are measured, but the strength and range of the magnetic exchange interactions are not known.

We have studied the case where the spin manifold ${\mathcal A}={\mathcal S}_2$ is that
of a three-component spin (unit vector), but other manifolds could be investigated using the same approach. For instance, {\it nematic regular orders} would be obtained with ${\mathcal A}={\mathcal S}_2/\mathbb Z_2$ and $S_S=SO(3)$.

\acknowledgments
 We thank L. Pierre for enlightening discussions on spiral states and V. Pasquier for interesting input on geometrical considerations and for mentioning Ref.~\onlinecite{coxeter}. This research was supported in part by the National Science Foundation under Grant No. PHY05-51164.

\appendix

\section{Powder-averaged structure factors of regular states}
\label{App:powder}

Equal time spin-spin correlations partially characterize a spin state and are independent of the energetic properties of the system.
Equal time structure factors can thus be analytically calculated on regular states to form a set of reference neutron scattering results.
They can be used to analyze measurements done on compounds with unknown GS.
We define the equal time structure factor $S(\mathbf{Q})$ of a state as
\begin{equation}
S(\mathbf{Q}) \propto \sum_{i,j}e^{-i\mathbf Q (\mathbf x_i -\mathbf x_j)}\mathbf S_i\cdot \mathbf S_j,
\end{equation}
where $\mathbf x_i$ is the position vector of the site $i$.
The proportionality factor is adjusted to verify the sum rule $\sum_\mathbf{Q}S(\mathbf{Q})=1$.
For perfect long-range orders, $S(\mathbf{Q})$ is zero everywhere except for a finite number of $\mathbf{Q}$ where Bragg peaks are present.
They are broadened when chemical defects, non zero temperature or quantum fluctuations are taken into account.

When only powders are realisable, one can measure the powder equal time structure factor $S(|\mathbf{Q}|)$.
It is the average of
$S(|\mathbf{Q}|\sin\theta(\mathbf u\cos\psi+\mathbf v\sin\psi))$
over all the possible 3d orientations of $\mathbf{Q}$, where $\theta$, $\psi$ are the spherical coordinates angles of $\mathbf Q$ in the orthonormal basis $(\mathbf u, \mathbf v, \mathbf u\land \mathbf v)$ with $\mathbf u$, $\mathbf v$ in the sample plane.
Thus
\begin{equation}
S(|\mathbf{Q}|) \propto
\int d^2\mathbf q \frac{\Theta(|\mathbf{Q}|-|\mathbf q|)}{|\mathbf{Q}|\sqrt{|\mathbf{Q}|^2-|\mathbf{q}|^2}}
S(\mathbf{q}),
\end{equation}
where $\Theta$ is the Heaviside step function and $\mathbf q$ browses the reciprocal 2d space.

\begin{figure}[here!]
\begin{center}
 \subfigure[\;F state\label{fig:struct_tri_ferro}]{
   \includegraphics[trim = 10mm 45mm 15mm 48mm, clip,width=.15\textwidth]{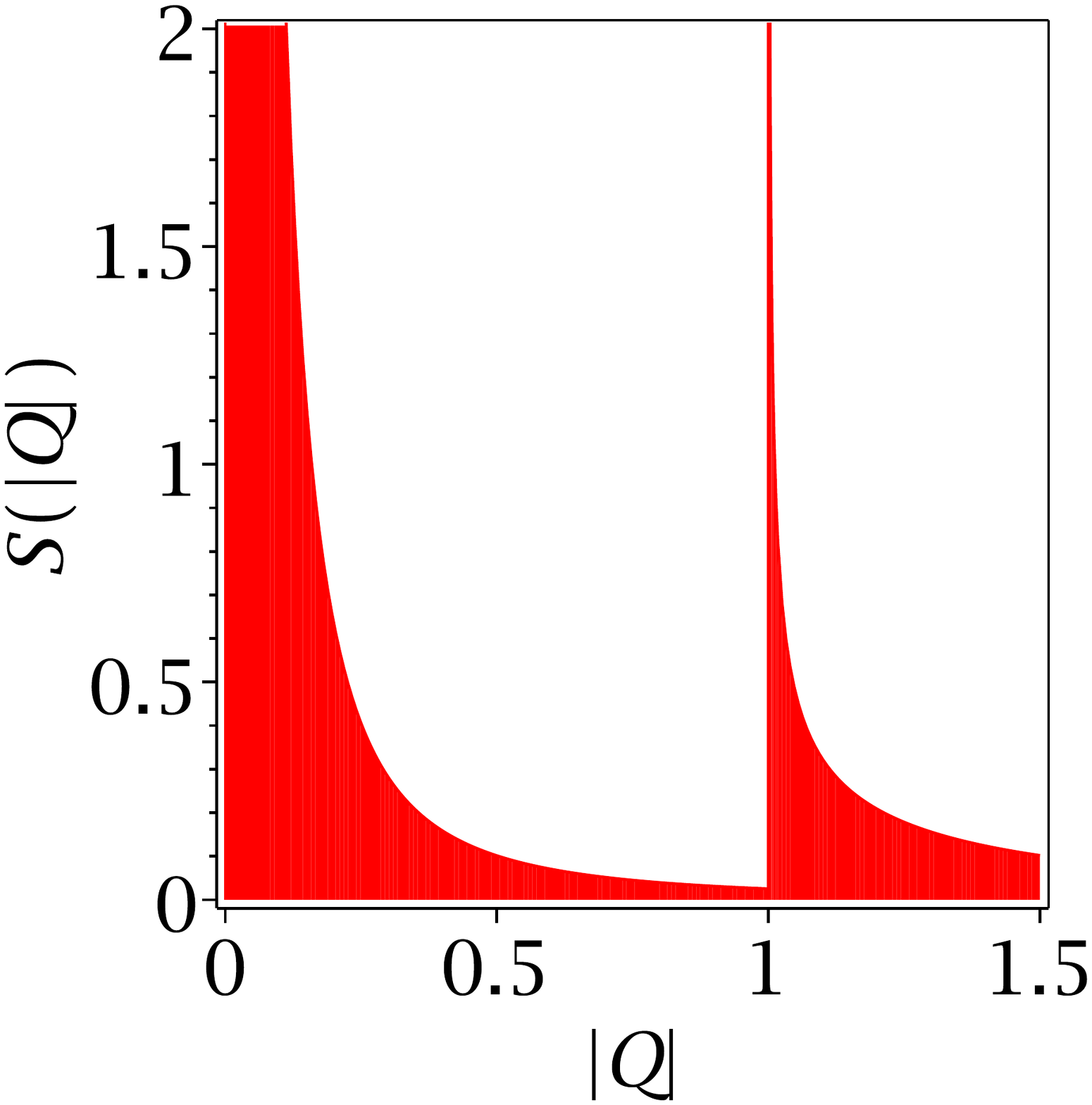}}
 \subfigure[\;Tetrahedral state\label{fig:struct_tri_tetra}]{
   \includegraphics[trim = 10mm 45mm 15mm 48mm, clip,width=.15\textwidth]{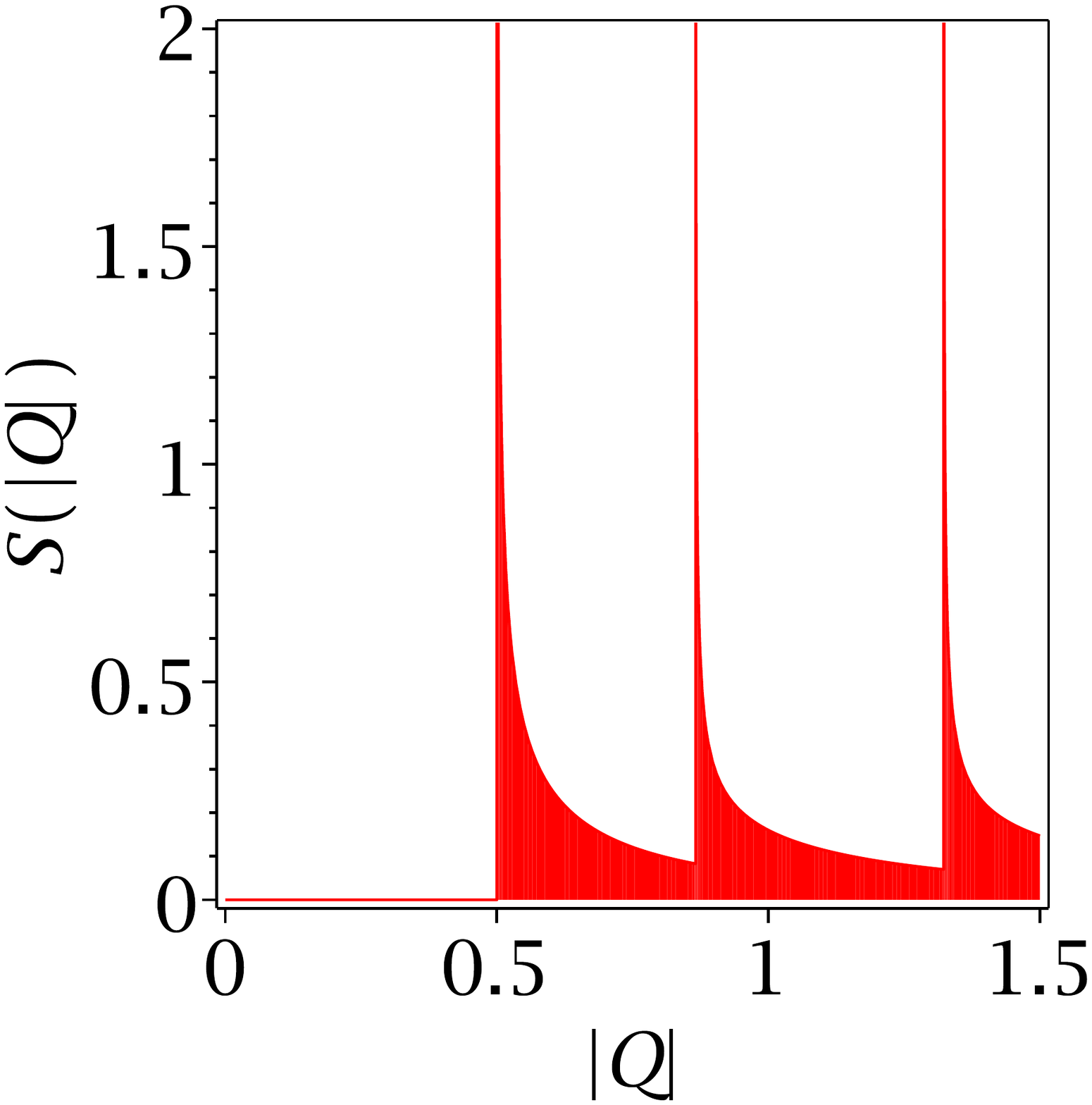}}
 \subfigure[\;Coplanar state\label{fig:struct_tri_tri}]{
   \includegraphics[trim = 10mm 45mm 15mm 48mm, clip,width=.15\textwidth]{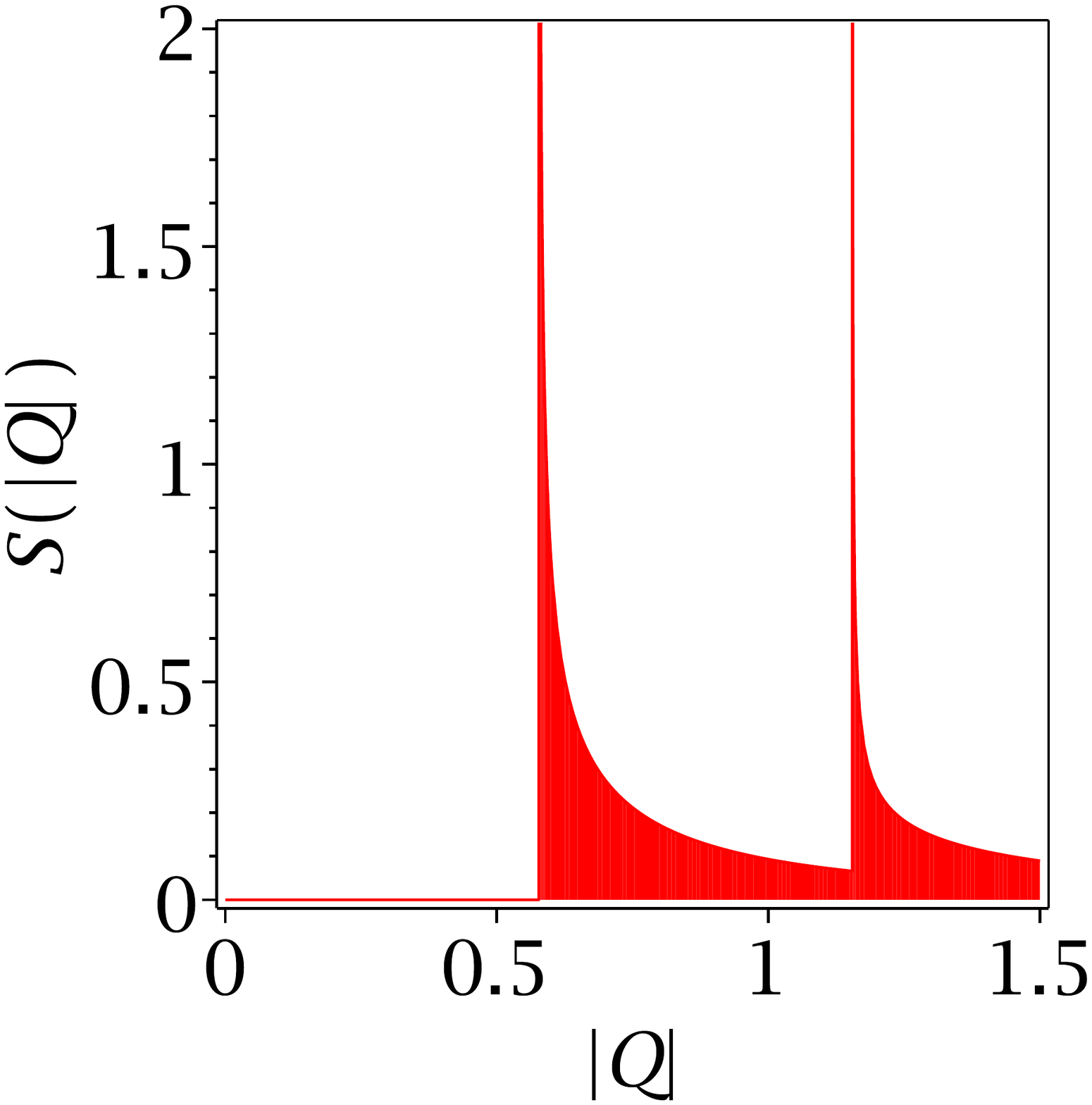}}
 \caption{Powder-averaged equal time structure factors $S(|\mathbf Q|)$ of the regular states on the triangular lattice ($|\mathbf Q|$ is in units of $2\pi$, $S(|\mathbf Q|)$ in arbitrary units).}
 \label{fig:struct_tri}
\end{center}
\end{figure}

\begin{figure}[here!]
\begin{center}
 \subfigure[\;F state\label{fig:struct_kag_ferro}]{
   \includegraphics[trim = 10mm 45mm 15mm 48mm, clip,width=.15\textwidth]{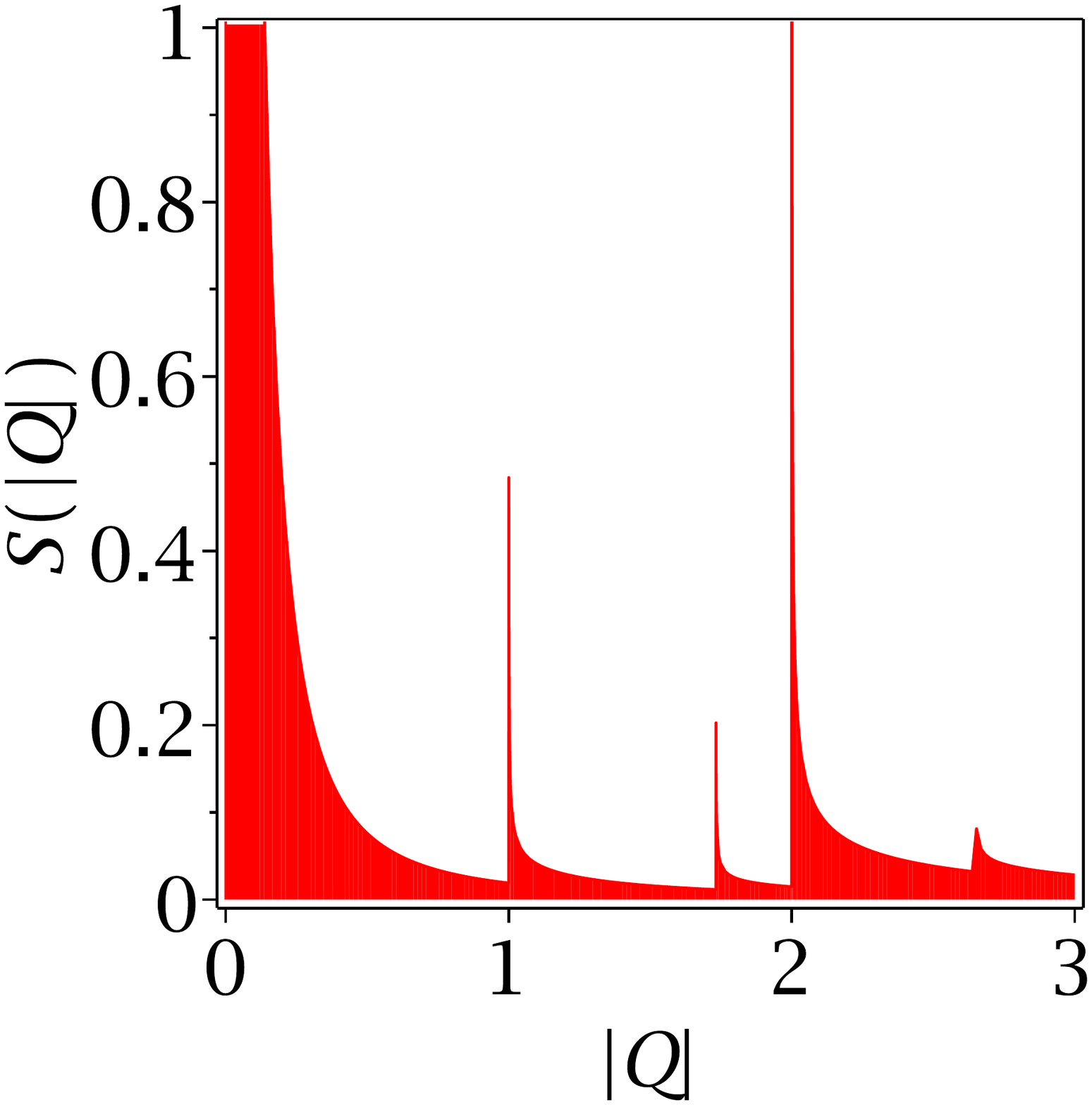}}
 \subfigure[\;$\mathbf q=\mathbf 0$ state\label{fig:struct_kag_q0}]{
   \includegraphics[trim = 10mm 45mm 15mm 48mm, clip,width=.15\textwidth]{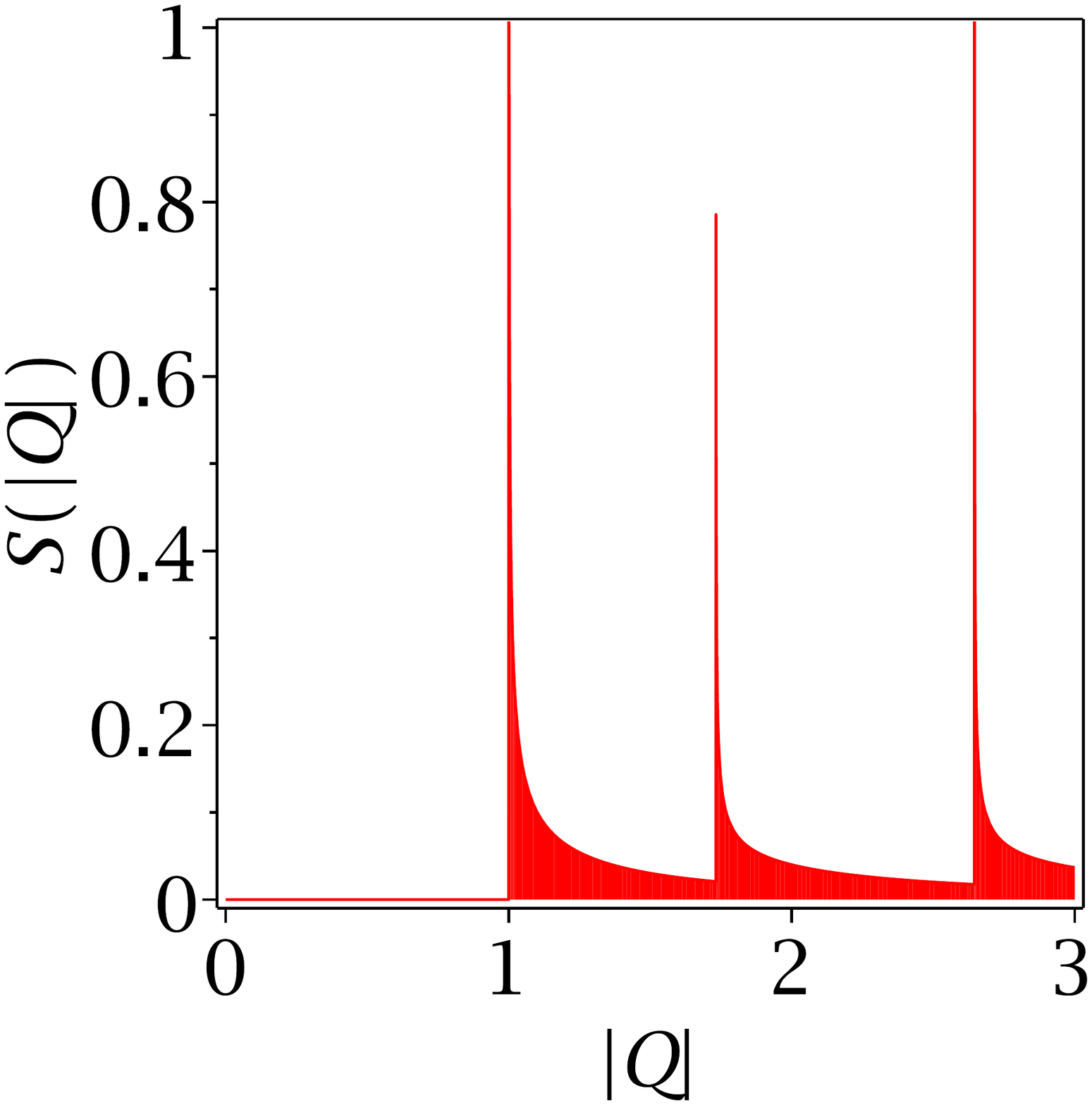}}
 \subfigure[\;$\sqrt{3}\times\sqrt{3}$ state\label{fig:struct_kag_sqrt3}]{
   \includegraphics[trim = 10mm 45mm 15mm 48mm, clip,width=.15\textwidth]{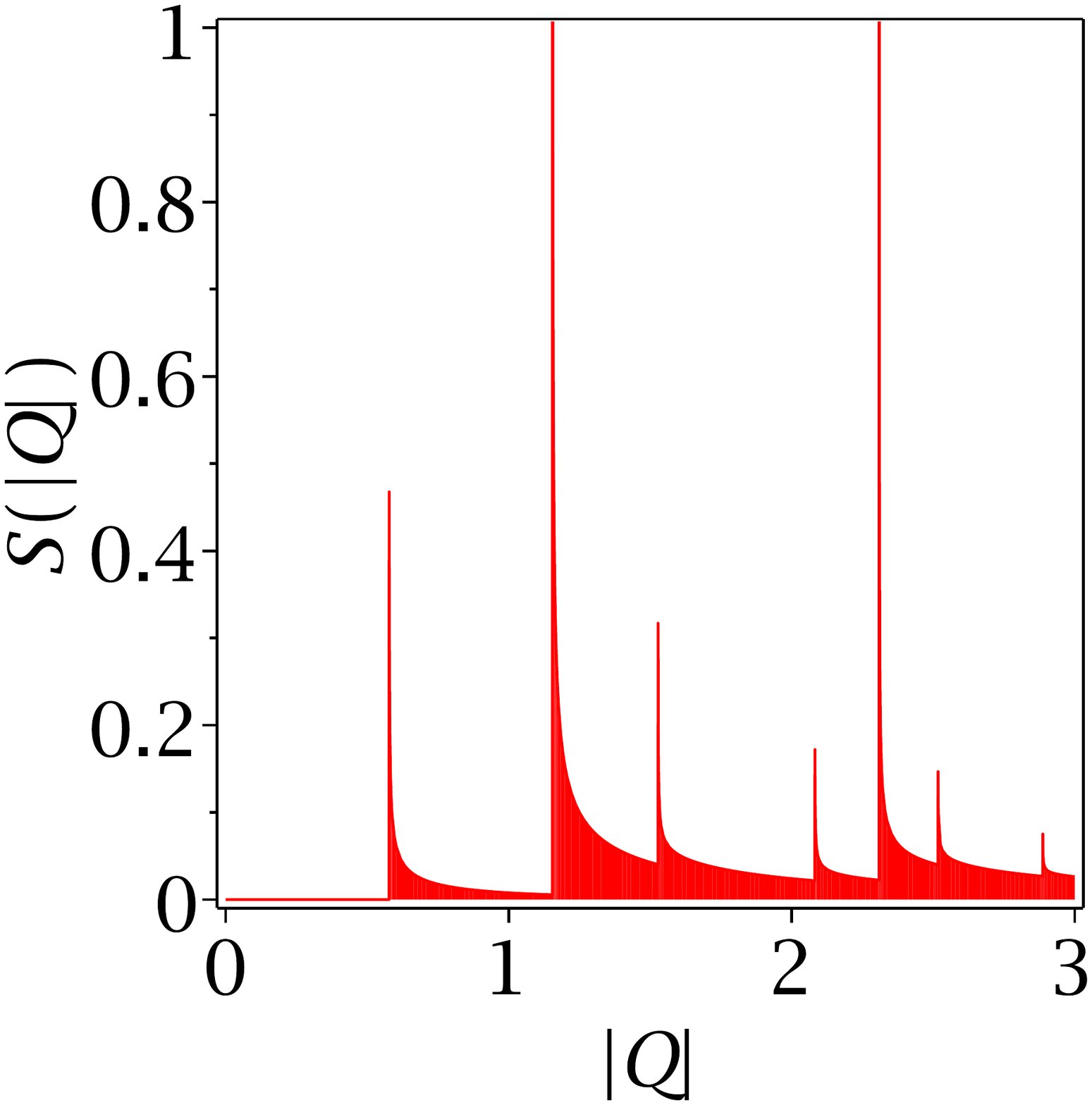}}
 \subfigure[\;Octahedral state\label{fig:struct_kag_6ssr}]{
   \includegraphics[trim = 10mm 45mm 15mm 48mm, clip,width=.15\textwidth]{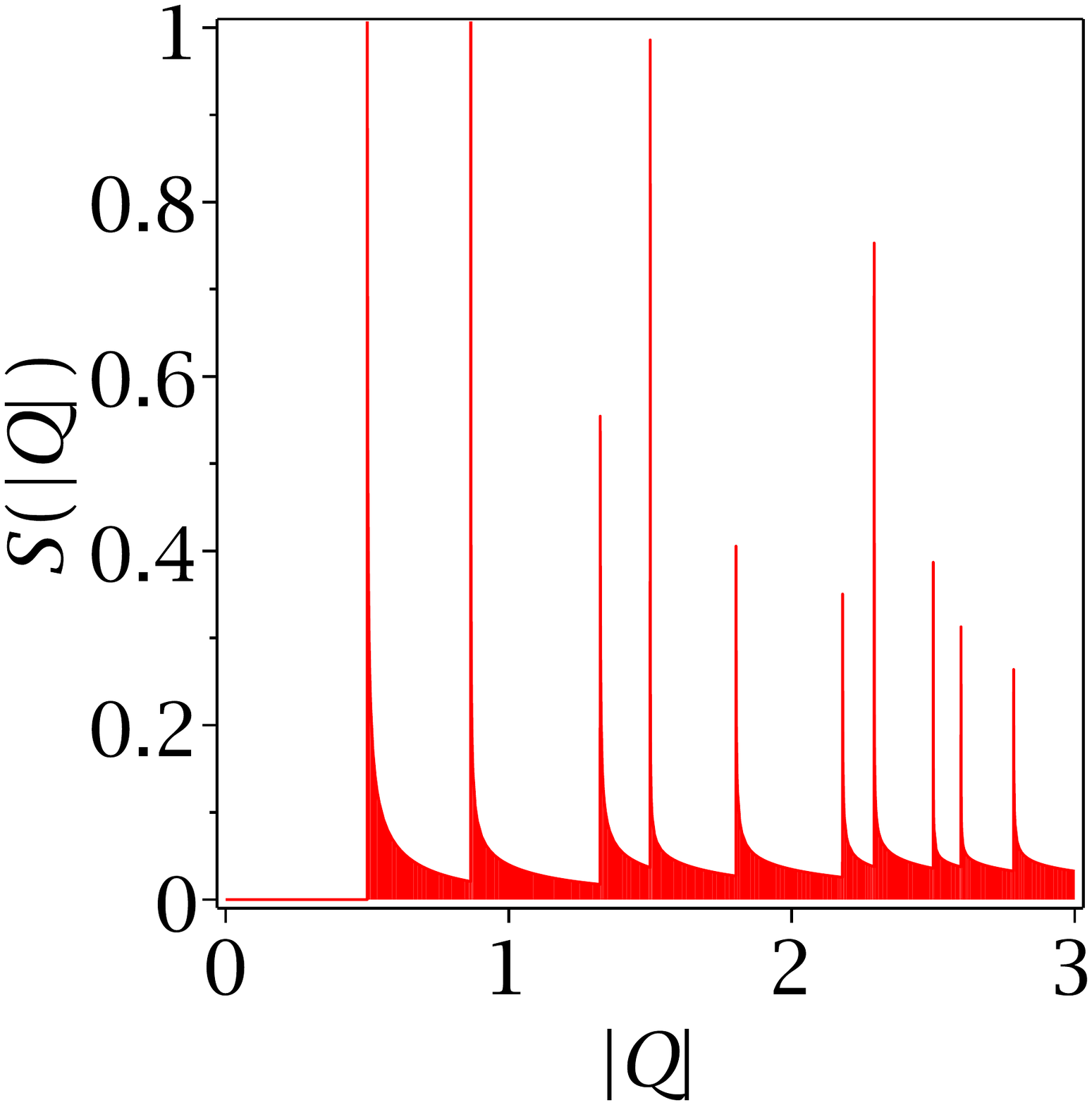}}
 \subfigure[\;Cuboc1 state\label{fig:struct_kag_12ssr}]{
   \includegraphics[trim = 10mm 45mm 15mm 48mm, clip,width=.15\textwidth]{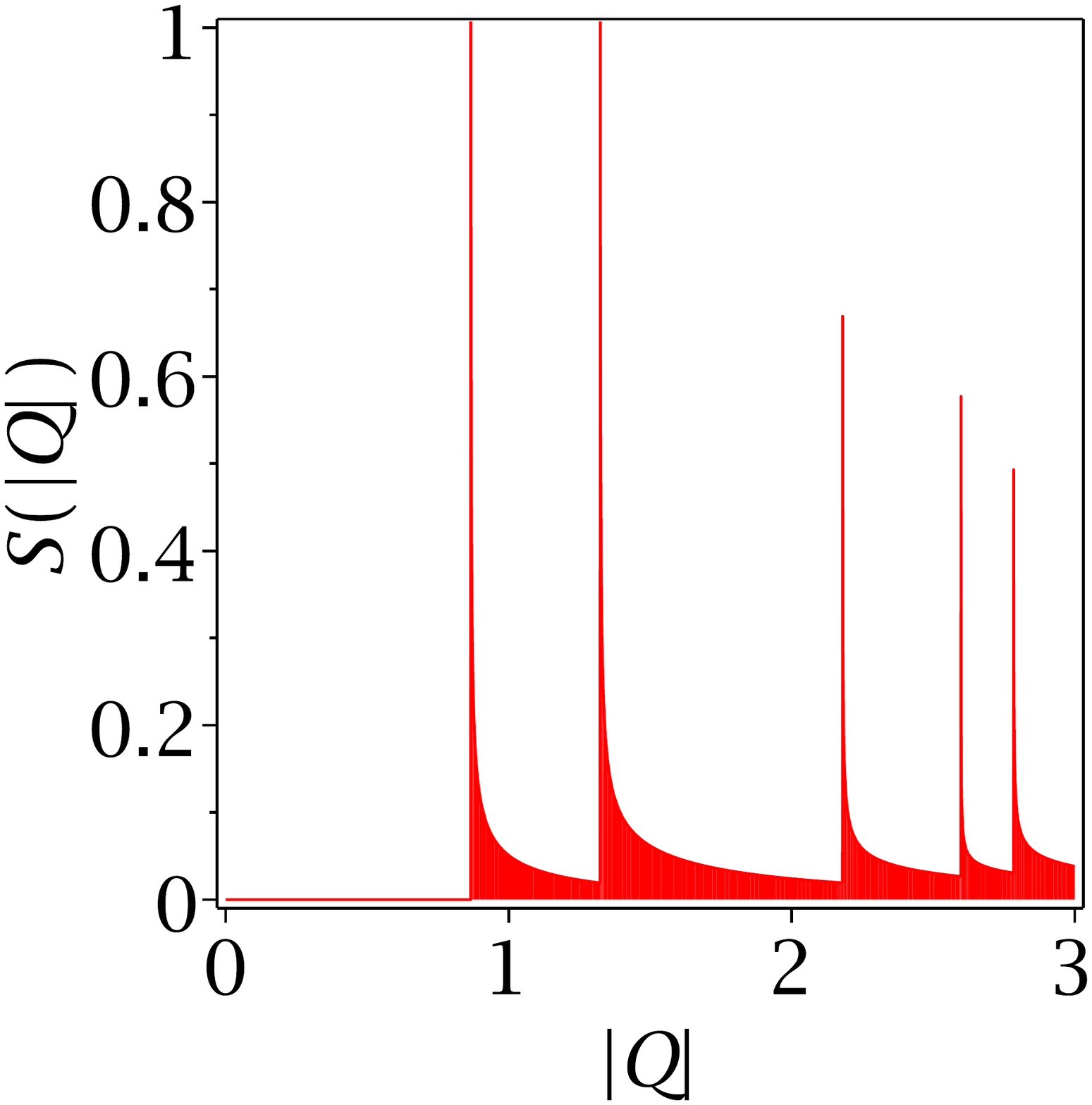}}
 \subfigure[\;Cuboc2 state\label{fig:struct_kag_12ssrJCD}]{
   \includegraphics[trim = 10mm 45mm 15mm 48mm, clip,width=.15\textwidth]{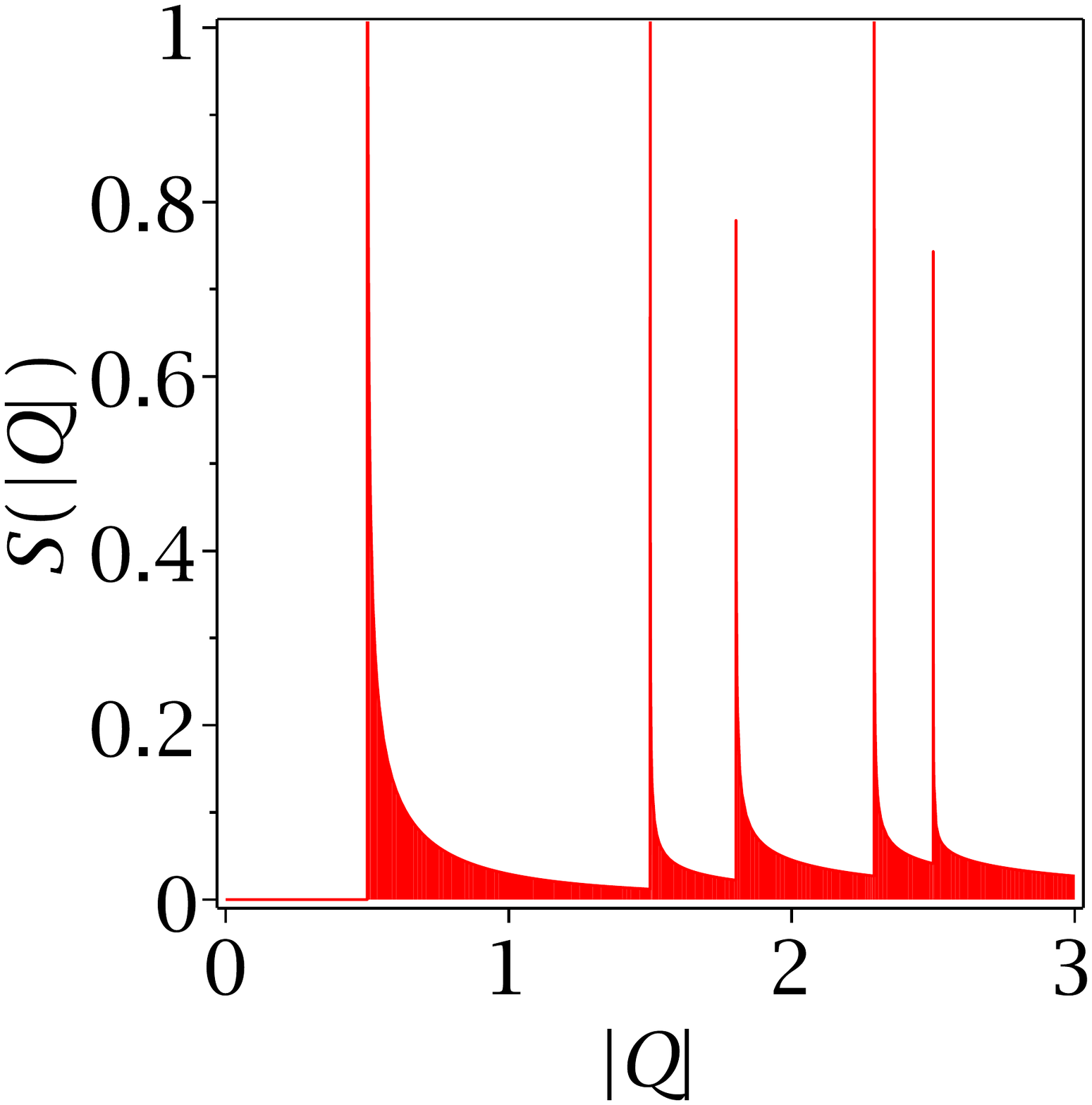}}
 \caption{Powder-averaged equal time structure factors $S(|\mathbf Q|)$ of the regular states on the kagome lattice ($|\mathbf Q|$ is in units of $2\pi$, $S(|\mathbf Q|)$ in arbitrary units). }
 \label{fig:struct_kag}
\end{center}
\end{figure}

The equal time structure factors $S(\mathbf{Q})$ were given in Fig.~\ref{fig:reg_tri}, \ref{fig:reg_kag}, \ref{fig:reg_hexa} and \ref{fig:reg_carre} for the regular states on the triangular, kagome, honeycomb and square lattices. The powder-averaged equal time structure factors $S(|\mathbf{Q}|)$ on the triangular and kagome lattices are shown in Fig.~\ref{fig:struct_tri} and \ref{fig:struct_kag}.

\section{Analogy with Wen's Projective symmetry groups (quantum spin models)}
\label{App:PSG_quantique}

For quantum spin-$\frac{1}{2}$ Heisenberg models, a standard mean-field approximation consists in expressing the spin operators  in term of fermionic operators $f_{i\alpha}$, where $i$ is a lattice site and $\alpha=\uparrow, \downarrow$ is the spin $\pm1/2$.
A mean-field decoupling based on some bond parameters $\eta_{ij}$ and $\xi_{ij}$ (notations and details to be found in in Ref.~\onlinecite{Wen_PSG}) can then be performed to make the Hamiltonian quadratic in the fermionic operators.

This theory has a local $SU(2)$ gauge invariance. The set of gauge transformations is denoted by $\Phi$.
Physical quantities, which can be expressed using spin operators, are unaffected by a gauge transformation, although $\eta_{ij}$ and $\xi_{ij}$ are generally modified.
A mean-field state is characterized by a set of $\eta_{ij}$ and $\xi_{ij}$ values, called  Ansatz.
Two mean-field states do have the same physical observables if they are related by a gauge transformation.
The group of transformations (lattice, gauge and combined transformations) that do not modify an Ansatz is called the projective symmetry group (PSG).
Its subgroup of pure gauge transformations is called the invariance gauge group (IGG).\cite{Wen_PSG}

One may be interested in states for which all the physical quantities are invariant under the lattice symmetries. To classify these ``uniform''  states, one can first fix the IGG and then
look for the ``algebraic'' PSG which obey the constraints  derived from  the algebraic structure of lattice symmetry group $S_L$.\cite{Wen_PSG}
The actual Ans\"atze can then be constructed.

Clearly, there is a close correspondence between the construction of regular states discussed in this paper, and that of symmetric Ans\"atze. This correspondence is summarized in Tab.~\ref{tab:reg_quantum}.

\begin{table}[h]
\renewcommand{\arraystretch}{2.1}
\begin{center}
 \begin{tabular}{|c||c|c|}
\hline
& \parbox{.15\textwidth}{Classical spin models}
& \parbox{.15\textwidth}{Quantum Mean-field} \\
\hline
\hline
State
& Regular state
&   \parbox{.15\textwidth}{Physically symmetric Ansatz}
\\
\hline
\parbox{.15\textwidth}{Internal symmetry group}
& \parbox{.15\textwidth}{$S_S$ (global spin rotation, etc.)}
&\parbox{.15\textwidth}{ $\Phi$ (local gauge transformations)} \\
\hline
\parbox{.15\textwidth}{Symmetry group\\ of a state}
& $H_c$
& PSG \\
\hline
\parbox{.15\textwidth}{Unbroken internal symmetries}
& $H_c^S$
& IGG\\
\hline
 \end{tabular}
\caption{\label{tab:reg_quantum}Analogy between the construction of regular states and that of symmetric Ans\"atze in \onlinecite{Wen_PSG}.}
\end{center}
\end{table}

\bibliographystyle{apsrev}

\end{document}